\documentclass[10pt, twocolumn]{revtex4}
\usepackage[dvipdfmx]{graphicx}
\usepackage{amsmath}
\usepackage{amssymb}

\usepackage{dcolumn}
\usepackage{bm}

\usepackage{color}

\newcommand{\mcal}[1]{\mathcal{#1}}
\newcommand{\mb}[1]{\mathbb{#1}}

\usepackage{hyperref}
\usepackage[normalem]{ulem}

\begin{document}

\preprint{APS/123-QED}

\title{Theory for Optimal Estimation and Control under Resource Limitations\\ and Its Applications to Biological Information Processing and Decision-Making}


\author{Takehiro Tottori$^{1,2}$}
\author{Tetsuya J. Kobayashi$^{2}$}
\affiliation{
$^{1}$Laboratory for Neural Computation and Adaptation, RIKEN Center for Brain Science, 2-1 Hirosawa, Wako, Saitama 351-0198, Japan\\
$^{2}$Institute of Industrial Science, The University of Tokyo, 4-6-1 Komaba, Meguro, Tokyo 153-8505, Japan
}

\date{\today}

\begin{abstract}
Despite being optimized, the information processing of biological organisms exhibits significant variability in its complexity and capability. 
One potential source of this diversity is the limitation of resources required for information processing. 
However, we lack a theoretical framework that comprehends the relationship between biological information processing and resource limitations and integrates it with decision-making conduced downstream of the information processing. 
In this paper, we propose a novel optimal estimation and control theory that accounts for the resource limitations inherent in biological systems. 
This theory explicitly formulates the memory that organisms can store and operate and obtains optimal memory dynamics using optimal control theory. 
This approach takes account of various resource limitations, such as memory capacity, intrinsic noise, and energy cost, and unifies state estimation and control. 
We apply this theory to minimal models of biological information processing and decision-making under resource limitations and find that such limitations induce discontinuous and non-monotonic phase transitions between memory-less and memory-based strategies. 
Therefore, this theory establishes a comprehensive framework for addressing biological information processing and decision-making under resource limitations, revealing the rich and complex behaviors that arise from resource limitations.
\end{abstract}

\maketitle

\section{Introduction}\label{sec: intro}
 \begin{figure}
 	\includegraphics[width=90mm]{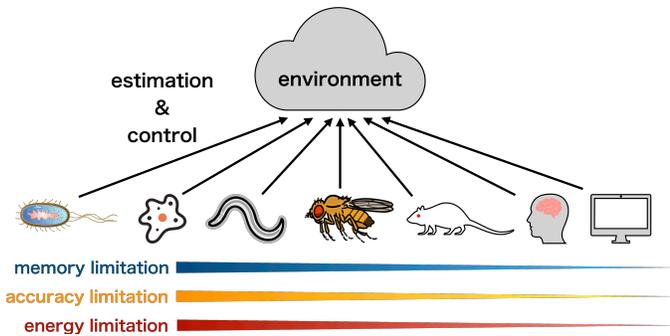}
 	\caption{\label{fig: evolution}
 	Schematic diagram of diverse information processing and resource limitations. 
	Resource limitations become more pronounced in primitive organisms such as bacteria. 
	A part of this figure is from TogoTV ($\copyright$ 2016 DBCLS TogoTV, CC-BY-4.0\href{https://creativecommons.org/licenses/by/4.0/}{https://creativecommons.org/licenses/by/4.0/}).
 	}
\end{figure}

A wide variety of organisms adapt to dynamic environments through sensory information processing. 
Recent studies have demonstrated that the information processing is nearly optimal not only in humans \cite{ernst_humans_2002,weiss_motion_2002,kording_bayesian_2004,stocker_noise_2006,burge_statistical_2008} but also in simpler organisms such as bacteria \cite{bialek_physical_2005,celani_bacterial_2010,mattingly_Escherichia_2021,nakamura_connection_2021,nakamura_optimal_2022}. 
Despite being optimized, the information processing of organisms is highly diverse, as evidenced by the significant differences between the information processing of humans and that of bacteria. 
This raises the following fundamental question: What causes the diversity of biological information processing?
One potential cause of this diversity could be the limitations of resources available for information processing in organisms. 
While human brains consume a substantial amount of energy and process information through billions of neurons \cite{laughlin_metabolic_1998,laughlin_energy_2001,lennie_cost_2003,niven_energy_2008,rolfe_cellular_1997,ames_cns_2000,aiello_expensive-tissue_1995,isler_metabolic_2006,fonseca-azevedo_metabolic_2012,balasubramanian_brain_2021}, bacteria have access to extremely limited energy and rely on chemical reactions involving only a small number of molecules to process information \cite{elowitz_stochastic_2002,swain_intrinsic_2002,paulsson_summing_2004,lestas_fundamental_2010,taniguchi_quantifying_2010,shibata_noisy_2005,ueda_stochastic_2007,kobayashi_dynamics_2011,govern_optimal_2014,govern_energy_2014,lang_thermodynamics_2014,bryant_physical_2023,harvey_universal_2023,ohga_thermodynamic_2023}. 
The wide variation in resource limitations may provide a unifying perspective on the diversity observed in biological information processing (Fig. \ref{fig: evolution}). 

\begin{figure*}
 	\includegraphics[width=170mm]{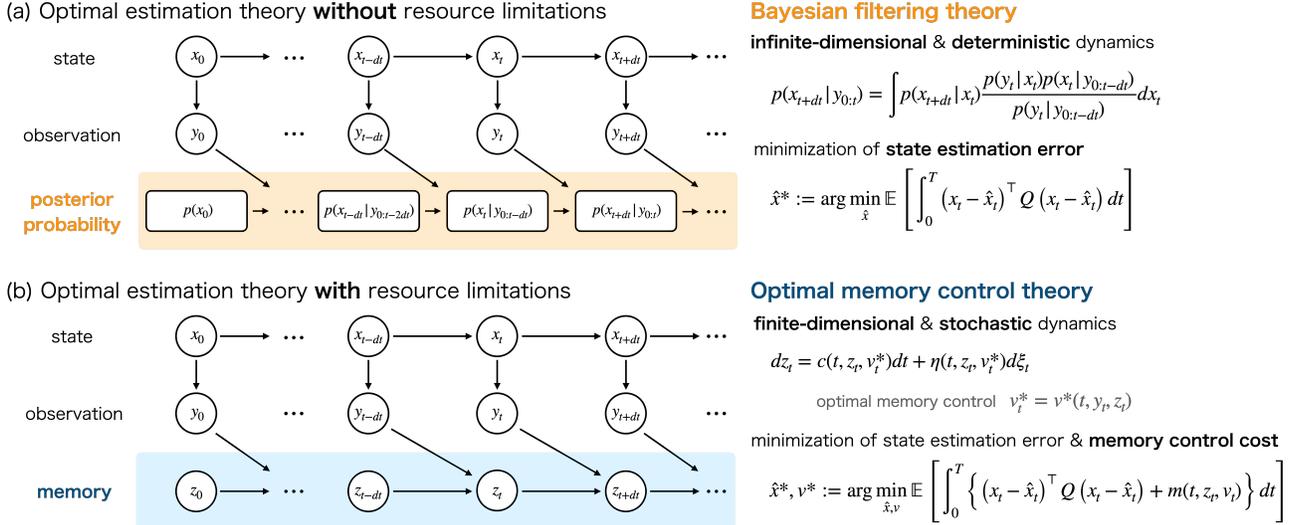}
 	\caption{\label{fig: BFTvsOCT}
 	Schematic diagrams of (a) the conventional optimal estimation theory without resource limitations and (b) the proposed optimal estimation theory with resource limitations. 
	(a) In the conventional Bayesian filtering theory, an agent compresses the information about past observation history $y_{0:t-dt}:=\{y_{0},...,y_{t-dt}\}$ into posterior probability $p(x_{t}|y_{0:t-dt})$ and estimates environmental state $x_{t}$ from current observation $y_{t}$ and posterior probability $p(x_{t}|y_{0:t-dt})$. 
	To obtain posterior probability $p(x_{t}|y_{0:t-dt})$, the agent needs to compute infinite-dimensional and deterministic Bayesian update. 
	In addition, optimal state estimator function $\hat{x}$ is determined by minimizing only the state estimation error. 
	These facts ignore the resource limitations of the agent. 
	(b) In the optimal memory control theory, the agent compresses the information about past observation history $y_{0:t-dt}$ into finite-dimensional memory $z_{t}$ and estimates environmental state $x_{t}$ from current observation $y_{t}$ and memory $z_{t}$. 
	The memory dynamics includes a noise process, $\xi_{t}$, which represents the intrinsic noise of the agent.  
	Furthermore, the memory dynamics is optimized by optimal memory control $v_{t}^{*}$ rather than by Bayes' theorem. 
	Optimal state estimator function $\hat{x}^{*}$ and optimal memory control function $v^{*}$ are determined by minimizing not only the state estimation error but also the memory control cost. 
	This problem formulation allows us to take into account the resource limitations of the agent. 
 	}
\end{figure*}

However, a comprehensive theoretical framework for investigating information processing under resource limitations has yet to be developed. 
Bayesian filtering theory is a well-established method for estimating dynamic environments from noisy observation histories and has numerous engineering applications (Fig. \ref{fig: BFTvsOCT}(a)) \cite{kallianpur_stochastic_1980,chen_bayesian_2003,jazwinski_stochastic_2007,bain_fundamentals_2009}. 
As exemplified by the Bayesian brain hypothesis \cite{pouget_information_2000,knill_bayesian_2004,doya_bayesian_2006,kording_bayesian_2006,pouget_probabilistic_2013,sohn_neural_2021,lange_bayesian_2023}, this theory is regarded as a fundamental principle of neural information processing, and there is a substantial body of research aimed at explaining brain functions from the perspective of Bayesian filtering \cite{deneve_optimal_2007,wilson_neural_2009,hiratani_redundancy_2018,hiratani_rapid_2020,aitchison_synaptic_2021,kutschireiter_bayesian_2023}. 
The approach used in the Bayesian brain hypothesis has recently applied to bacteria without brains to investigate the problem whether and how bacteria can perform Bayesian filtering through intracellular chemical reactions  \cite{andrews_optimal_2006,kobayashi_implementation_2010,kobayashi_dynamics_2011,hinczewski_cellular_2014,zechner_molecular_2016,mora_physical_2019,husain_kalman-like_2019,nakamura_connection_2021,novak_bayesian_2021,auconi_gradient_2022,rode_information_2024,nakamura_gradient_2024}. 
While Bayesian filtering provides statistically optimal estimations, it fails to take into account the resource limitations inherent in biological information processing. 
 
In particular, there are three key resource limitations that Bayesian filtering theory cannot capture. 
These limitations become especially inevitable in primitive organisms such as bacteria. 
The first limitation is memory capacity. 
To perform Bayesian filtering, organisms need to store posterior probabilities. 
However, this task is challenging for organisms with limited memory capacity, as the posterior probabilities become infinite-dimensional even when the environmental states are finite-dimensional. 
While brains can approximate infinite-dimensional posterior probabilities using numerous neurons \cite{pouget_information_2000,knill_bayesian_2004,pouget_probabilistic_2013,zemel_probabilistic_1998,ma_bayesian_2006}, it is far more challenging for bacteria to represent these probabilities with only a small number of molecules. 
In fact, previous studies implementing Bayesian filtering through intracellular chemical reactions are limited to special cases where the infinite-dimensional posterior probabilities can be reduced to finite-dimensional sufficient statistics to circumvent this issue \cite{andrews_optimal_2006,kobayashi_implementation_2010,kobayashi_dynamics_2011,hinczewski_cellular_2014,zechner_molecular_2016,mora_physical_2019,husain_kalman-like_2019,nakamura_connection_2021,novak_bayesian_2021,auconi_gradient_2022,rode_information_2024,nakamura_gradient_2024}. 

The second limitation is the accuracy of information processing.  
Biological information processing based on either neural firing or intracellular chemical reactions often involves high levels of stochasticity \cite{elowitz_stochastic_2002,swain_intrinsic_2002,paulsson_summing_2004,lestas_fundamental_2010,taniguchi_quantifying_2010,shibata_noisy_2005,ueda_stochastic_2007,kobayashi_dynamics_2011,faisal_noise_2008,ribrault_stochasticity_2011,mcdonnell_benefits_2011}, making it difficult to accurately compute the Bayesian update of posterior probabilities. 
Enhancing the accuracy of biological information processing requires increasing the number of neurons or molecules involved in Bayesian filtering, which in turn demands greater resource investment. 
In previous studies, while extrinsic noise arising from environments and observations has been considered, intrinsic noise within internal systems performing Bayesian filtering has often been neglected \cite{andrews_optimal_2006,kobayashi_implementation_2010,kobayashi_dynamics_2011,hinczewski_cellular_2014,zechner_molecular_2016,mora_physical_2019,husain_kalman-like_2019,nakamura_connection_2021,novak_bayesian_2021,auconi_gradient_2022,rode_information_2024,nakamura_gradient_2024}. 
However, this assumption is often unrealistic, particularly in the context of cellular information processing. 

The third limitation is the cost associated with information processing. 
Since biological information processing is realized through physical processes such as neural firing and chemical reactions, it inevitably requires costs \cite{govern_optimal_2014,govern_energy_2014,lang_thermodynamics_2014,bryant_physical_2023,harvey_universal_2023,ohga_thermodynamic_2023,laughlin_metabolic_1998,laughlin_energy_2001,lennie_cost_2003,niven_energy_2008,rolfe_cellular_1997,ames_cns_2000,aiello_expensive-tissue_1995,isler_metabolic_2006,fonseca-azevedo_metabolic_2012,balasubramanian_brain_2021}. 
Experimental evolution studies indicate that excessive investment of energy or other costs in information processing can reduce the fitness of organisms \cite{dukas_costs_1999,mery_fitness_2003,mery_cost_2005,burger_learning_2008,kolss_reduced_2008,burns_costs_2010}. 
Therefore, it is crucial for organisms not only to minimize estimation errors, but also to reduce cost the required for these computations. 
In fact, some studies have discussed the trade-off between estimation errors and the costs implicitly via the information bottleneck method \cite{tishby_information_2000,chechik_information_2005,creutzig_past-future_2009,fox_minimum-information_2016,fox_minimum-information_2016-1,sachdeva_optimal_2021,tjalma_trade-offs_2023,bauer_information_2023}. 
Also other studies considered the energy costs from the viewpoints of thermodynamics \cite{govern_optimal_2014,govern_energy_2014,lang_thermodynamics_2014,bryant_physical_2023,harvey_universal_2023,ohga_thermodynamic_2023}.
However, their connections to Bayesian filtering theory are still elusive. 

In summary, Bayesian filtering theory fails to consider resource limitations inherent in biological systems, such as memory capacity, intrinsic noise, and energy cost. 
This highlights the need for a new theoretical framework to elucidate the relationship between information processing and resource limitations in biological systems.

In addition to estimating environmental states, organisms must control their behaviors based on these estimations. 
The decision-making of organisms has been extensively studied using optimal control theory \cite{todorov_optimal_2002,todorov_optimality_2004,todorov_stochastic_2005,kromer_decision_2018,calascibetta_optimal_2023}. 
The combination of Bayesian filtering and optimal control is one approach to understanding decision-making under uncertain observations \cite{wonham_separation_1968,davis_dynamic_1973,bensoussan_stochastic_1992,yong_stochastic_1999,nisio_stochastic_2015,bensoussan_estimation_2018,wang_introduction_2018}. 
However, this combination complicates the optimization process, resulting in studies being limited to simple cases \cite{li_iterative_2006,li_iterative_2007,nakamura_optimal_2022,verano_olfactory_2023}. 
Therefore, there is also a need for a new theoretical framework that integrates estimation and control in a more comprehensive manner.

To address the issues outlined above, we propose a new optimal estimation theory that takes into account resource limitations (Fig. \ref{fig: BFTvsOCT}(b)). 
While this theory builds upon our previously proposed framework \cite{tottori_memory-limited_2022,tottori_forward-backward_2023,tottori_decentralized_2023}, the formulation is slightly modified for biological plausibility. 
This theory explicitly formulates finite-dimensional and stochastic memories that organisms can store and operate, and optimizes the embedding of information from past observation histories into these memories using optimal control theory. 
This approach accepts a more general objective function that includes not only estimation errors but also energy costs. 
Therefore, this theory achieves optimal estimations under resource limitations by replacing Bayesian filtering of posterior probabilities with the optimal control of memories. 

We apply this theory to a minimal model of biological information processing and investigate optimal estimation strategies under resource limitations. 
Although our minimal model considers a simple optimal estimation problem with linear Gaussian dynamics and quadratic costs, it exhibits discontinuous and non-monotonic phase transitions between memory-less and memory-based estimation strategies depending on available resources. 
Such phase transitions do not occur in the conventional Bayesian filtering that disregards resource limitations. 
Therefore, our analysis suggests that resource limitations may drive a punctuated and complex evolution of biological information processing, leading to qualitatively diverse behaviors across organisms. 

This theory also provides a unified framework for both state estimation and control by utilizing memory control instead of Bayesian filtering. 
We demonstrate the effectiveness of this approach by applying it to a minimal model of biological information processing and decision-making. 
This model also exhibits discontinuous phase transitions between memory-less and memory-based control strategies. 
Yet, non-monotonic phase transitions do not appear with respect to sensory noise, while it occurs with respect to environmental uncertainty. 
These findings suggest that while non-monotonicity may depend on specific models and parameters, discontinuity is a fundamental characteristic under resource limitations. 

This paper is organized as follows: 
In Sec. \ref{sec: CTOE}, we briefly review Bayesian filtering theory and explain how this theory fails to capture the resource limitations inherent in biological systems. 
In Sec. \ref{sec: NTOE}, we introduce a new optimal estimation theory that can take into account these resource limitations. 
In Sec. \ref{sec: ABIP}, we apply this theory to a minimal model of biological information processing and investigate optimal estimation strategies under resource limitations. 
In Sec. \ref{sec: NTOEC}, we extend the optimal estimation theory to incorporate optimal control. 
In Sec. \ref{sec: ABDM}, we apply this extended theory to a minimal model of biological information processing and decision-making. 
In Sec. \ref{sec: discussion}, we discuss future directions and extensions of our work. 

In Appendix \ref{sec: OSC} and \ref{sec: OMSC}, we derive the optimal solutions for the optimal estimation and control problem under resource limitations. 
Specifically, Appendix \ref{sec: OSC} and \ref{sec: OMSC} addresses the cases without and with memory, respectively. 
In Appendix \ref{sec: LQG}, we discuss the Linear-Quadratic-Gaussian (LQG) problem in this theory. 
In Appendix \ref{sec: ABIPA} and  \ref{sec: ABDMA}, we further analyze the minimal models of biological information processing and decision-making. 
In Appendix \ref{sec: ML-POSC}, we compare this work with our previous works \cite{tottori_memory-limited_2022,tottori_forward-backward_2023,tottori_decentralized_2023}. 

\section{Optimal Estimation Theory without Resource Limitations}\label{sec: CTOE}
In this section, we briefly review the conventional optimal theory for estimating dynamic environments, Bayesian filtering theory (Fig. \ref{fig: BFTvsOCT}(a)) \cite{kallianpur_stochastic_1980,chen_bayesian_2003,jazwinski_stochastic_2007,bain_fundamentals_2009}. 
This theory considers an environment and an agent. 
The state of the environment at time $t\in[0,T]$ is denoted by $x_{t}\in\mb{R}^{d_{x}}$, which evolves by the following stochastic differential equation (SDE): 
\begin{align}
	dx_{t}=b(t,x_{t})dt+\sigma(t,x_{t})d\omega_{t}, 
	\label{eq: CTOE-state SDE}
\end{align}
where initial state $x_{0}$ follows $p_{0}(x_{0})$ and $\omega_{t}\in\mb{R}^{d_{\omega}}$ is a standard Wiener process. 
The agent cannot completely observe environmental state $x_{t}$. 
Instead, it obtains noisy observation $y_{t}\in\mb{R}^{d_{y}}$, which is generated from probability density function $p_{t}(y_{t}|x_{t})$ as follows: 
\begin{align}
	y_{t}\sim p_{t}(y_{t}|x_{t}). 
	\label{eq: CTOE-observation PDF}
\end{align}
The agent estimates environmental state $x_{t}$ from noisy observation history $y_{t},y_{t-dt},...,y_{0}$. 
The state estimator of the agent at time $t$ is generally given by
\begin{align}
	\hat{x}_{t}=\hat{x}(t,y_{t},y_{t-dt},...,y_{0}). 
	\label{eq: CTOE-state estimator}
\end{align}
The objective function the agent minimizes is the following state estimation error: 
\begin{align}
	J[\hat{x}]:=\mb{E}\left[\int_{0}^{T}\left(x_{t}-\hat{x}_{t}\right)^{\top}Q\left(x_{t}-\hat{x}_{t}\right)dt\right],
	\label{eq: CTOE-OF}
\end{align}
where $Q\succ O$. 
Bayesian filtering theory considers a problem to find optimal state estimator function $\hat{x}^{*}$ that minimizes state estimation error $J[\hat{x}]$: 
\begin{align}
	\hat{x}^{*}:=\arg\min_{\hat{x}}J[\hat{x}]. 
	\label{eq: CTOE-OP}
\end{align}

By solving Eq. (\ref{eq: CTOE-OP}), optimal state estimator function $\hat{x}^{*}$ is obtained by the conditional expectation of state $x_{t}$ given observation history $y_{t},y_{t-dt},...,y_{0}$: 
\begin{align}
	\hat{x}^{*}(t,y_{t},y_{t-dt},...,y_{0})=\mathbb{E}_{p(x_{t}|y_{t},y_{t-dt},...,y_{0})}\left[x_{t}\right]. 
	\label{eq: CTOE-optimal state estimator tmp}
\end{align}
In order to obtain optimal state estimator $\hat{x}_{t}^{*}=\hat{x}^{*}(t,y_{t},y_{t-dt},...,y_{0})$, the agent needs to memorize past observation history $y_{t-dt},...,y_{0}$ and it becomes harder and harder as time progresses. 
Bayesian filtering theory resolves this problem by compressing past observation history $y_{t-dt},...,y_{0}$ into the posterior probability density function 
\begin{align}
	p(x_{t}|y_{t-dt},...,y_{0})\in\mcal{P}[\mb{R}^{d_{x}}], 
	\label{eq: CTOE-posterior probability}
\end{align}
where $\mcal{P}[\mb{R}^{d_{x}}]$ indicates the space of a probability density function on $\mb{R}^{d_{x}}$. 
The optimal state estimator is given by 
\begin{align}
	\hat{x}_{t}^{*}&=\mathbb{E}_{p(x_{t}|y_{t},p(x_{t}|y_{t-dt},...,y_{0}))}\left[x_{t}\right], 
	\label{eq: CTOE-optimal state estimator}
\end{align}
which means that the agent estimates environmental state $x_{t}$ from observation $y_{t}$ and posterior probability density function $p(x_{t}|y_{t-dt},...,y_{0})$. 
This compression does not impair the performance because $p(x_{t}|y_{t-dt},...,y_{0})$ is the sufficient statistic of $y_{t-dt},...,y_{0}$ for estimating $x_{t}$. 
$p(x_{t}|y_{t-dt},...,y_{0})$ can be updated sequentially by using Bayes' theorem: 
\begin{align}
	&p(x_{t+dt}|y_{t},y_{t-dt},...,y_{0})=\nonumber\\
	&\int p(x_{t+dt}|x_{t})\frac{p(y_{t}|x_{t})p(x_{t}|y_{t-dt},...,y_{0})}{p(y_{t}|y_{t-dt},...,y_{0})}dx_{t}. 
	\label{eq: CTOE-Bayesian filtering theory}
\end{align}
Therefore,  in Bayesian filtering theory, the agent obtains optimal state estimator $\hat{x}_{t}^{*}$ [Eq. (\ref{eq: CTOE-optimal state estimator})] by computing the Bayesian update of the posterior probability [Eq. (\ref{eq: CTOE-Bayesian filtering theory})]. 

However, Bayesian filtering is an ideal theory for optimal estimation and overlooks the resource limitations inherent in organisms. 
This oversight prevents organisms from fully implementing Bayesian filtering. 
For example, organisms cannot completely store infinite-dimensional posterior probability $p(x_{t}|y_{t-dt},...,y_{0})$ because their memory capacities are finite \cite{andrews_optimal_2006,kobayashi_implementation_2010,kobayashi_dynamics_2011,hinczewski_cellular_2014,zechner_molecular_2016,mora_physical_2019,husain_kalman-like_2019,nakamura_connection_2021,novak_bayesian_2021,auconi_gradient_2022,rode_information_2024,nakamura_gradient_2024}. 
Additionally, organisms find it challenging to accurately compute the Bayesian update of the posterior probability [Eq. (\ref{eq: CTOE-Bayesian filtering theory})] due to the intrinsic noise in their information processing \cite{elowitz_stochastic_2002,swain_intrinsic_2002,paulsson_summing_2004,lestas_fundamental_2010,taniguchi_quantifying_2010,shibata_noisy_2005,ueda_stochastic_2007,kobayashi_dynamics_2011,faisal_noise_2008,ribrault_stochasticity_2011,mcdonnell_benefits_2011}. 
Furthermore, organisms need to minimize not only the state estimation error but also the energy consumption associated with the estimation process, as their available energy are severely limited \cite{govern_optimal_2014,govern_energy_2014,lang_thermodynamics_2014,bryant_physical_2023,harvey_universal_2023,ohga_thermodynamic_2023,laughlin_metabolic_1998,laughlin_energy_2001,lennie_cost_2003,niven_energy_2008,rolfe_cellular_1997,ames_cns_2000,aiello_expensive-tissue_1995,isler_metabolic_2006,fonseca-azevedo_metabolic_2012,balasubramanian_brain_2021}. 
These resource limitations become more pronounced and increasingly difficult to ignore in more primitive organisms such as bacteria. 
Therefore, a new optimal estimation theory that takes into account these limitations is necessary to understand the estimation strategies of organisms. 

\section{Optimal Estimation Theory with Resource Limitations}\label{sec: NTOE}
In this section, we present a novel optimal estimation theory that takes into account the resource limitations inherent in biological systems (Fig. \ref{fig: BFTvsOCT}(b)). 
This theory builds upon our previous works \cite{tottori_memory-limited_2022,tottori_forward-backward_2023,tottori_decentralized_2023}, but the problem formulation is slightly modified, which allows us to discuss biologically plausible situations (See Appendix \ref{sec: ML-POSC}). 
In the following, we first formulate the optimal estimation problem with resource limitations in Sec. \ref{subsec: NTOE-PF} and then present its optimal solution in Sec. \ref{subsec: NTOE-OS}. 

\subsection{Problem Formulation}\label{subsec: NTOE-PF}
In this subsection, we formulate the optimal estimation problem with resource limitations. 
The state $x_{t}$ and observation $y_{t}$ are identical to those in the Bayesian filtering problem, which are given by Eqs. (\ref{eq: CTOE-state SDE}) and (\ref{eq: CTOE-observation PDF}), respectively. 
In contrast, this problem explicitly formulates the finite-dimensional memory that the agent can store and operate,
\begin{align}
	z_{t}\in\mb{R}^{d_{z}},
\end{align}
where the dimension of the memory, $d_{z}$, is determined by the available memory size of the agent. 
The agent determines state estimator $\hat{x}_{t}$ based on observation $y_{t}$ and memory $z_{t}$ as follows: 
\begin{align}
	\hat{x}_{t}=\hat{x}(t,y_{t},z_{t}). 
	\label{eq: NTOE-state estimator}
\end{align}
Compared to Eq. (\ref{eq: CTOE-state estimator}), infinite-dimensional past observation history $y_{t-dt},...,y_{0}$ is replaced with finite-dimensional memory $z_{t}$, which allows for the implementation in organisms with finite memory capacities. 

Memory $z_{t}$ is assumed to evolve by the following SDE: 
\begin{align}
	dz_{t}=c(t,z_{t},v_{t})dt+\eta(t,z_{t},v_{t})d\xi_{t}, 
	\label{eq: NTOE-memory SDE}
\end{align}
where initial memory $z_{0}$ follows $p_{0}(z_{0})$, $v_{t}\in\mb{R}^{d_{v}}$ is the memory control of the agent, and $\xi_{t}\in\mb{R}^{d_{\xi}}$ is a standard Wiener process. 
The agent determines memory control $v_{t}$ based on observation $y_{t}$ and memory $z_{t}$: 
\begin{align}
	v_{t}=v(t,y_{t},z_{t}). 
	\label{eq: NTOE-memory control}
\end{align}
While the dynamics of the posterior probability [Eq. (\ref{eq: CTOE-Bayesian filtering theory})] is deterministic given observation $y_{t}$, that of the memory [Eq. (\ref{eq: NTOE-memory SDE})] is stochastic due to the intrinsic noise, $\xi_{t}$. 

A key challenge in this problem lies in how to optimally compress the information about past observation history $y_{t-dt},...,y_{0}$ into memory $z_{t}$. 
Bayesian filtering theory optimally compresses the past observation information into infinite-dimensional and deterministic posterior probability $p(x_{t}|y_{t-dt},...,y_{0})$ using Bayes' theorem. 
However, this optimization method cannot be applied to memory $z_{t}$ because it is finite-dimensional and stochastic. 
We address this issue by using optimal control theory: 
The agent optimally compresses the past observation information into memory $z_{t}$ by optimizing memory control $v_{t}$. 
This idea leads to the simultaneous optimization of state estimator function $\hat{x}$ and memory control function $v$ as follows: 
\begin{align}
	\hat{x}^{*},v^{*}:=\arg\min_{\hat{x},v}J[\hat{x},v].  
	\label{eq: NTOE-OP}
\end{align}
Objective function $J[\hat{x},v]$ is given by the following expected cumulative cost function: 
\begin{align}
	J[\hat{x},v]:=
	\mb{E}\left[\int_{0}^{T}f(t,x_{t},z_{t},\hat{x}_{t},v_{t})dt+g(x_{T},z_{T})\right], 
	\label{eq: NTOE-OF}
\end{align}
where $f(t,x_{t},z_{t},\hat{x}_{t},v_{t})$ and $g(x_{T},z_{T})$ are the running and terminal cost functions, respectively. 

For comparison with Bayesian filtering theory, we consider the following cost functions as a special case:  
\begin{align}
	&f:=\left(x_{t}-\hat{x}_{t}\right)^{\top}Q\left(x_{t}-\hat{x}_{t}\right)+m(t,z_{t},v_{t}),\label{eq: NTOE-OF-f}\\
	&g:=0,\label{eq: NTOE-OF-g}
\end{align}
where $m(t,z_{t},v_{t})$ is the cost function for memory $z_{t}$ and memory control $v_{t}$. 
These cost functions indicate that our theory can minimize not only the state estimation error (the first term of Eq. (\ref{eq: NTOE-OF-f})) but also the energy cost associated with the estimation process (the second term of Eq. (\ref{eq: NTOE-OF-f})). 

In summary, this theory achieves optimal estimation under resource limitations, including memory capacity, intrinsic noise, and energy cost, by replacing the Bayesian filtering of the posterior probability with the optimal control of the memory.

\subsection{Optimal Solution}\label{subsec: NTOE-OS}
The optimal estimation problem with resource limitations can be solved by using Pontryagin's minimum principle on the probability density function space, which has been recently employed in mean-field stochastic optimal control problems \cite{bensoussan_master_2015,bensoussan_interpretation_2017,carmona_probabilistic_2018,carmona_probabilistic_2018-1} and our previous works \cite{tottori_memory-limited_2022,tottori_forward-backward_2023,tottori_decentralized_2023}. 
In this subsection, we skip the derivation and only show the main result. 
The details of the derivation are shown in Appendix \ref{sec: OSC} and \ref{sec: OMSC}. 

Optimal state estimator function $\hat{x}^{*}$ and optimal memory control function $v^{*}$ are obtained by minimizing the conditional expected Hamiltonian: 
\begin{align}
	&\hat{x}^{*}(t,y,z),v^{*}(t,y,z)=\nonumber\\
	&\arg\min_{\hat{x},v}\mb{E}_{p_{t}(x|y,z)}\left[\mcal{H}(t,x,z,\hat{x},v,w)\right]. 
	\label{eq: NTOE-optimal control function}
\end{align}
Hamiltonian $\mcal{H}$ is defined by 
\begin{align}
	\mcal{H}(t,x,z,\hat{x},v,w):=f(t,x,z,\hat{x},v)+\mcal{L}_{v}w(t,x,z), 
\end{align}
where $f(t,x,z,\hat{x},v)$ is the current cost and $\mcal{L}_{v}w(t,x,z)$ represents the future cost. 
$\mcal{L}_{v}$ is the time-backward diffusion operator for state $x$ and memory $z$, which is defined by
\begin{align}
	&\mcal{L}_{v}w(t,x,z):=\sum_{i=1}^{d_{x}}b_{i}(t,x)\frac{\partial w(t,x,z)}{\partial x_{i}}\nonumber\\
	&+\sum_{i=1}^{d_{z}}c_{i}(t,z,v)\frac{\partial w(t,x,z)}{\partial z_{i}}\nonumber\\
	&+\frac{1}{2}\sum_{i,j=1}^{d_{x}}\sigma_{i}(t,x)\sigma_{j}(t,x)\frac{\partial^{2} w(t,x,z)}{\partial x_{i}\partial x_{j}}\nonumber\\
	&+\frac{1}{2}\sum_{i,j=1}^{d_{z}}\eta_{i}(t,z,v)\eta_{j}(t,z,v)\frac{\partial^{2} w(t,x,z)}{\partial z_{i}\partial z_{j}}, 
\end{align}
and $w(t,x,z)$ is the value function, which represents the expected cumulative cost starting from time $t$ with state $x$ and memory $z$ to terminal time $T$. 
Value function $w(t,x,z)$ is obtained by solving the following time-backward partial differential equation, Hamilton-Jacobi-Bellman (HJB) equation: 
\begin{align}
	-\frac{\partial w(t,x,z)}{\partial t}=\mb{E}_{p_{t}(y|x)}\left[\mcal{H}(t,x,z,\hat{x}^{*},v^{*},w)\right], 
	\label{eq: NTOE-HJB eq}
\end{align}
where the terminal condition is given by $w(T,x,z)=g(x,z)$. 
In the optimal solution [Eq. (\ref{eq: NTOE-optimal control function})], conditional expectation $\mb{E}_{p_{t}(x|y,z)}\left[\cdot\right]$ appears, which is consistent with the fact that the agent needs to estimate state $x$ from observation $y$ and memory $z$. 
Conditional probability density function $p_{t}(x|y,z)$ is calculated as follows: 
\begin{align}
	p_{t}(x|y,z)=\frac{p_{t}(y|x)p_{t}(x,z)}{\int p_{t}(y|x)p_{t}(x,z)dx}.
\end{align}
Probability density function $p_{t}(x,z):=p(t,x,z)$ is obtained by solving the following time-forward partial differential equation, Fokker-Planck (FP) equation: 
\begin{align}
	\frac{\partial p(t,x,z)}{\partial t}=\check{\mcal{L}}_{v^{*}}^{\dag}p(t,x,z), 
	\label{eq: NTOE-FP eq}
\end{align}
where the initial condition is given by $p(0,x,z)=p_{0}(x)p_{0}(z)$. 
$\check{\mcal{L}}_{v}^{\dag}$ is the expected time-forward diffusion operator for state $x$ and memory $z$, which is defined by
\begin{align}
	&\check{\mcal{L}}_{v}^{\dag}p(t,x,z):=-\sum_{i=1}^{d_{x}}\frac{\partial b_{i}(t,x)p(t,x,z)}{\partial x_{i}}\nonumber\\
	&-\sum_{i=1}^{d_{z}}\frac{\partial \mb{E}_{p_{t}(y|x)}\left[c_{i}(t,z,v)\right]p(t,x,z)}{\partial z_{i}}\nonumber\\
	&+\frac{1}{2}\sum_{i,j=1}^{d_{x}}\frac{\partial^{2} \sigma_{i}(t,x)\sigma_{j}(t,x)p(t,x,z)}{\partial x_{i}\partial x_{j}}\nonumber\\
	&+\frac{1}{2}\sum_{i,j=1}^{d_{z}}\frac{\partial^{2} \mb{E}_{p_{t}(y|x)}\left[\eta_{i}(t,z,v)\eta_{j}(t,z,v)\right]p(t,x,z)}{\partial z_{i}\partial z_{j}}, 
\end{align}
where $\mb{E}_{p_{t}(y|x)}\left[\cdot\right]$ is necessary because memory control $v$ depends on observation $y$. 
Therefore, optimal state estimator function $\hat{x}^{*}$ and optimal memory control function $v^{*}$ are obtained by jointly solving HJB equation [Eq. (\ref{eq: NTOE-HJB eq})] and FP equation [Eq. (\ref{eq: NTOE-FP eq})]. 
 
It should be noted that organisms do not need to solve HJB and FP equations [Eqs. (\ref{eq: NTOE-HJB eq}) and (\ref{eq: NTOE-FP eq})] because they may obtain optimal solutions $\hat{x}^{*}$ and $v^{*}$ as a result of evolution. 
Therefore, it is not necessary to consider the resource limitations of organisms for the computation of optimal solutions $\hat{x}^{*}$ and $v^{*}$. 
In contrast, it is necessary to consider their resource limitations associated with the dynamics of memory $z_{t}$ because memory $z_{t}$ depends on real-time observation $y_{t}$ such as olfactory, auditory, and visual sensing information. 

If cost function $f$ is given by Eq. (\ref{eq: NTOE-OF-f}), optimal state estimator function $\hat{x}^{*}$ [Eq. (\ref{eq: NTOE-optimal control function})] is given by 
\begin{align}
	\hat{x}^{*}(t,y_{t},z_{t})=\mathbb{E}_{p(x_{t}|y_{t},z_{t})}\left[x_{t}\right]. 
	\label{eq: NTOE-optimal state estimator}
\end{align}
A comparison of Eqs. (\ref{eq: CTOE-optimal state estimator tmp}) and (\ref{eq: NTOE-optimal state estimator}) clarifies that our theory replaces infinite-dimensional past observation history $y_{t-dt},...,y_{0}$ with finite-dimensional memory $z_{t}$. 

\section{Application to Biological Information Processing}\label{sec: ABIP}
\begin{figure*}
	\includegraphics[width=170mm]{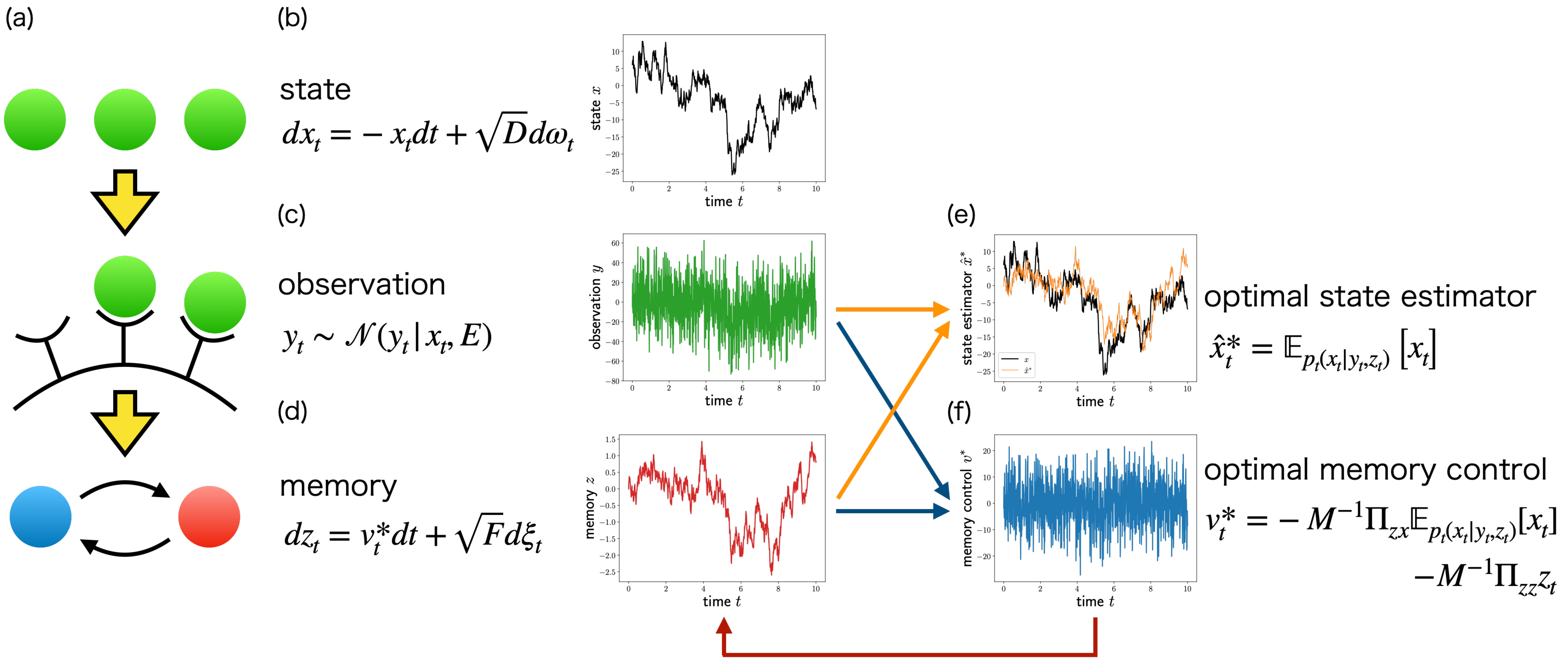}
	\caption{\label{fig: ABIP-diagram}
	Schematic diagram of a minimal model of biological information processing. 
	(a) A cell achieves adaptive behavior by accurately perceiving environmental ligands through receptor activities and intracellular chemical reactions. 
	The environmental ligands, receptor activities, and intracellular chemical reactions correspond to the state $x_{t}\in\mb{R}$, observation $y_{t}\in\mb{R}$, and memory $z_{t}\in\mb{R}$, respectively. 
	(b-f) Stochastic simulation of state $x_{t}$ (b), observation $y_{t}$ (c), memory $z_{t}$ (d), optimal state estimator $\hat{x}_{t}^{*}$ (e), and optimal memory control $v_{t}^{*}$ (f). 
	Optimal state estimator $\hat{x}_{t}^{*}$ and optimal memory control $v_{t}^{*}$ are determined based on observation $y_{t}$ and memory $z_{t}$ (orange and blue arrows, respectively). 
	Optimal memory control $v_{t}^{*}$ shapes the optimal dynamics of memory $z_{t}$ (red arrow). 
	For comparison, true state $x_{t}$ (black line) and estimated state $\hat{x}_{t}^{*}$ (orange line) are plotted on the same panel (e). 
	The parameters are $D=100$, $E=500$, $F=1$, $Q=10$, and $M=1$.  
	}
\end{figure*}

In this section, we apply our theory to a minimal model of biological information processing (Fig. \ref{fig: ABIP-diagram}(a)) to investigate optimal estimation strategies under resource limitations. 
This section is organized as follows: 
In Sec. \ref{subsec: ABIP-PF}, we formulate a minimal model of biological information processing. 
In Sec. \ref{subsec: ABIP-OS}, we show its optimal solution. 
In Sec. \ref{subsec: ABIP-SS}, we demonstrate the validity of the optimal solution by conducting a stochastic simulation. 
In Sec. \ref{subsec: ABIP-TMS}, we investigate the parameter dependence of the optimal solution and find discontinuous phase transitions between memory-less and memory-based estimation strategies. 
In Sec. \ref{subsec: ABIP-TEC}, we examine the phase transitions by analyzing the landscapes of the objective function, the state estimation error, and the memory control cost. 
In Sec. \ref{subsec: ABIP-FA}, we further investigate the relationships between parameters in the phase transitions of optimal estimation strategies. 

\subsection{Problem Formulation}\label{subsec: ABIP-PF}
In this subsection, we formulate the minimal model of biological information processing (Fig. \ref{fig: ABIP-diagram}(a)). 
We illustrate this model using cellular information processing as an example following previous studies \cite{kobayashi_implementation_2010,zechner_molecular_2016,mora_physical_2019}, but it should be noted that this model is general and can be applied to the information processing of other systems such as neurons and robots. 

A cell has to accurately perceive environmental state $x_{t}\in\mb{R}$ to achieve various physiological functions such as chemotaxis, growth, and differentiation. 
Environmental state $x_{t}\in\mb{R}$ is assumed to evolve by the following Ornstein-Uhlenbeck process: 
\begin{align}
	dx_{t}&=-x_{t}dt+\sqrt{D}d\omega_{t},
	\label{eq: ABIP-state SDE}
\end{align}
where initial state $x_{0}$ follows standard Gaussian distribution $\mcal{N}(x_{0}|0,1)$, and $\omega_{t}\in\mb{R}$ is a standard Wiener process. 
$D\geq0$ represents the intensity of the state noise. 

The cell observes environmental state $x_{t}$ through receptor activities, which correspond to observation $y_{t}\in\mb{R}$. 
Observation $y_{t}$ is generated from the following Gaussian distribution: 
\begin{align}
	y_{t}\sim\mcal{N}(y_{t}|x_{t},E),
	\label{eq: ABIP-observation SDE}
\end{align}
where $E\geq0$ represents the intensity of the observation noise. 

The cell memorizes the information about past observation history $y_{t-dt},...,y_{0}$ through intracellular chemical molecules, which correspond to memory $z_{t}\in\mb{R}$. 
Memory $z_{t}\in\mb{R}$ evolves by the following SDE: 
\begin{align}
	dz_{t}&=v_{t}dt+\sqrt{F}d\xi_{t}, 
	\label{eq: ABIP-memory SDE}
\end{align}
where initial memory $z_{0}$ follows standard Gaussian distribution $\mcal{N}(z_{0}|0,1)$, $v_{t}\in\mb{R}$ is the memory control, and $\xi_{t}\in\mb{R}$ is a standard Wiener process. 
$F\geq0$ represents the intensity of the memory noise. 

The memory dynamics corresponds to intracellular chemical reactions in cellular information processing. 
$v_{t}$ and $F$ represent the intensity and the noise of intracellular chemical reactions, respectively. 
If a cell could invest a large number of molecules in information processing, intracellular chemical reactions would become almost deterministic, resulting in $F=0$. 
However, in most cases, a cell can only invest a small number of molecules in their intracellular chemical reactions due to their resource limitations, which inevitably leads to significant memory noise $F>0$ \cite{elowitz_stochastic_2002,swain_intrinsic_2002,paulsson_summing_2004,lestas_fundamental_2010,taniguchi_quantifying_2010,shibata_noisy_2005,ueda_stochastic_2007,kobayashi_dynamics_2011}. 
While the conventional Bayesian filtering theory takes into account extrinsic noise such as state noise $D$ and observation noise $E$, it fails to capture intrinsic noise such as memory noise $F$. 

A cell determines state estimator $\hat{x}_{t}\in\mb{R}$ and memory control $v_{t}\in\mb{R}$ based on observation $y_{t}$ and memory $z_{t}$, such that $\hat{x}_{t}=\hat{x}(t,y_{t},z_{t})$ and $v_{t}=v(t,y_{t},z_{t})$. 
These functions are simultaneously optimized as follows: 
\begin{align}
	\hat{x}^{*},v^{*}:=\arg\min_{\hat{x},v}J[\hat{x},v]. 
\end{align}
We use an objective function $J[\hat{x},v]$ defined as follows: 
\begin{align}
	&J[\hat{x},v]:=\nonumber\\
	&\lim_{T\to\infty}\frac{1}{T}\mb{E}\left[\int_{0}^{T}\left(Q(x_{t}-\hat{x}_{t})^{2}+Mv_{t}^{2}\right)dt\right],
	\label{eq: ABIP-OF}
\end{align}
where $Q>0$ and $M>0$ are weighting parameters. 
The long-time limit $T\to\infty$ means that this problem focuses on the case where the probability density function of state $x$ and memory $z$ is stationary, but it should be noted that state $x$ and memory $z$ themselves are non-stationary. 

The first term of the cost function in Eq. (\ref{eq: ABIP-OF}) is a quadratic function of the difference between true state $x_{t}$ and estimated state $\hat{x}_{t}$, which represents the state estimation error. 
The second term of the cost function in Eq. (\ref{eq: ABIP-OF}) is a quadratic function of memory control $v_{t}$, representing the memory control cost. 
As the absolute value of memory control $v_{t}$ increases, the relative impact of memory noise decreases in the memory dynamics [Eq. (\ref{eq: ABIP-memory SDE})], which enhances memory effectiveness and improves state estimation. 
On the other hand, a larger absolute value of memory control $v_{t}$ corresponds to more intense intracellular chemical reactions, leading to greater energy consumption. 
Therefore, the objective function [Eq. (\ref{eq: ABIP-OF})] expresses the trade-off between estimation errors and energy costs. 

\subsection{Optimal Solution}\label{subsec: ABIP-OS}
The minimal model of biological information processing is composed of the linear Gaussian state, observation, and memory [Eqs. (\ref{eq: ABIP-state SDE})--(\ref{eq: ABIP-memory SDE})] and the quadratic costs [Eq. (\ref{eq: ABIP-OF})], which corresponds to a Linear-Quadratic-Gaussian (LQG) problem  \cite{bensoussan_stochastic_1992,yong_stochastic_1999,nisio_stochastic_2015,bensoussan_estimation_2018}. 
This setup allows for a semi-analytical calculation of optimal state estimator function $\hat{x}^{*}$ and optimal memory control function $v^{*}$ [Eq. (\ref{eq: NTOE-optimal control function})] as follows (See Appendix \ref{sec: LQG} and \ref{sec: ABIPA}): 
\begin{align}
	\hat{x}^{*}(t,y,z)&=\mb{E}_{p_{t}(x|y,z)}\left[x\right],\label{eq: ABIP-optimal state estimator}\\
	v^{*}(t,y,z)&=-M^{-1}\Pi_{zx}\mb{E}_{p_{t}(x|y,z)}\left[x\right]-M^{-1}\Pi_{zz}z.\label{eq: ABIP-optimal memory control}
\end{align}
Optimal state estimator function $\hat{x}^{*}$ is the conditional expectation of state $x$ given observation $y$ and memory $z$, which reflects the fact that a cell needs to estimate state $x$ from observation $y$ and memory $z$. 
Optimal memory control function $v^{*}$ is a linear function of conditional expected state $\mb{E}_{p_{t}(x|y,z)}\left[x\right]$ and memory $z$, and there are two control gains: $\Pi_{zx}$ and $\Pi_{zz}$. 
Control gain $\Pi_{zx}$ determines how to encode the information from observation $y$ to memory $z$ because $\mb{E}_{p_{t}(x|y,z)}\left[x\right]$ contains information about $y$. 
In fact, if $\Pi_{zx}=0$, observation $y$ does not appear in the optimal memory control function [Eq. (\ref{eq: ABIP-optimal memory control})] and the information is not encoded from observation $y$ to memory $z$. 
In this case, since memory $z$ has no information about state $x$, the optimal state estimator function reduces to $\hat{x}^{*}(t,y)=\mb{E}_{p_{t}(x|y)}\left[x\right]$, resulting in a memory-less estimation. 
Conversely, if $\Pi_{zx}\neq 0$, memory $z$ accumulates information about state $x$ through observation $y$, and optimal state estimator function $\hat{x}^{*}$ depends not only on observation $y$ but also on memory $z$ as in Eq. (\ref{eq: ABIP-optimal state estimator}), resulting in a memory-based estimation. 
The other control gain, $\Pi_{zz}$, is a negative feedback control gain for memory $z$, which suppresses the divergence and the stochasticity of the memory. 
When the weight of the memory control cost $M$ is higher, the absolute value of optimal memory control function $v^{*}$ is lower. 

In this problem, since $p_{t}(x|y,z)$ is a Gaussian distribution, $\mb{E}_{p_{t}(x|y,z)}\left[x\right]$ becomes a linear function of observation $y$ and memory $z$ as follows: 
\begin{align}
	\mb{E}_{p_{t}(x|y,z)}\left[x\right]=K_{xy}y+K_{xz}z. 
	\label{eq: ABIP-conditional expectation by estimation gain}
\end{align}
The estimation gains, $K_{xy}$ and $K_{xz}$, and the control gains, $\Pi_{zx}$ and $\Pi_{zz}$, are nonlinear functions of $D$, $E$, $F$, $Q$, and $M$, which are obtained from HJB and FP equations (See Appendix \ref{sec: LQG} and \ref{sec: ABIPA}). 

\subsection{Stochastic Simulation}\label{subsec: ABIP-SS}
We conduct a stochastic simulation to validate the behavior of optimal state estimator function $\hat{x}^{*}$ and memory control function $v^{*}$ [Eqs. (\ref{eq: ABIP-optimal state estimator}) and (\ref{eq: ABIP-optimal memory control})] (Fig. \ref{fig: ABIP-diagram}(b-f)). 
A cell estimates environmental state $x_{t}$ (Fig. \ref{fig: ABIP-diagram}(b)) from noisy observation $y_{t}$ (Fig. \ref{fig: ABIP-diagram}(c)) and limited memory $z_{t}$ (Fig. \ref{fig: ABIP-diagram}(d)). 
The dynamics of memory $z_{t}$ is optimized by optimal memory control $v_{t}^{*}$ (Fig. \ref{fig: ABIP-diagram}(f)). 
Despite the high noise in observation $y_{t}$, optimal state estimator $\hat{x}_{t}^{*}$ accurately recovers true state $x_{t}$ (Fig. \ref{fig: ABIP-diagram}(e)). 
This result demonstrates that accurate state estimation is achieved with limited memory without relying on Bayesian filtering. 

The optimal memory dynamics (Fig. \ref{fig: ABIP-diagram}(d)) is similar to the state dynamics (Fig. \ref{fig: ABIP-diagram}(b)). 
This is because memory $z$ has a linear relationship with the conditional expected state $\mb{E}_{p_{t}(x|y,z)}\left[x\right]$, as shown in Eq. (\ref{eq: ABIP-conditional expectation by estimation gain}). 
This property of the optimal memory dynamics is similar to that of Kalman filter \cite{kalman_new_1960,chui_kalman_2017,bensoussan_estimation_2018,kallianpur_stochastic_1980,chen_bayesian_2003,jazwinski_stochastic_2007,bain_fundamentals_2009}. 
In fact, we have proven that the optimal memory dynamics is equivalent to Kalman filter when the LQG problem has no resource limitations \cite{tottori_memory-limited_2022}. 
However, when the LQG problem has resource limitations, the optimal memory dynamics deviates from Kalman filter. 
In particular, as shown in the next subsection, when the observation noise is lower and the memory noise is higher, the optimal memory control becomes zero and the optimal memory dynamics becomes diffusion dynamics, which is entirely different from Kalman filter. 

\subsection{Phase Transition}\label{subsec: ABIP-TMS}
\begin{figure*}
	\includegraphics[width=170mm]{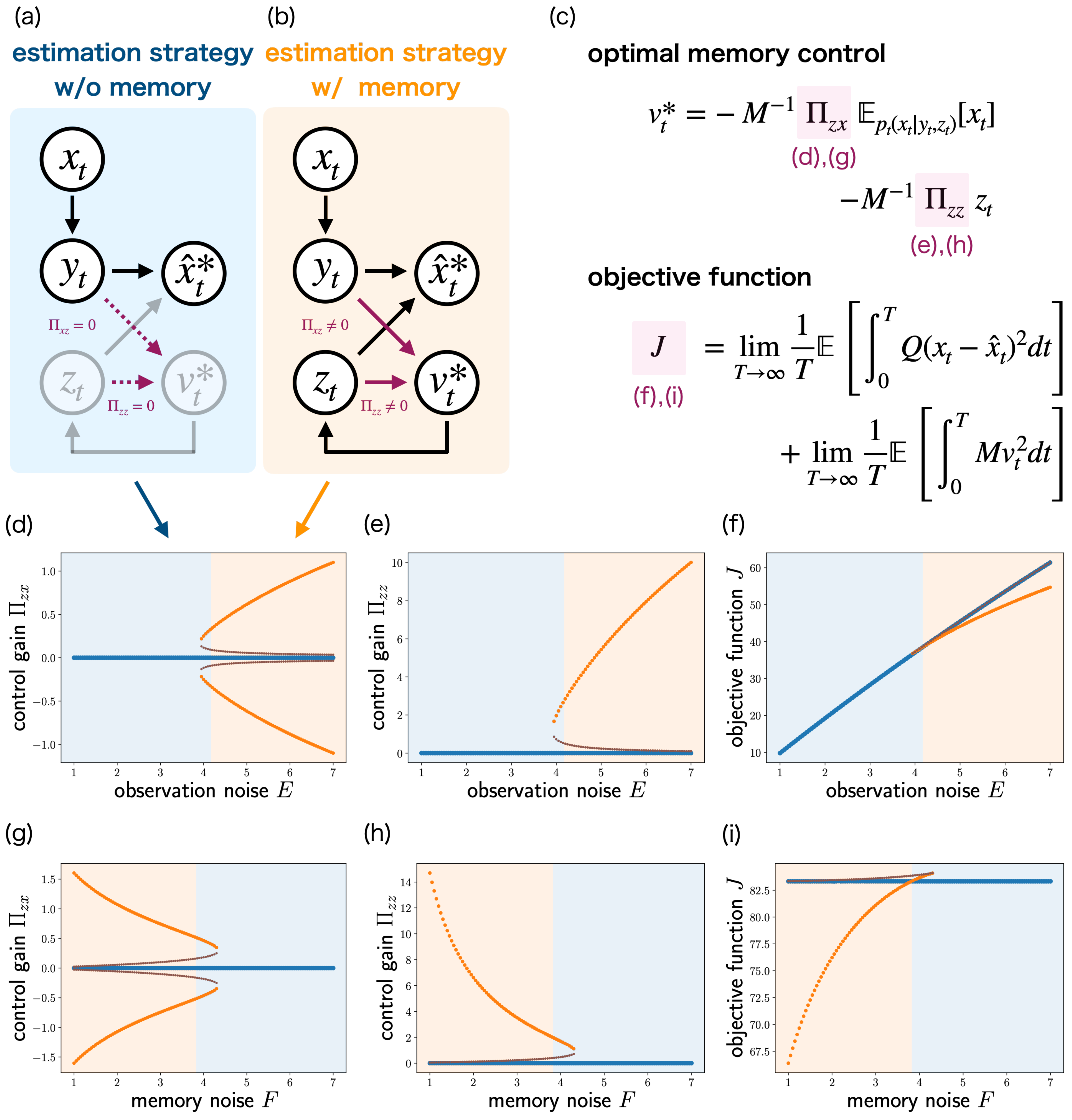}
	\caption{\label{fig: ABIP-TMS}
	Phase transition in optimal estimation strategy. 
	(a,b) Schematic diagrams of estimation strategies without memory (a) and with memory (b). 
	(a) When $\Pi_{xz}=\Pi_{zz}=0$, the past observation information is not encoded into memory $z_{t}$, and state $x_{t}$ is estimated only from observation $y_{t}$. 
	(b) When $\Pi_{xz},\Pi_{zz}\neq0$, the past observation information is encoded into memory $z_{t}$, and state $x_{t}$ is estimated from observation $y_{t}$ and memory $z_{t}$. 
	(c) Expressions of optimal memory control $v_{t}^{*}$ and objective function $J$. 
	(d--i) Memory control gains $\Pi_{zx}$ and $\Pi_{zz}$ and associated objective function $J$ with respect to observation noise $E$ (d--f) and memory noise $F$ (g--i). 
	The blue, brown, and orange dots are the solutions of Pontryagin's minimum principle, corresponding to zero, intermediate, and high memory controls, respectively. 
	The blue dots are optimal in the blue region, while the orange dots are optimal in the orange region. 
	There are no regions where the brown dots are optimal. 
	The parameters in (d--f) are $D=100$, $F=1$, $Q=10$, and $M=1$, while those in (g--i) are $D=100$, $E=10$, $Q=10$, and $M=1$.  
	}
\end{figure*}

In this subsection, we examine the parameter dependence of the optimal estimation strategy under resource limitations (Fig. \ref{fig: ABIP-TMS}). 
The optimal estimation strategy is characterized by the memory control gains, $\Pi_{zx}$ and $\Pi_{zz}$, in the optimal memory control function, $v^{*}$ (Fig. \ref{fig: ABIP-TMS}(c)).  
If $\Pi_{xz}=\Pi_{zz}=0$, the past observation information is not encoded into memory $z_{t}$, and state $x_{t}$ is estimated only from observation $y_{t}$, which corresponds to an estimation strategy without memory (Fig. \ref{fig: ABIP-TMS}(a)). 
Conversely, if $\Pi_{xz},\Pi_{zz}\neq0$, memory $z_{t}$ is regulated based on the past observation information, and state $x_{t}$ is estimated not only from observation $y_{t}$ but also from memory $z_{t}$, which corresponds to an estimation strategy with memory (Fig. \ref{fig: ABIP-TMS}(b)). 
In the following, we investigate the parameter dependence of the optimal memory control gains. 

We first analyze the dependence of the optimal memory control gains on observation noise $E$ (Fig. \ref{fig: ABIP-TMS}(d--f)). 
The memory control gains, $\Pi_{zx}$ and $\Pi_{zz}$, are obtained by solving HJB and FP equations (See Appendix \ref{sec: ABIPA} for further details), which corresponds to Pontryagin's minimum principle on the probability density function space. 
Given that Pontryagin's minimum principle is equivalent to the method of Lagrange multipliers (See Appendix \ref{subsec: OSC-MLM}), this method identifies not only optimal solutions but also extremal solutions (Fig. \ref{fig: ABIP-TMS}(d,e)(blue, brown, and orange dots)). 
The optimal memory control gains are identified by the minimizer of the objective function (Fig. \ref{fig: ABIP-TMS}(f)). 

When observation noise $E$ is low, the solution obtained from Pontryagin's minimum principle is only $\Pi_{zx}=\Pi_{zz}=0$, which is the optimal solution (Fig. \ref{fig: ABIP-TMS}(d--f)(blue dots)). 
This result indicates that the estimation strategy without memory is optimal when the current observation is sufficiently reliable. 
The current observation contains enough information for the accurate state estimation, making it optimal to save on the memory control cost. 
As observation noise $E$ increases, the solutions where $\Pi_{zx},\Pi_{zz}\neq0$ discontinuously emerge (Fig. \ref{fig: ABIP-TMS}(d--f)(brown and orange dots)), while the solution of $\Pi_{zx}=\Pi_{zz}=0$ remains optimal (Fig. \ref{fig: ABIP-TMS}(d--f)(blue dots)). 
As observation noise $E$ further increases, the optimal solution switches from $\Pi_{zx}=\Pi_{zz}=0$ (Fig. \ref{fig: ABIP-TMS}(d--f)(blue dots)) to $\Pi_{zx},\Pi_{zz}\neq0$ (Fig. \ref{fig: ABIP-TMS}(d--f)(orange dots)), demonstrating that the estimation strategy with memory is optimal. 
This result suggests that the reduction in the state estimation error through the memory control outweighs the memory control cost  when the observation is less reliable. 

We next analyze the dependence of the optimal memory control gains on memory noise $F$ (Fig. \ref{fig: ABIP-TMS}(g--i)).  
The discontinuous phase transition is also observed for memory noise $F$, but the direction of the transition is different. 
While the estimation strategy with memory is optimal when memory noise $F$ is low, that without memory becomes optimal as memory noise $F$ increases. 
This result indicates that the higher intrinsic noise caused by resource limitations renders memory less useful. 

The discontinuous phase transition also occurs for the other parameters, i.e., the intensity of the state noise $D$, the weight of the state estimation error $Q$, and the weight of the memory control cost $M$ (See Appendix \ref{sec: ABIPA}, Fig. \ref{fig: ABIPA-TMS}). 
The phase transition for state noise $D$ and state estimation error $Q$ is similar to that for observation noise $E$. 
This result suggests that the estimation strategy with memory is optimal when the state estimation is either challenging or significant. 
In contrast, the phase transition for memory control cost $M$ is similar to that for memory noise $F$, indicating that resource limitations, such as the noise and cost associated with intrinsic information processing, render the estimation strategy without memory optimal. 

Our minimal model considers linear Gaussian dynamics and quadratic cost functions, avoiding unnecessary nonlinearities that could be potential sources of complex behaviors. 
Nevertheless, our model exhibits a discontinuous phase transition in the optimal estimation strategy. 
This phenomenon is not observed in the conventional Bayesian filtering theory, which always uses past observation information for accurate state estimation. 
It is consistent with the fact that Bayesian filtering does not account for resource limitations, corresponding to $F\to0$ and $M\to0$. 
Therefore, our analysis suggests that resource limitations may drive the punctuated evolution of biological information processing, leading to qualitatively different behaviors across organisms. 

\subsection{Landscape Analysis}\label{subsec: ABIP-TEC}
\begin{figure*}
	\includegraphics[width=170mm]{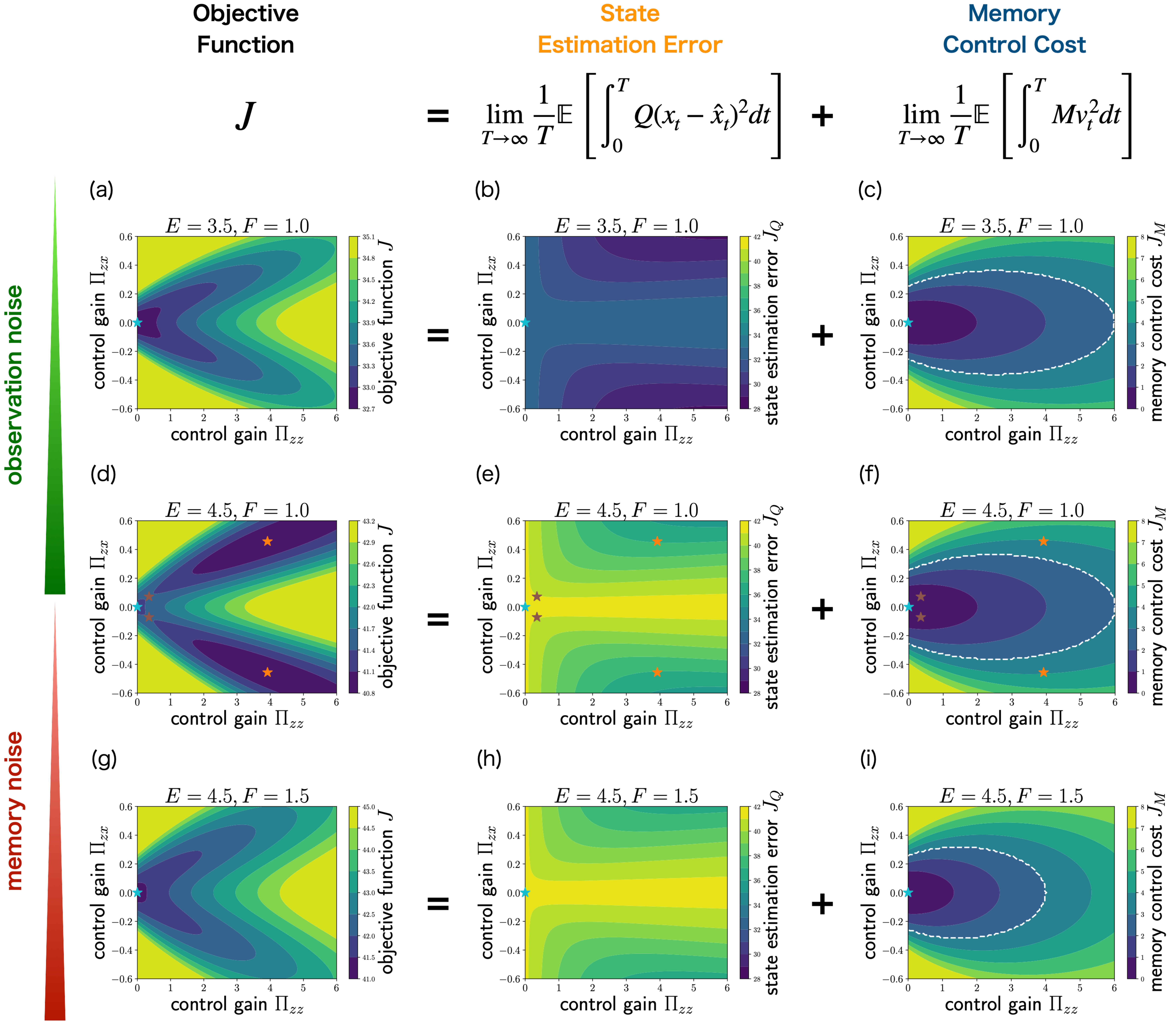}
	\caption{\label{fig: ABIP-TEC}
	Objective function $J$ (a,d,g), state estimation error $J_{Q}$ (b,e,h), and memory control cost $J_{M}$ (c,f,i) with respect to memory control gains $\Pi_{zx}$ and $\Pi_{zz}$ under different observation noise $E$ and memory noise $F$. 
	The blue, brown, and orange stars are the solutions of Pontryagin's minimum principle. 
	The white dashed curves represent $J_{M}=3$, which clarify the changes in $J_{M}$. 
	The rest of the parameters are $D=100$, $Q=10$, and $M=1$. 
	}
\end{figure*}

In order to examine the phase transition in the optimal estimation strategy from another perspective, we analyze the landscape of objective function $J$ with respect to memory control gains $\Pi_{xz}$ and $\Pi_{zz}$ (Fig. \ref{fig: ABIP-TEC}(a,d,g)). 
When observation noise $E$ is low, $\Pi_{xz}=\Pi_{zz}=0$ is the global minimum (Fig. \ref{fig: ABIP-TEC}(a)). 
As observation noise $E$ increases, $\Pi_{xz}=\Pi_{zz}=0$ shifts to a local minimum, and the global minima emerge at $\Pi_{xz},\Pi_{zz}\neq0$ (Fig. \ref{fig: ABIP-TEC}(d)). 
Increasing memory noise $F$ further causes the extrema at $\Pi_{xz},\Pi_{zz}\neq0$ to disappear, returning $\Pi_{xz}=\Pi_{zz}=0$ as the global minimum (Fig. \ref{fig: ABIP-TEC}(g)). 
These results are consistent with the previous subsection. 
We also confirm that the solutions of Pontryagin's minimum principle appear at the extrema (Fig. \ref{fig: ABIP-TEC}(a,d,g)(blue, brown, orange stars)).

In order to gain additional insights into the phase transition in the optimal estimation strategy, we decompose objective function $J$ into state estimation error $J_{Q}$ and memory control cost $J_{M}$, and analyze their landscapes (Fig. \ref{fig: ABIP-TEC}(b,e,h) and (c,f,i)). 
State estimation error $J_{Q}$ and memory control cost $J_{M}$ are defined as follows: 
\begin{align}
	J_{Q}&:=\lim_{T\to\infty}\frac{1}{T}\mb{E}\left[\int_{0}^{T}Q(x_{t}-\hat{x}_{t})^{2}dt\right],\\
	J_{M}&:=\lim_{T\to\infty}\frac{1}{T}\mb{E}\left[\int_{0}^{T}Mv_{t}^{2}dt\right]. 
\end{align}

There is a trade-off between state estimation error $J_{Q}$ and memory control cost $J_{M}$ with respect to memory control gains $\Pi_{zx}$ and $\Pi_{zz}$. 
As memory control gains $\Pi_{zx}$ and $\Pi_{zz}$ increase, state estimation error $J_{Q}$ decreases (Fig. \ref{fig: ABIP-TEC}(b,e,h)), whereas memory control cost $J_{M}$ increases (Fig. \ref{fig: ABIP-TEC}(c,f,i)). 
Therefore, the estimation strategy with memory prioritizes reducing the state estimation error, whereas the estimation strategy without memory prioritizes minimizing the memory control cost. 

When observation noise $E$ increases, state estimation error $J_{Q}$ increases (Fig. \ref{fig: ABIP-TEC}(b,e)), while memory control cost $J_{M}$ remains almost unchanged (Fig. \ref{fig: ABIP-TEC}(c,f)), which makes the estimation strategy with memory optimal to reduce the state estimation error. 
In contrast, when memory noise $F$ increases, memory control cost $J_{M}$ increases (Fig. \ref{fig: ABIP-TEC}(f,i)) compared to state estimation error $J_{Q}$ (Fig. \ref{fig: ABIP-TEC}(e,h)). 
This may be because a large memory fluctuation leads to a higher memory feedback control. 
This result renders the estimation strategy without memory optimal for minimizing the memory control cost.  
Therefore, the phase transition in the optimal estimation strategy may be caused by the trade-off between the state estimation error and the memory control cost. 

\subsection{Relationships among Parameters}\label{subsec: ABIP-FA}
\begin{figure*}
	\includegraphics[width=170mm]{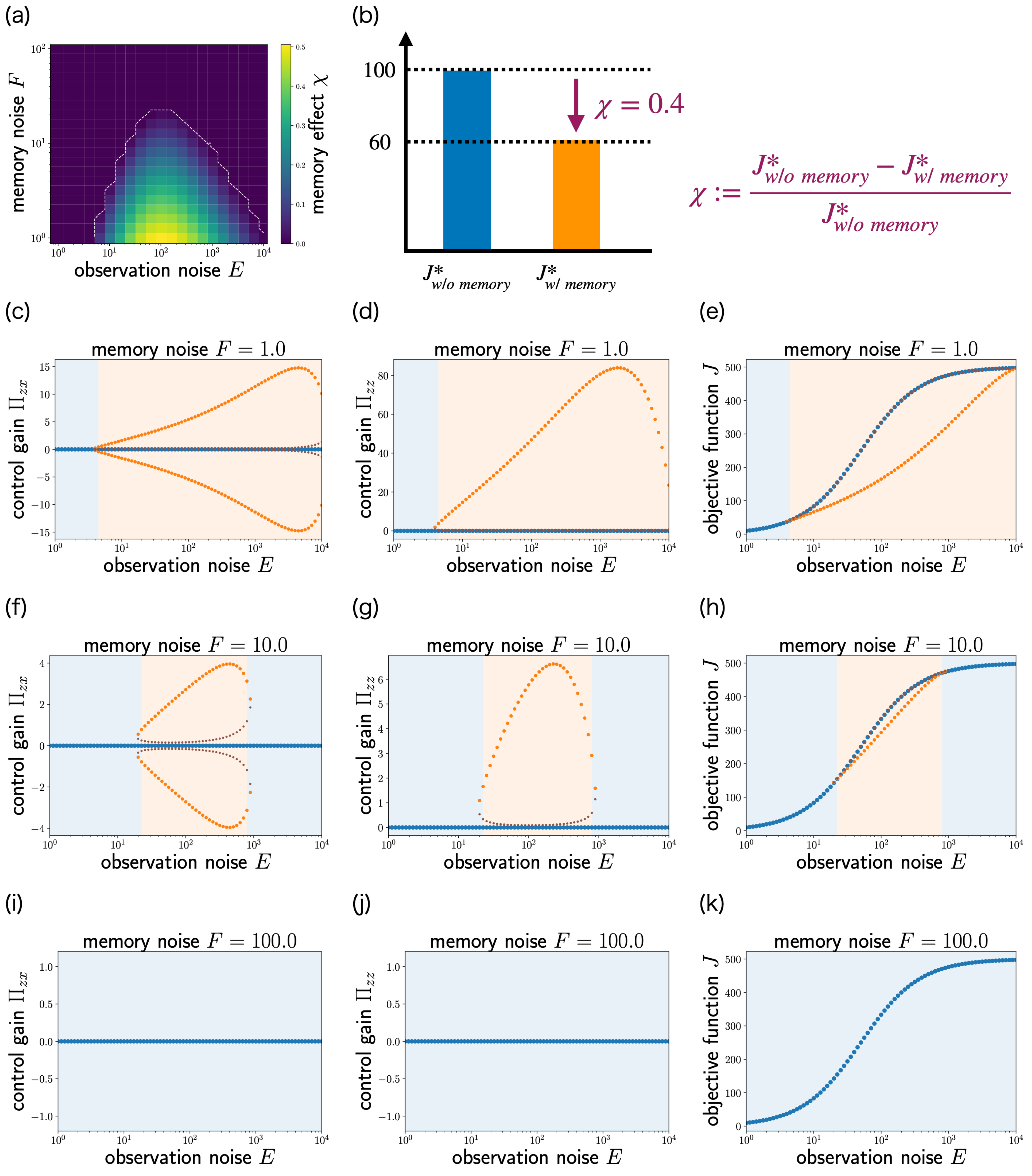}
	\caption{\label{fig: ABIP-TMS-2param}
	(a) Memory effect $\chi$ as a function of observation noise $E$ and memory noise $F$. 
	The white dashed curve marks the boundary between $\chi=0$ and $\chi>0$. 
	(b) Illustration of the definition of memory effect $\chi$. 
	(c--k) Memory control gains $\Pi_{zx}$ and $\Pi_{zz}$ and associated objective function $J$ with respect to observation noise $E$ under different memory noise $F$: 
	lower memory noise (c--e), intermediate memory noise (f--h), and higher memory noise (i--k). 
	The blue, brown, and orange dots are the solutions of Pontryagin's minimum principle, corresponding to zero, intermediate, and high memory controls, respectively. 
	The blue dots are optimal in the blue region, while the orange dots are optimal in the orange region. 
	The rest of the parameters are $D=100$, $Q=10$, and $M=1$.  
	}
\end{figure*}
\begin{figure*}
	\includegraphics[width=170mm]{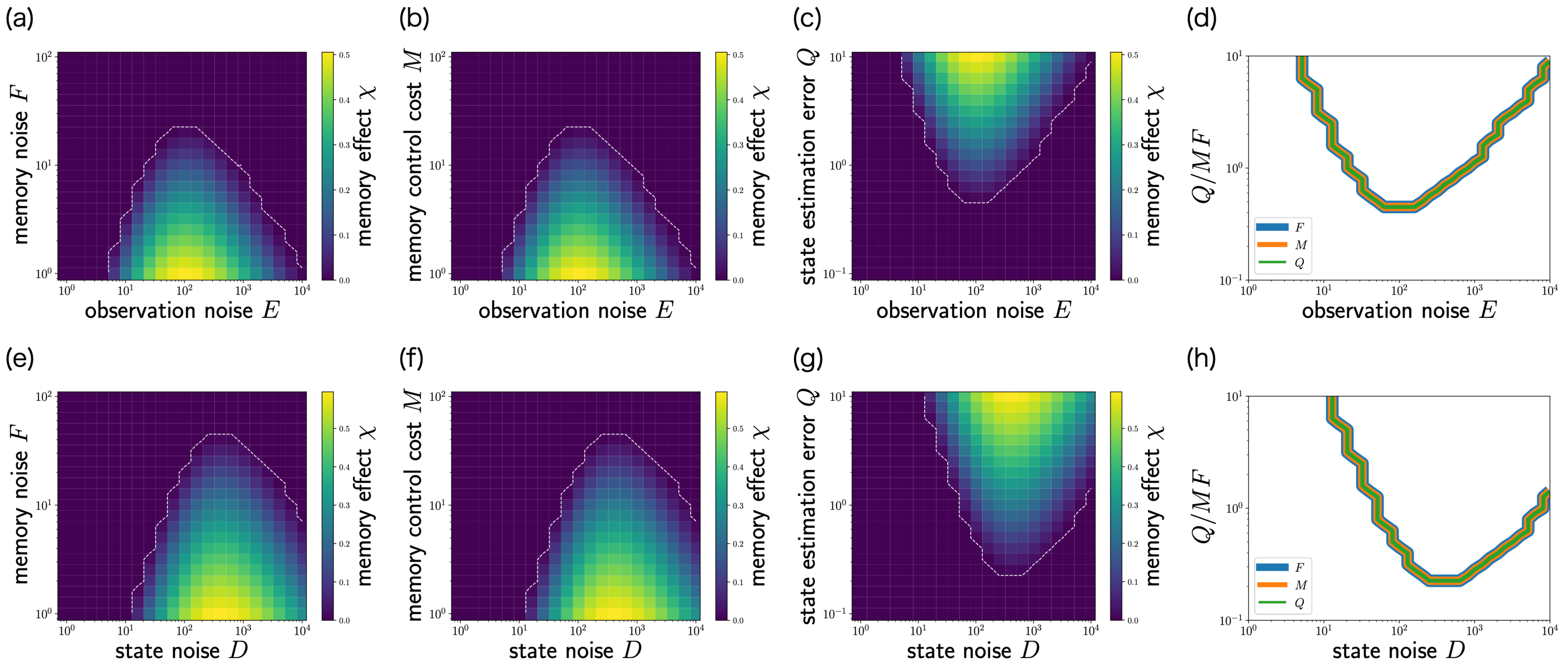}
	\caption{\label{fig: ABIP-scaling}
	(a--c) Memory effect $\chi$ as a function of observation noise $E$ and memory noise $F$ (a), memory control cost $M$ (b), and state estimation error $Q$ (c). 
	(d) Phase boundaries as functions of $E$ and $Q/MF$. 
	The blue, orange, and green curves correspond to the phase boundaries in (a), (b), and (c), respectively. 
	(e--h) Similar behaviors are observed for state noise $D$. 
	The parameters not indicated in the figures are set to $D=100$, $E=100$, $F=1$, $Q=10$, and $M=1$. 
	}
\end{figure*}

We further investigate the relationships among parameters in the phase transition of the optimal estimation strategy. 
In order to quantify the performance improvement induced by the memory, we define the memory effect
\begin{align}
	\chi:= \frac{J_{\rm w/o\ memory}^{*}-J_{\rm w/\ memory}^{*}}{J_{\rm w/o\ memory}^{*}},
\end{align}
where $J_{\rm w/\ memory}^{*}$ and $J_{\rm w/o\ memory}^{*}$ are the minimum values of objective function $J$ in the systems with and without memory, respectively (Fig. \ref{fig: ABIP-TMS-2param}(b)). 
$J_{\rm w/\ memory}^{*}\geq0$ and $J_{\rm w/o\ memory}^{*}\geq0$ obviously hold because the cost function is defined as a quadratic function with a minimum of 0. 
Moreover, $J_{\rm w/\ memory}^{*}\leq J_{\rm w/o\ memory}^{*}$ always holds because the system with memory becomes equivalent to the system without memory by not controlling the memory. 
Consequently, $0 \leq \chi \leq 1$ holds. 
A higher memory effect $\chi$ indicates a greater performance improvement owing to the memory. 
Additionally, memory effect $\chi$ is useful for analyzing the phase transition. 
The estimation strategy without memory is optimal when $\chi = 0$, whereas that with memory is optimal when $\chi>0$. 
The boundary between $\chi = 0$ and $\chi > 0$ corresponds to the phase boundary between the memory-less and memory-based estimation strategies.

Figure \ref{fig: ABIP-TMS-2param}(a) visualizes memory effect $\chi$ with respect to observation noise $E$ and memory noise $F$. 
When memory noise $F$ is low, only a phase transition from the memory-less to memory-based estimation strategies occurs as observation noise $E$ increases (Fig. \ref{fig: ABIP-TMS-2param}(c--e)). 
In contrast, as memory noise $F$ increases, phase transitions occur not only from the memory-less to memory-based estimation strategies, but also back to the memory-less estimation strategy as observation noise $E$ increases (Fig. \ref{fig: ABIP-TMS-2param}(f--h)). 
It is consistent with the fact that memory effect $\chi$ is maximized at an intermediate level of observation noise $E$ (Fig. \ref{fig: ABIP-TMS-2param}(a)). 
As memory noise $F$ increases further, the phase transition to the memory-based estimation strategy disappears with respect to observation noise $E$ (Fig. \ref{fig: ABIP-TMS-2param}(i--k)). 
Therefore, despite the simplicity of our model, it exhibits not only discontinuous but also non-monotonic phase transitions due to resource limitations. 

We also investigate memory effect $\chi$ with respect to the other parameters, i.e., state noise $D$, state estimation error $Q$, and memory control cost $M$, in addition to observation noise $E$ and memory noise $F$ (Fig. \ref{fig: ABIP-scaling}(a--c,e--g)). 
With respect to observation noise $E$ and state noise $D$, memory effect $\chi$ is maximized at an intermediate level, which exhibits non-monotonic phase transitions. 
In contrast, for memory noise $F$, state estimation error $Q$, and memory control cost $M$, memory effect $\chi$ monotonically changes, resulting in monotonic phase transitions. 
Additionally, we find that the phase boundaries for $Q$, $M$, and $F$ are nearly identical when plotted against $Q/MF$ (Fig. \ref{fig: ABIP-scaling}(d,h)), suggesting that increasing $Q$ is nearly equivalent to decreasing $M$ and $F$ in terms of the phase transition. 

\section{Optimal Estimation and Control Theory with Resource Limitations}\label{sec: NTOEC}
Our theory achieves optimal state estimation through optimal memory control rather than Bayesian filtering. 
This approach allows us to extend our theory straightforwardly to include optimal state control \cite{tottori_memory-limited_2022,tottori_forward-backward_2023,tottori_decentralized_2023}. 
In this section, we introduce the unified theory for optimal estimation and control under resource limitations. 

In this theory, the agent not only estimates environmental state $x_{t}$ but also controls it. 
Thus, the dynamics of environmental state $x_{t}$ is given by the following SDE: 
\begin{align}
	dx_{t}=b(t,x_{t},u_{t})dt+\sigma(t,x_{t},u_{t})d\omega_{t}, \label{eq: NTOEC-state SDE}
\end{align}
where initial state $x_{0}$ follows $p_{0}(x_{0})$, $\omega_{t}\in\mb{R}^{d_{\omega}}$ is a standard Wiener process, and $u_{t}\in\mb{R}^{d_{u}}$ is the state control of the agent. 
Unlike the state dynamics in Sec. \ref{sec: NTOE} [Eq. (\ref{eq: CTOE-state SDE})],  Eq. (\ref{eq: NTOEC-state SDE}) depends on state control $u_{t}$, which allows the agent to achieve desired state $x_{t}$ through state control $u_{t}$. 
The agent determines state control $u_{t}$ based on observation $y_{t}$ and memory $z_{t}$ as follows: 
\begin{align}
	u_{t}&=u(t,y_{t},z_{t}).\label{eq: NTOEC-state control}
\end{align}
Observation $y_{t}$, memory $z_{t}$, and memory control $v_{t}$ are the same as those in Sec. \ref{sec: NTOE}, given by Eqs. (\ref{eq: CTOE-observation PDF}), (\ref{eq: NTOE-memory SDE}), and (\ref{eq: NTOE-memory control}), respectively.
The objective function is given by the following expected cumulative cost function: 
\begin{align}
	J[u,v]:=\mb{E}\left[\int_{0}^{T}f(t,x_{t},z_{t},u_{t},v_{t})dt+g(x_{T},z_{T})\right], 
\end{align}
where $f$ and $g$ are the running and terminal cost functions, respectively. 
Compared with the objective function in Sec. \ref{sec: NTOE} [Eq. (\ref{eq: NTOE-OF})], state estimator function $\hat{x}$ is replaced by state control function $u$. 
State control function $u$ and memory control function $v$ are simultaneously optimized as follows: 
\begin{align}
	u^{*},v^{*}:=\arg\min_{u,v}J[u,v]. 
\end{align}

This optimal estimation and control problem can be solved in almost the same way as the optimal estimation problem in Sec. \ref{sec: NTOE}, and the optimal control functions are obtained by minimizing the conditional expected Hamiltonian: 
\begin{align}
	&u^{*}(t,y,z),v^{*}(t,y,z)=\nonumber\\
	&\arg\min_{u,v}\mb{E}_{p_{t}(x|y,z)}\left[\mcal{H}(t,x,z,u,v,w)\right], 
	\label{eq: NTOEC-optimal control function}
\end{align}
where $w$ and $p$ are the solutions of HJB and FP equations, respectively. 
The details are shown in Appendix \ref{sec: OSC} and \ref{sec: OMSC}. 
This result demonstrates that our theory provides a unified approach to optimal state estimation and optimal state control.

\section{Application to Biological Information Processing and Decision-Making}\label{sec: ABDM}
\begin{figure*}
	\includegraphics[width=170mm]{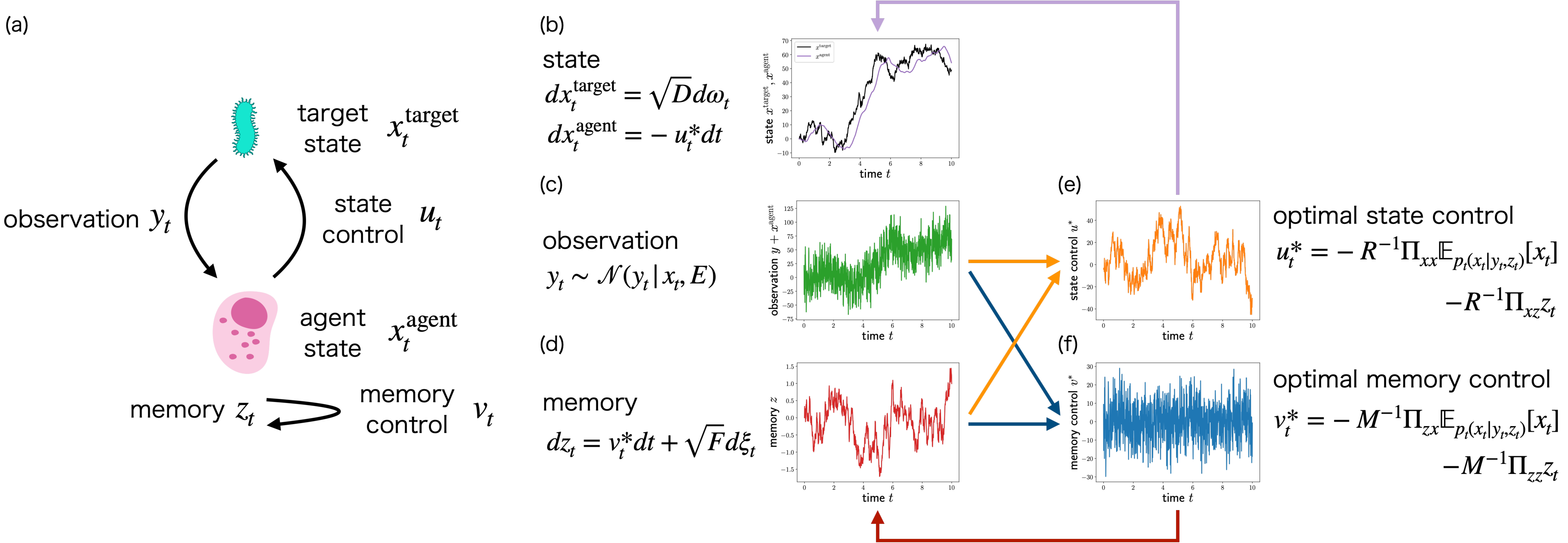}
	\caption{\label{fig: ABDM-diagram}
	Schematic diagram of a target tracking problem that is a minimal model of biological information processing and decision-making. 
	(a) In the target tracking problem, the agent estimates the position of the target $x_{t}^{\rm target}$ from noisy observation $y_{t}$ and limited memory $z_{t}$ and tracks it by controlling its own position $x_{t}^{\rm agent}$ through state control $u_{t}$. 
	The agent also controls its own memory $z_{t}$ through memory control $v_{t}$ to compress the information from past observation history $y_{t-dt},...,y_{0}$ to memory $z_{t}$. 
	(b-f) Stochastic simulation of the target and agent states (b), observation (c), memory (d), optimal state control (e), and optimal memory control (f). 
	Optimal state control $u_{t}^{*}$ and optimal memory control $v_{t}^{*}$ are determined based on observation $y_{t}$ and memory $z_{t}$ (orange and blue arrows, respectively) and they shape the optimal dynamics of agent state $x_{t}^{\rm agent}$ and memory $z_{t}$, respectively (purple and red arrows, respectively). 
	For comparison, target state $x_{t}^{\rm target}$ (black line) and agent state $x_{t}^{\rm agent}$ (purple line) are plotted on the same panel (b). 
	The parameters are $D=100$, $E=500$, $F=1$, $Q=10$, $R=1$, and $M=1$.  
	}
\end{figure*}

In this section, we apply the optimal estimation and control theory with resource limitations to a target tracking problem, which is a minimal model of biological information processing and decision-making (Fig. \ref{fig: ABDM-diagram}(a)). 
This section is organized as follows: 
In Sec. \ref{subsec: ABDM-PF}, we formulate a target tracking problem. 
In Sec. \ref{subsec: ABDM-OS}, we show its optimal solution and validate it by conducting a stochastic simulation. 
In Sec. \ref{subsec: ABDM-TMS}, we investigate the parameter dependence of the optimal solution and find the discontinuous phase transition between memory-less and memory-based control strategies. 
We also explore the relationships between parameters in this phase transition.

\subsection{Problem Formulation}\label{subsec: ABDM-PF}
We first formulate a target tracking problem as a minimal model of biological information processing and decision-making (Fig. \ref{fig: ABDM-diagram}(a)). 
In the target tracking problem, an agent estimates the position of a target from a noisy observation and controls its own body to track it. 
The target tracking problem is prevalent in many biological phenomena, such as cellular chemotaxis \cite{krummel_t_2016,nakamura_optimal_2022,sowinski_semantic_2023,rode_information_2024}, insect olfactory search  \cite{heinonen_optimal_2023,verano_olfactory_2023,celani_olfactory_2024}, and animal foraging behavior \cite{viswanathan_optimizing_1999,benichou_optimal_2005,viswanathan_levy_2008,benichou_intermittent_2011,hein_natural_2016}. 

The positions of the target and the agent are denoted by $x_{t}^{\rm target}\in\mb{R}$ and $x_{t}^{\rm agent}\in\mb{R}$, respectively. 
The position of the target evolves by the following diffusion process: 
\begin{align}
	dx_{t}^{\rm target}&=\sqrt{D}d\omega_{t},\label{eq: ABDM-target state SDE}
\end{align}
where initial target state $x_{0}^{\rm target}$ follows standard Gaussian distribution $\mcal{N}(x_{0}^{\rm target}|0,1)$ and $\omega_{t}\in\mb{R}$ is a standard Wiener process. 
$D\geq0$ represents the intensity of the state noise. 
The agent controls its own position as follows: 
\begin{align}
	dx_{t}^{\rm agent}&=-u_{t}dt,\label{eq: ABDM-agent state SDE}
\end{align}
where $x_{0}^{\rm agent}=0$ and $u_{t}\in\mb{R}$ is the state control of the agent. 
The target tracking problem is concerned only with the difference between target state $x_{t}^{\rm target}$ and agent state $x_{t}^{\rm agent}$. 
Thus, we define the state by $x_{t}:=x_{t}^{\rm target}-x_{t}^{\rm agent}$, which evolves by the following SDE: 
\begin{align}
	dx_{t}&=u_{t}dt+\sqrt{D}d\omega_{t},\label{eq: ABDM-state SDE}
\end{align}
where initial state $x_{0}$ follows standard Gaussian distribution $\mcal{N}(x_{0}|0,1)$. 

The agent cannot completely observe state $x_{t}$ and instead obtains noisy observation $y_{t}\in\mb{R}$ that is generated from the following Gaussian distribution: 
\begin{align}
	y_{t}\sim\mcal{N}(y_{t}|x_{t},E),\label{eq: ABDM-observation SDE}
\end{align}
where $E\geq0$ represents the intensity of the observation noise. 

The memory of the agent, $z_{t}\in\mb{R}$, evolves by the following SDE: 
\begin{align}
	dz_{t}&=v_{t}dt+\sqrt{F}d\xi_{t}, 
\end{align}
where initial memory $z_{0}$ follows standard Gaussian distribution $\mcal{N}(z_{0}|0,1)$, $v_{t}\in\mb{R}$ is the memory control of the agent, and $\xi_{t}\in\mb{R}$ is a standard Wiener process. 
$F\geq0$ represents the intensity of the memory noise. 

State control $u_{t}$ and memory control $v_{t}$ are determined based on observation $y_{t}$ and memory $z_{t}$ as $u_{t}=u(t,y_{t},z_{t})$ and $v_{t}=v(t,y_{t},z_{t})$, respectively. 

The objective function the agent minimizes is given by
\begin{align}
	&J[u,v]:=\nonumber\\
	&\lim_{T\to\infty}\frac{1}{T}\mb{E}\left[\int_{0}^{T}\left(Qx_{t}^{2}+Ru_{t}^{2}+Mv_{t}^{2}\right)dt\right], 
\end{align}
where $Q>0$, $R>0$, and $M>0$. 
The first term of the cost function is the distance between the target and the agent, the second term is the state control cost, and the third term is the memory control cost. 
$Q$, $R$, and $M$ are parameters representing the weights for the state tracking error, the state control cost, and the memory control cost, respectively. 

The target tracking problem is to find optimal state control function $u^{*}$ and optimal memory control function $v^{*}$ that minimize objective function $J[u,v]$: 
\begin{align}
	u^{*},v^{*}:=\arg\min_{u,v}J[u,v], 
\end{align}
which means that the agent minimizes the distance to the target with smaller control costs. 

\subsection{Optimal Solution}\label{subsec: ABDM-OS}
The target tracking problem also corresponds to the LQG problem \cite{bensoussan_stochastic_1992,yong_stochastic_1999,nisio_stochastic_2015,bensoussan_estimation_2018} and optimal control functions $u^{*}$ and $v^{*}$ [Eq. (\ref{eq: NTOEC-optimal control function})] are semi-analitically calculated as follows (See Appendix \ref{sec: ABDMA}): 
\begin{align}
	u^{*}(t,y,z)&=-R^{-1}\Pi_{xx}\mb{E}_{p_{t}(x|y,z)}\left[x\right]-R^{-1}\Pi_{xz}z,\label{eq: ABDM-optimal state control}\\
	v^{*}(t,y,z)&=-M^{-1}\Pi_{zx}\mb{E}_{p_{t}(x|y,z)}\left[x\right]-M^{-1}\Pi_{zz}z.\label{eq: ABDM-optimal memory control}
\end{align}
While optimal memory control function $v^{*}$ is the same as that in the target estimation problem [Eq. (\ref{eq: ABIP-optimal memory control})], optimal state estimator function $\hat{x}^{*}$ is replaced by optimal state control function $u^{*}$ in the target tracking problem. 
State control gains $\Pi_{xx}$ and $\Pi_{xz}$ represent how to control state $x_{t}$ based on conditionally expected state $\mb{E}_{p_{t}(x|y,z)}\left[x\right]$ and memory $z_{t}$, respectively. 
The higher the weight of the state control cost $R$, the smaller the absolute value of optimal state control $u_{t}^{*}$. 

By using optimal state estimator function $\hat{x}^{*}(t,y,z)=\mb{E}_{p_{t}(x|y,z)}\left[x\right]$ in the target estimation problem [Eq. (\ref{eq: ABIP-optimal state estimator})], optimal control functions $u^{*}$ and $v^{*}$ are rewritten as follows: 
\begin{align}
	u^{*}(t,y,z)&=-R^{-1}\Pi_{xx}\hat{x}^{*}(t,y,z)-R^{-1}\Pi_{xz}z,\\
	v^{*}(t,y,z)&=-M^{-1}\Pi_{zx}\hat{x}^{*}(t,y,z)-M^{-1}\Pi_{zz}z.
\end{align}
These equations clarify that the optimal controls are determined based on the optimal state estimator.

In the target tracking problem, $p_{t}(x|y,z)$ is a Gaussian distribution, and $\mb{E}_{p_{t}(x|y,z)}\left[x\right]$ is a linear function of observation $y$ and memory $z$ as follows: 
\begin{align}
	\mb{E}_{p_{t}(x|y,z)}\left[x\right]=K_{xy}y+K_{xz}z, 
\end{align}
which is the same as Eq. (\ref{eq: ABIP-conditional expectation by estimation gain}) in the target estimation problem. 
The estimation gains, $K_{xy}$ and $K_{xz}$, and the control gains, $\Pi_{xx}$, $\Pi_{xz}$, $\Pi_{zx}$, and $\Pi_{zz}$, are nonlinear functions of $D$, $E$, $F$, $Q$, $R$, and $M$, which are obtained from HJB and FP equations. 
We note that $\Pi_{xz}=\Pi_{zx}$ holds, which means that the control from the memory to the state is identical to that from the state to the memory.  
The details are shown in Appendix \ref{sec: ABDMA}. 

To verify the performance of the optimal control functions, we conduct a stochastic simulation  (Fig. \ref{fig: ABDM-diagram}(b-f)). 
Despite the high observation noise, the agent tracks the target accurately (Fig. \ref{fig: ABDM-diagram}(b)). 

\subsection{Phase Transition}\label{subsec: ABDM-TMS}
\begin{figure*}
	\includegraphics[width=170mm]{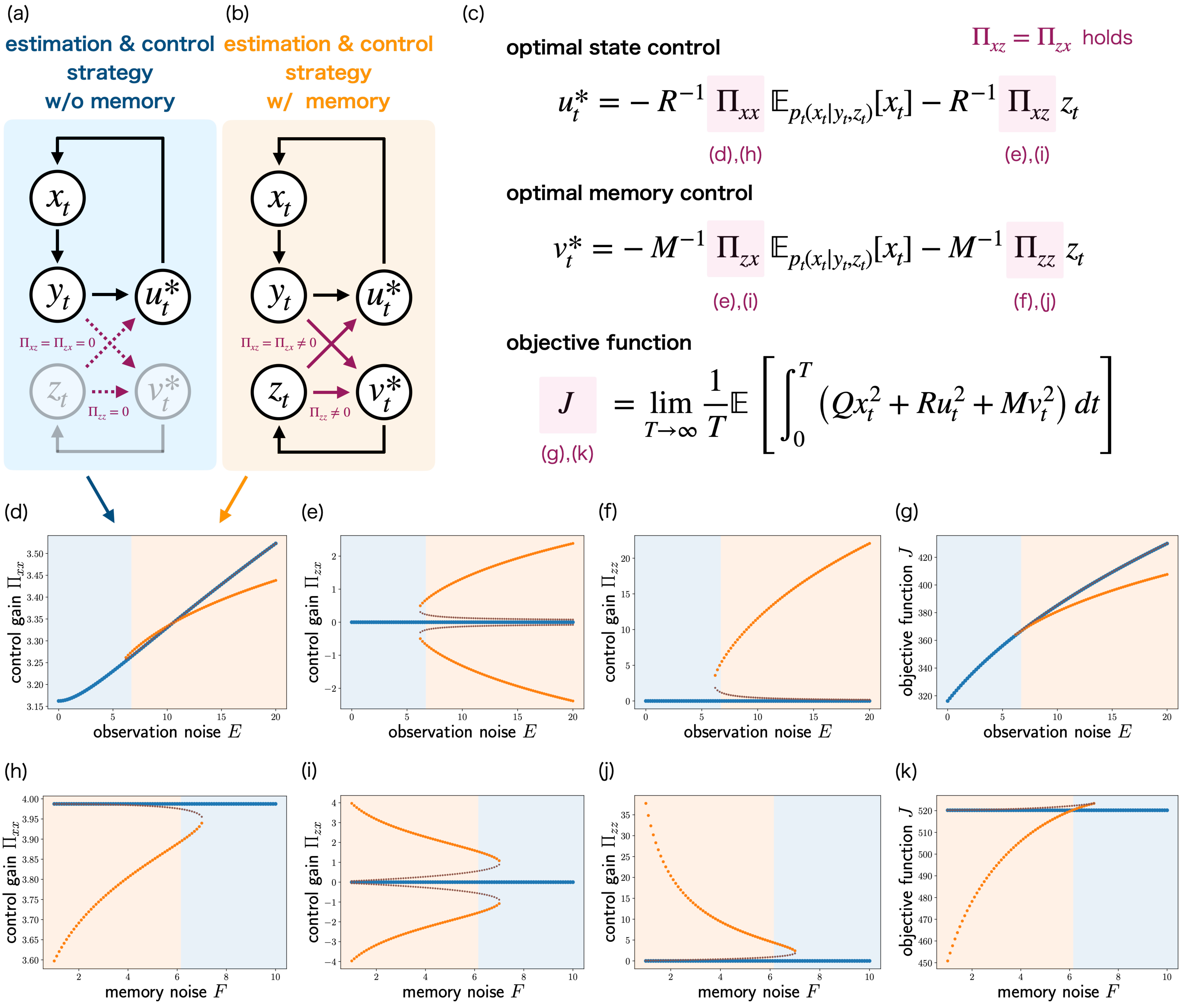}
	\caption{\label{fig: ABDM-TMS}
	Phase transition in optimal estimation and control strategy. 
	(a,b) Schematic diagrams of estimation and control strategies without memory (a) and with memory (b). 
	(a) When $\Pi_{zx}=\Pi_{zz}=0$, optimal state control $u_{t}$ is determined based only on current observation $y_{t}$. 
	(b) When $\Pi_{zx},\Pi_{zz}\neq0$, optimal state control $u_{t}$ is determined based not only on current observation $y_{t}$ but also on past memory $z_{t}$. 
	(c) Expressions of optimal state control function $u^{*}$, optimal memory control function $v^{*}$, and objective function $J$. 
	We note that $\Pi_{xz}=\Pi_{zx}$ holds. 
	(d--k) Control gains $\Pi_{xx}$, $\Pi_{zx}$, and $\Pi_{zz}$ and associated objective function $J$ with respect to observation noise $E$ (d--g) and memory noise $F$ (h--k). 
	$\Pi_{xz}$ is not visualized because $\Pi_{xz}=\Pi_{zx}$ holds. 
	The blue, brown, and orange dots are the solutions of Pontryagin's minimum principle, corresponding to zero, intermediate, and high memory controls, respectively. 
	The blue dots are optimal in the blue region, while the orange dots are optimal in the orange region. 
	There are no regions where the brown dots are optimal. 
	The parameters in (d--g) are $D=100$, $F=1$, $Q=10$, $R=1$, and $M=1$, while those in (i--k) are $D=100$, $E=50$, $Q=10$, $R=1$, and $M=1$.  
	}
\end{figure*}
\begin{figure*}
	\includegraphics[width=170mm]{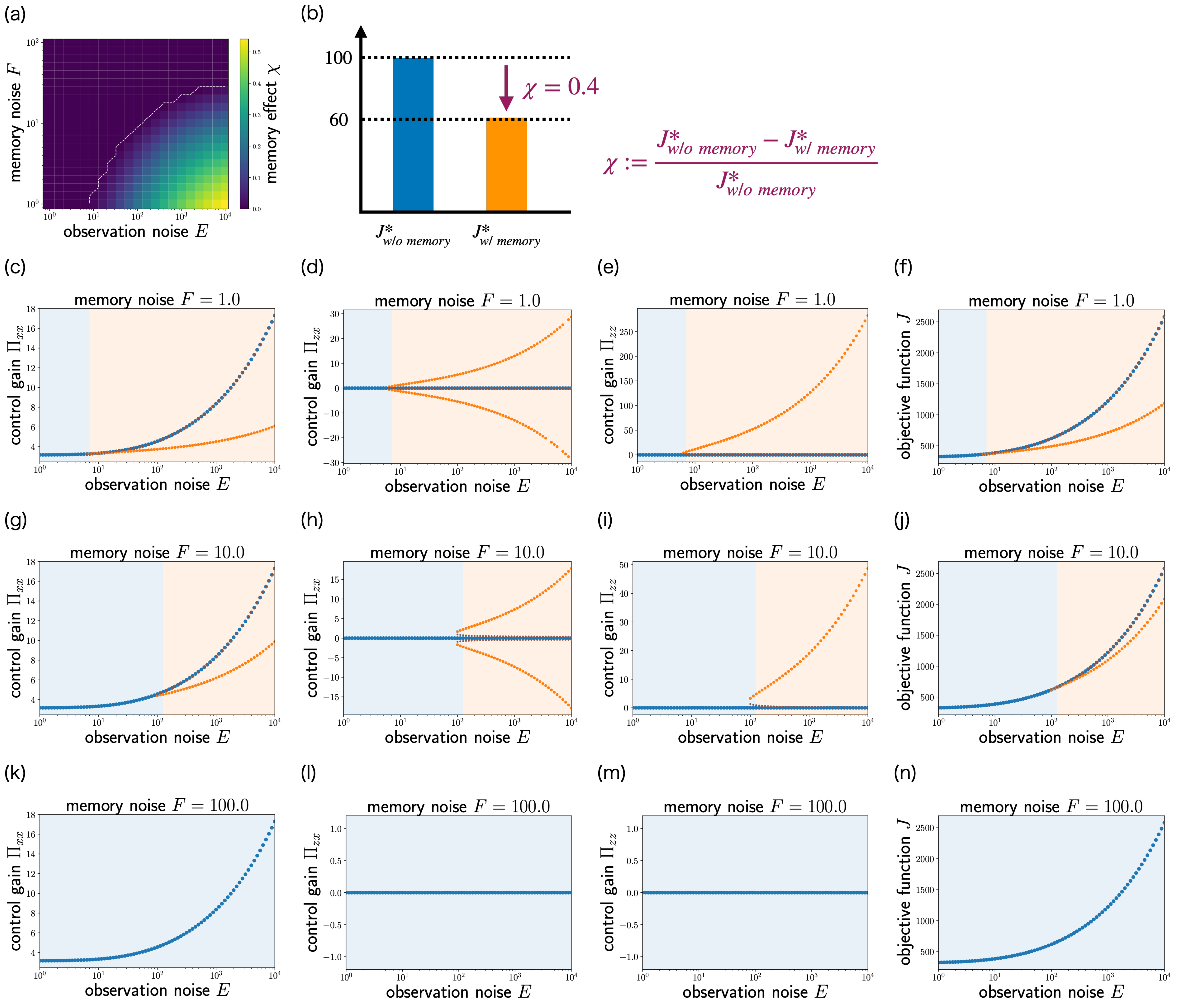}
	\caption{\label{fig: ABDM-TMS-2param}
	(a) Memory effect $\chi$ as a function of observation noise $E$ and memory noise $F$. 
	The white dashed curve marks the boundary between $\chi=0$ and $\chi>0$. 
	(b) Illustration of the definition of memory effect $\chi$. 
	(c--n) Control gains $\Pi_{xx}$, $\Pi_{zx}$, and $\Pi_{zz}$ and associated objective function $J$ with respect to observation noise $E$ under different memory noise $F$: lower memory noise (c--f), intermediate memory noise (g--j), and higher memory noise (k--n). 
	The blue, brown, and orange dots are the solutions of Pontryagin's minimum principle, corresponding to zero, intermediate, and high memory controls, respectively. 
	The blue dots are optimal in the blue region, while the orange dots are optimal in the orange region. 
	The rest of the parameters are $D=100$, $Q=10$, $R=1$, and $M=1$.  
	}
\end{figure*}
\begin{figure*}
	\includegraphics[width=170mm]{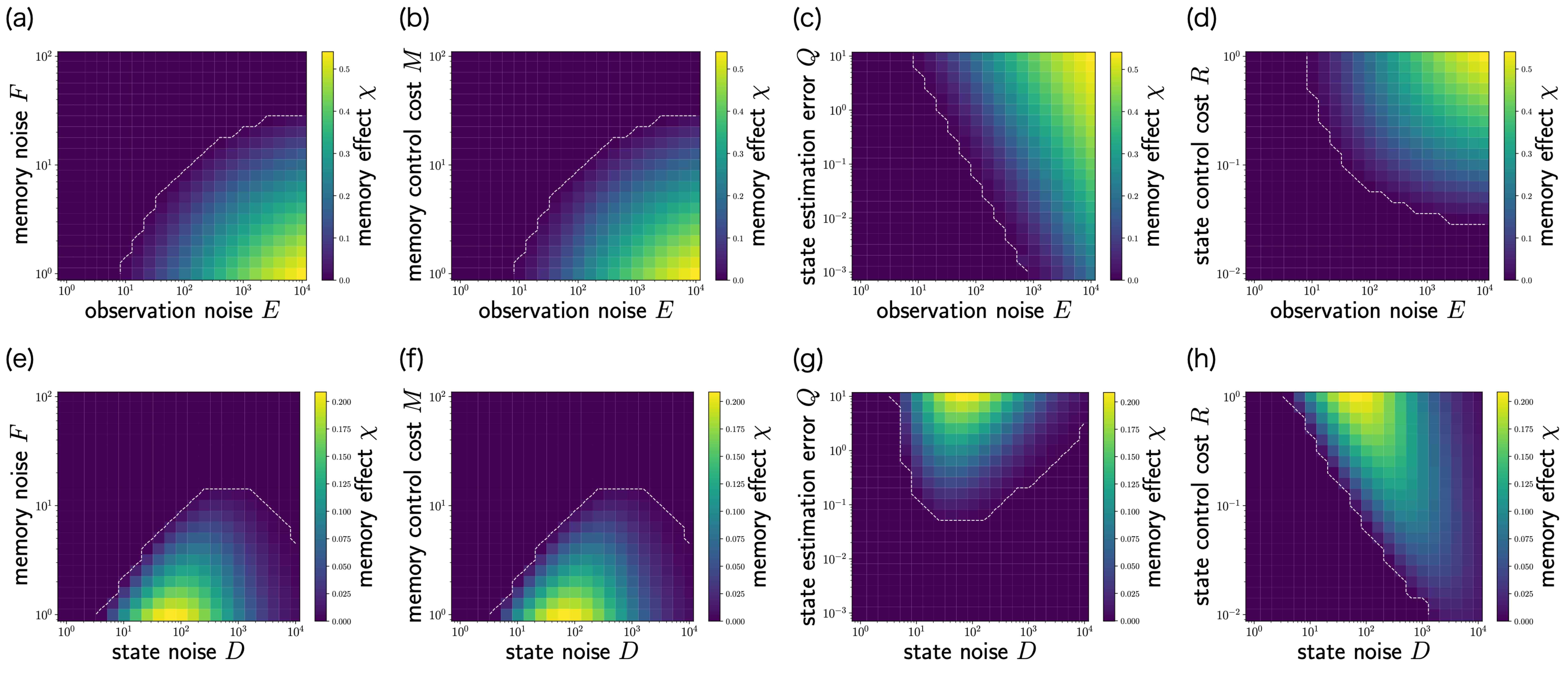}
	\caption{\label{fig: ABDM-scaling}
	Memory effect $\chi$ with respect to the indicated parameters. 
	The white dashed curves mark the boundary between $\chi=0$ and $\chi>0$. 
	The parameters not indicated in the figures are set to $D=100$, $E=100$, $F=1$, $Q=10$, $R=1$, and $M=1$. 
	}
\end{figure*}

We examine the parameter dependence of the optimal estimation and control strategy under resource limitations (Fig. \ref{fig: ABDM-TMS}). 
Similarly to the optimal estimation strategy, the optimal estimation and control strategy is also characterized by the memory control gains, $\Pi_{zx}$ and $\Pi_{zz}$. 
When $\Pi_{zx}=\Pi_{zz}=0$, the past observation information is not encoded to memory $z_{t}$, and the state estimator is reduced from $\mb{E}_{p_{t}(x|y,z)}\left[x\right]$ to $\mb{E}_{p_{t}(x|y)}\left[x\right]$. 
Furthermore, with $\Pi_{xz}=\Pi_{zx}=0$, the direct control from memory $z_{t}$ to state $x_{t}$ also disappears. 
Thus, optimal state control $u_{t}^{*}$ is determined based solely on current observation $y_{t}$, corresponding to the estimation and control strategy without memory (Fig. \ref{fig: ABDM-TMS}(a)). 
In contrast, when $\Pi_{zx},\Pi_{zz}\neq0$, optimal state control $u_{t}^{*}$ is determined based on both observation $y_{t}$ and memory $z_{t}$, resulting in the estimation and control strategy with memory (Fig. \ref{fig: ABDM-TMS}(b)). 

As observation noise $E$ increases, optimal memory control gains $\Pi_{zx}$ and $\Pi_{zz}$ undergo a discontinuous shift from zero to nonzero values (Fig. \ref{fig: ABDM-TMS}(d--g)), indicating that the memory-based control is optimal when the current observation is less reliable (Fig. \ref{fig: ABDM-TMS}(b)). 
Conversely, as memory noise $F$ increases, optimal memory control gains $\Pi_{zx}$ and $\Pi_{zz}$ discontinuously shift from nonzero values to zero (Fig. \ref{fig: ABDM-TMS}(h--k)), indicating that the memory-less control is optimal when the intrinsic information processing is highly stochastic (Fig. \ref{fig: ABDM-TMS}(a)). 
The discontinuous phase transition also occurs for the other parameters, i.e., state noise $D$, state tracking error $Q$, state control cost $R$, and memory control cost $M$ (See Appendix \ref{sec: ABDMA}, Fig. \ref{fig: ABDMA-TMS}). 
While the phase transition for $D$, $Q$, and $R$ is similar to that for $E$, the phase transition for $M$ is similar to that for $F$. 
These results suggest that the discontinuous phase transition is an inherent feature of biological information processing and decision-making under resource limitations. 

We further investigate the relationship between observation noise $E$ and memory noise $F$ in the optimal estimation and control strategy (Fig. \ref{fig: ABDM-TMS-2param}).  
Unlike the target estimation problem, the target tracking problem does not exhibit the non-monotonic phase transition with respect to observation noise $E$. 
Memory effect $\chi$ monotonically increases with respect to observation noise $E$ (Fig. \ref{fig: ABDM-TMS-2param}(a)), and the phase transition from memory-based to memory-less control strategies does not occur even when observation noise $E$ is extremely high and memory noise $F$ is at an intermediate level (Fig. \ref{fig: ABDM-TMS-2param}(g--j)). 

This difference may be attributed to the fact that in the target tracking problem, the objective function diverges to infinity when the agent fails in estimation and control. 
In the target estimation problem, even if the agent neglects the sensing information and fails in estimation, the upper bound of the state estimation error is limited to the finite variance of the environmental state (Fig. \ref{fig: ABIP-TMS-2param}(e,h,k)), because it considers the environmental state dynamics that are stable near the origin without state control. 
As a result, the agent does not need to control the memory when the observation noise is extremely high, leading to the non-monotonic phase transition with respect to the observation noise. 
By contrast, in the target tracking problem, if the agent neglects the sensing information in estimation and control, the agent completely loses track of the target, and the state tracking error diverges to infinity (Fig. \ref{fig: ABDM-TMS-2param}(f,j,n)). 
Therefore, even when the observation noise is extremely high, the agent cannot afford to give up controlling the memory to avoid the diverging tracking error. 

We also explore memory effect $\chi$ with respect to state noise $D$, state tracking error $Q$, state control cost $R$, and memory control cost $M$, in addition to observation noise $E$ and memory noise $F$ (Fig. \ref{fig: ABDM-scaling}). 
While increasing observation noise $E$ does not lead to a non-monotonic phase transition (Fig. \ref{fig: ABDM-scaling}(a--d)), increasing state noise $D$ exhibits such a transition (Fig. \ref{fig: ABDM-scaling}(e--h)). 
This may be because, even if state noise $D$ is extremely high, accurate state estimation can still be achieved solely through current observation $y_{t}$ as long as observation noise $E$ is low. 
Moreover, while increasing memory noise $F$ and memory control cost $M$ decreases memory effect $\chi$ (Fig. \ref{fig: ABDM-scaling}(a,b,e,f)), increasing state tracing error $Q$ and state control cost $R$ has the opposite effect, increasing memory effect $\chi$ (Fig. \ref{fig: ABDM-scaling}(c,d,g,h)). 
The phase boundaries for memory noise $F$ and memory control cost $M$ are similar, suggesting a scaling relationship between them, such as $MF$. 
On the other hand, since the phase boundaries for state tracking error $Q$ and state control cost $R$ differ from those for memory noise $F$ and memory control cost $M$, it suggests that a scaling relationship, such as $Q/MF$ or $R/MF$, does not hold in the target tracking problem. 
These results indicate that while the discontinuity in the phase transition is a fundamental characteristic under resource limitations, the non-monotonicity and the scaling relationship are model- and parameter-dependent. 

\section{Discussion}\label{sec: discussion}
In this paper, we have proposed a comprehensive theoretical framework for addressing biological information processing and decision-making under resource limitations. 
Instead of relying on Bayesian filtering of posterior probabilities, our theory employs the optimal control of memories that organisms can store and operate, which achieves optimal estimation under resource limitations. 
Additionally, since this theory realizes state estimation through memory control, it can be straightforwardly extended to include state control, thereby allowing for an integrated approach towards biological information processing and decision-making. 
We applied this theory to minimal models of biological information processing and decision-making and investigated optimal estimation and control strategies under resource limitations. 
Our analyses reveal that, despite the simplicity of these models, resource limitations induce discontinuous phase transitions between memory-less and memory-based estimations and controls. 
Therefore, resource limitations may contribute to the rich diversity observed in biological information processing and decision-making.

For the discontinuous phase transitions we observed,  a key question is whether such phase transitions also occur in real biological systems. 
One possible example of this phenomenon can be found in the circadian rhythms of the cyanobacteria, {\it S. elongatus} and {\it P. marinus} \cite{johnson_timing_2017,monti_robustness_2018,pittayakanchit_biophysical_2018,seki_evolution_2022}. 
In the circadian rhythms of cyanobacteria, it is crucial to estimate the time of day by processing sunlight information through intracellular chemical reactions. 
The circadian rhythm of {\it S. elongatus} exhibits a self-sustained oscillation due to a protein called KaiA, which functions as a negative feedback controller of intracellular chemical reactions \cite{ishiura_expression_1998,iwasaki_kaia-stimulated_2002,nakajima_reconstitution_2005,tomita_no_2005,kawamoto_damped_2020}. 
In contrast, {\it P. marinus} has lost the KaiA protein through evolutionary processes \cite{dvornyk_origin_2003}, and the intracellular chemical reaction exhibits a damped oscillation rather than a self-sustained one \cite{holtzendorff_genome_2008,axmann_biochemical_2009,mullineaux_rolex_2009,ma_evolution_2016,chew_high_2018}.
We hypothesize that the KaiA protein regulating the circadian rhythms of cyanobacteria corresponds to the memory control in our theoretical framework. 
In this context, the self-sustained oscillation of {\it S. elongatus} may represent a complex information processing with memory control, while the damped oscillation of {\it P. marinus} may correspond to a simple information processing without memory control. 
In fact, {\it P. marinus} has a much smaller cell volume than {\it S. elongatus}, leading to a higher level of intrinsic stochasticity due to the limited number of molecules available for information processing \cite{pittayakanchit_biophysical_2018}. 
Additionally, {\it S. elongatus} inhabits a wider range of environments than {\it P. marinus}, which likely requires more accurate estimations \cite{flombaum_present_2013}. 
These experimental observations suggest a potential alignment with our theoretical predictions.

Another example can be found in the dual-process theory of human information processing and decision-making \cite{epstein_integration_1994,muraven_self-control_1998,kahneman_perspective_2003,stanovich_who_1999}. 
This theory posits that human cognition operates through two distinct systems: System 1, which is unconscious, implicit, and intuitive, and System 2, which is conscious, explicit, and logical. 
Recent experiments have reported that these systems switch based on available energy resources, such as the level of glucose in the bloodstream \cite{masicampo_toward_2008,mcmahon_glucose_2010,wang_sweet_2010}. 
When the glucose level is low, System 1 dominates, whereas System 2 becomes more active as the glucose level increases. 
This transition is similar to the findings in our study. 
System 1 represents simple information processing and decision-making, prioritizing energy cost reduction, whereas System 2 involves more complex operations, enhancing performance even at the expense of higher energy costs. 
Therefore, the dual-process theory could serve as a good example of how resource limitations can lead to phase transitions in the complexity and capacity of information processing and decision-making.

Investigation of these phenomena requires a more complex and involved setup than those in this study. Our theory is general enough to include such high-dimensional and nonlinear cases, yet we restricted our analysis in this study to the one-dimensional and linear minimal models partly for efficient computation of phase diagrams.
To work on more complex cases, efficient numerical algorithms for solving high-dimensional HJB-FP equations should be developed. 
This is a challenging task. 
We may employ some techniques \cite{achdou_finite_2013,achdou_mean_2020,lauriere_numerical_2021} and neural network-based algorithms \cite{ruthotto_machine_2020,lin_alternating_2021,pham_mean-field_2023,cao_connecting_2024}, which have been developed in the context of mean-field games and control \cite{bensoussan_master_2015,bensoussan_interpretation_2017,carmona_probabilistic_2018,carmona_probabilistic_2018-1} to resolve this problem. 

If being equipped with these algorithms, our theory could address more realistic and complex biological situations where environmental states are higher-dimensional than biological memories \cite{sasagawa_prediction_2005,covert_achieving_2005,hasegawa_multidimensional_2018,hahn_dynamical_2023}. 
The conventional Bayesian filtering theory struggles to address such situations because the dimensions of biological memories are determined based on those of environmental states. 
Our theory addresses this issue by allowing the dimensions of biological memories to be determined independently of those of environmental states from which valuable insights would be gained on how organisms compress high-dimensional environmental information into low-dimensional biological memories.
While the dimensionality reduction in machine learning is one approach to this problem \cite{bishop_pattern_2006,van_der_maaten_dimensionality_2007,lee_nonlinear_2007}, the potential strength of our theory lies in the explicit consideration of the dynamics of biological memories. 
This would allow us to potentially reveal phenomena where diverse environmental information is encoded into varied dynamical patterns of biological memories, including sustained, transient, and oscillatory dynamics. 
Similar phenomena have recently been observed in cellular information processing, where different ligand information is encoded into the diverse dynamical concentration patterns of the same downstream molecule \cite{sasagawa_prediction_2005,covert_achieving_2005,hasegawa_multidimensional_2018,hahn_dynamical_2023}. 
Therefore, our theory could also serve as a foundation for dynamical information encoding in biological systems. 

Apart from future problems in numerical computation, one important theoretical challenge is to extend our theory from a single-agent system to a multi-agent system with communication. 
In nature, organisms often form groups, allowing for more complex and sophisticated information processing and decision-making via communication \cite{fancher_fundamental_2017,varennes_emergent_2017,erez_cell--cell_2020,vennettilli_multicellular_2020,tweedy_seeing_2020,pezzotta_chemotaxis_2018,durve_collective_2020,borra_optimal_2021}. 
We have achieved this extension in our previous work where we proposed a very similar yet biologically less useful theoretical framework \cite{tottori_decentralized_2023}. 
Therefore, our theory has the potential to provide an effective approach for investigating collective information processing and decision-making and associated phase transitions.

\begin{acknowledgments}
We thank Kenji Kashima, Simon K. Schnyder, Kento Nakamura, Shuhei A. Horiguchi, Masaki Kato, and Taro Toyoizumi for valuable discussions. 
The first author received a JSPS Research Fellowship (Grant Number 22KJ0557). 
This research was supported by JST CREST (Grant Number JPMJCR2011) and JSPS KAKENHI (Grant Number 19H05799, 24H02148, and 24H01465).
\end{acknowledgments}

\appendix
\onecolumngrid
\section{Observation-based Stochastic Control}\label{sec: OSC}
\begin{figure*}
	\includegraphics[width=170mm]{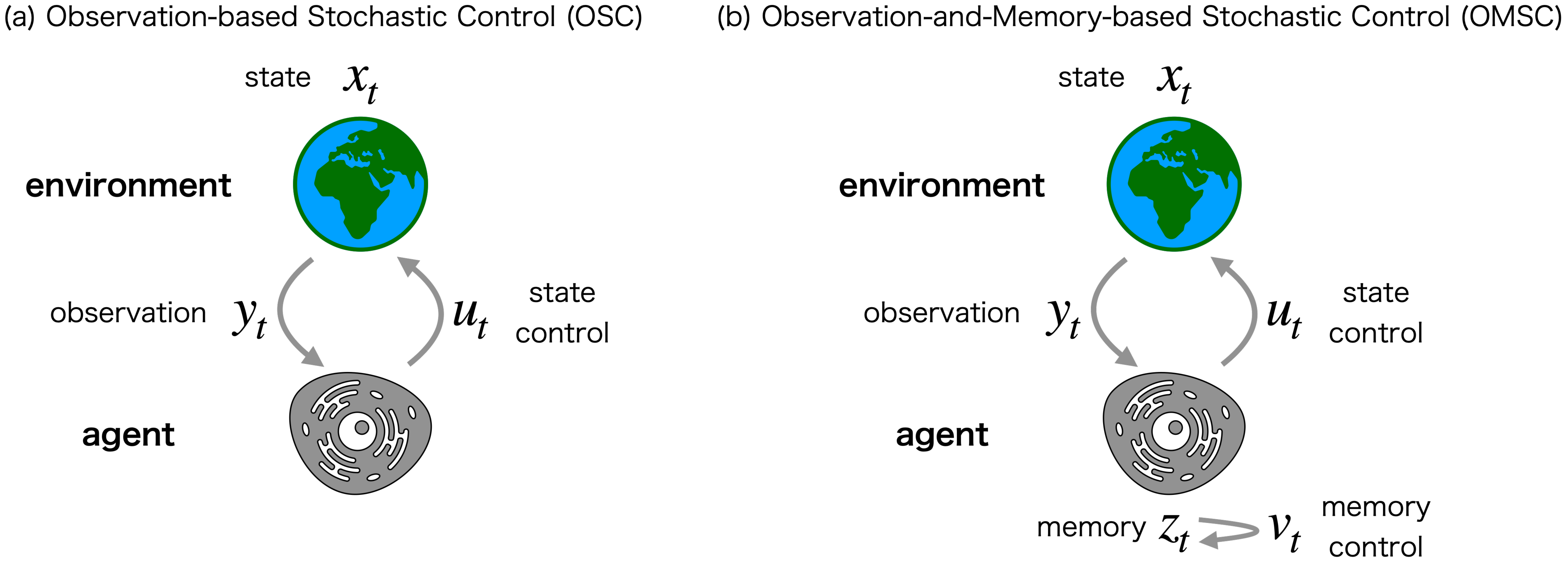}
	\caption{\label{fig: OSC-OMSC}
	Schematic diagrams of (a) observation-based stochastic control (OSC) and (b) observation-and-memory-based stochastic control (OMSC). 
	OSC and OMSC are optimal estimation and control problems without memory and with memory, respectively. 
	We discuss OSC and OMSC in Appendix \ref{sec: OSC} and \ref{sec: OMSC}, respectively. 
	}
\end{figure*}

The purpose of Appendix \ref{sec: OSC} and \ref{sec: OMSC} is to derive the optimal solutions of our theory presented in Sec. \ref{sec: NTOE} [Eq. (\ref{eq: NTOE-optimal control function})] and \ref{sec: NTOEC} [Eq. (\ref{eq: NTOEC-optimal control function})]. 
Appendix \ref{sec: OSC} serves as a preliminary section, where we consider the optimal estimation and control problem without memory (Fig. \ref{fig: OSC-OMSC}(a)). 
We refer to this problem as observation-based stochastic control (OSC) because it corresponds to a stochastic optimal control problem where the agent determines the control based only on the current observation. 
Appendix \ref{sec: OMSC} is the main section, where we address the optimal estimation and control problem with memory (Fig. \ref{fig: OSC-OMSC}(b)). 
This problem is referred to as observation-and-memory-based stochastic control (OMSC) because the agent determines the controls based on the current observation and memory. 
OMSC corresponds to the optimal estimation and control problem with resource limitations presented in Sec. \ref{sec: NTOEC}. 
Additionally, the optimal estimation problem with resource limitations presented in Sec. \ref{sec: NTOE} is a special case of OMSC.  
Thus, the purpose of Appendix \ref{sec: OSC} and \ref{sec: OMSC} reduces to deriving the optimal solution of OMSC. 
The optimal solution of OMSC can be derived in a similar manner to that of OSC because OMSC can be interpreted as a special case of OSC on an extended space (Fig. \ref{fig: OMSC-OSC}). 
Therefore, in this section, we aim to derive the optimal solution of OSC as a preparatory step for deriving that of OMSC. 
The optimal solution of OMSC is then derived in the following section, Appendix \ref{sec: OMSC}. 

It should be noted that OSC and OMSC build upon our  previously proposed theoretical framework, memory-limited partially observable stochastic control (ML-POSC) \cite{tottori_memory-limited_2022,tottori_forward-backward_2023,tottori_decentralized_2023}. 
ML-POSC is an optimal estimation and control theory with resource limitations. 
While OMSC is similar to ML-POSC, the problem formulation is modified, allowing for more biologically plausible setups. 
A comparison between OMSC and ML-POSC is provided in Appendix \ref{sec: ML-POSC}. 

This section is organized as follows: 
In Appendix \ref{subsec: OSC-PF}, we formulate OSC. 
In Appendix \ref{subsec: OSC-OS}, we present the optimal solution of OSC. 
In Appendix \ref{subsec: OSC-PMP}, we derive the optimal solution of OSC using Pontryagin's minimum principle on the probability density function space. 
In Appendix \ref{subsec: OSC-MLM}, we show that Pontryagin's minimum principle corresponds to the method of Lagrange multipliers. 

\subsection{Problem Formulation}\label{subsec: OSC-PF}
In this subsection, we formulate the optimal estimation and control problem without memory, i.e., observation-based stochastic control (OSC) (Fig. \ref{fig: OSC-OMSC}(a)). 
OSC considers an environment and an agent. 
The state of the environment at time $t\in[0,T]$ is denoted by $x_{t}\in\mb{R}^{d_{x}}$, which evolves by the following stochastic differential equation (SDE): 
\begin{align}
	dx_{t}=b(t,x_{t},u_{t})dt+\sigma(t,x_{t},u_{t})d\omega_{t}, 
	\label{eq: OSC-state SDE}
\end{align}
where initial state $x_{0}$ follows $p_{0}(x_{0})$, $u_{t}\in\mb{R}^{d_{u}}$ is the control of the agent, and $\omega_{t}\in\mb{R}^{d_{\omega}}$ is a standard Wiener process. 
The agent cannot completely observe environmental state $x_{t}$ and instead obtains noisy observation $y_{t}\in\mb{R}^{d_{y}}$, which is generated from probability density function $p_{t}(y_{t}|x_{t})$ as follows: 
\begin{align}
	y_{t}\sim p_{t}(y_{t}|x_{t}). 
	\label{eq: OSC-observation PDF}
\end{align}
The agent determines control $u_{t}$ based on observation $y_{t}$: 
\begin{align}
	u_{t}=u(t,y_{t}). 
	\label{eq: OSC-state control}
\end{align}
The objective function is given by the following expected cumulative cost function: 
\begin{align}
	J[u]:=\mb{E}\left[\int_{0}^{T}f(t,x_{t},u_{t})dt+g(x_{T})\right],
	\label{eq: OSC-OF}
\end{align}
where $f$ and $g$ are the running and terminal cost functions, respectively. 
OSC considers a problem to find optimal control function $u^{*}$ that minimizes objective function $J[u]$: 
\begin{align}
	u^{*}:=\arg\min_{u}J[u]. 
	\label{eq: OSC}
\end{align}

We refer to this stochastic optimal control problem as observation-based stochastic control (OSC) because the agent determines control $u_{t}$ based on observation $y_{t}$ [Eq. (\ref{eq: OSC-state control})]. 
If observation $y_{t}$ is identical to state $x_{t}$, i.e., $y_{t}=x_{t}$, this problem reduces to the most basic stochastic control, where the agent can completely observe state $x_{t}$ and determines control $u_{t}$ based on it as $u_{t}=u(t,x_{t})$ \cite{bellman_dynamic_1958,kushner_optimal_1962,fleming_deterministic_1975,yong_stochastic_1999,nisio_stochastic_2015,bensoussan_estimation_2018}. 
In contrast, in OSC, the agent cannot completely observe state $x_{t}$ and needs to estimate it from observation $y_{t}$. 
Additionally, the agent is not allowed to use information from past observation history $y_{t-dt},\cdots,y_{0}$ to estimate and control state $x_{t}$. 
Therefore, OSC is an optimal estimation and control problem without memory. 

\subsection{Optimal Solution}\label{subsec: OSC-OS}
In the conventional stochastic control, where the agent can completely observe state $x_{t}$, the optimal control function is obtained using Bellman's dynamic programming principle on the state space \cite{bellman_dynamic_1958,kushner_optimal_1962,fleming_deterministic_1975,yong_stochastic_1999,nisio_stochastic_2015,bensoussan_estimation_2018}. 
However, this method is not applicable to OSC because the agent cannot completely observe state $x_{t}$. 
To address this issue, we propose Pontryagin's minimum principle on the probability density function space, which has recently been used in mean-field stochastic optimal control problems \cite{bensoussan_master_2015,bensoussan_interpretation_2017,carmona_probabilistic_2018,carmona_probabilistic_2018-1} and in our previous works \cite{tottori_memory-limited_2022,tottori_forward-backward_2023,tottori_decentralized_2023}. 
In this subsection, we show the main result. 
The details of the derivation are shown in the following subsection, Appendix \ref{subsec: OSC-PMP}. 

The optimal control function of OSC is given by minimizing the conditional expected Hamiltonian: 
\begin{align}
	u^{*}(t,y)=\arg\min_{u}\mb{E}_{p_{t}(x|y)}\left[\mcal{H}(t,x,u,w)\right]. 
	\label{eq: OSC-OCF}
\end{align}
Hamiltonian $\mcal{H}$ is defined by
\begin{align}
	\mcal{H}(t,x,u,w):=f(t,x,u)+\mcal{L}_{u}w(t,x), 
	\label{eq: OSC-Hamiltonian}
\end{align}
where $f(t,x,u)$ is the current cost and $\mcal{L}_{u}w(t,x)$ represents the future cost. 
$\mcal{L}_{u}$ is the time-backward diffusion operator for state $x$, which is defined by
\begin{align}
	\mcal{L}_{u}w(t,x):=\sum_{i=1}^{d_{x}}b_{i}(t,x,u)\frac{\partial w(t,x)}{\partial x_{i}}
	+\frac{1}{2}\sum_{i,j=1}^{d_{x}}D_{ij}(t,x,u)\frac{\partial^{2} w(t,x)}{\partial x_{i}\partial x_{j}},
	\label{eq: OSC-time-backward diffusion operator}
\end{align}
where $D(t,x,u):=\sigma(t,x,u)\sigma^{\top}(t,x,u)$. 
$w(t,x)$ is the value function, which represents the expected cumulative cost from the starting point at time $t$ and state $x$ to the terminal time $T$. 
Value function $w(t,x)$ is obtained by solving the following time-backward partial differential equation, Hamilton-Jacobi-Bellman (HJB) equation: 
\begin{align}
	-\frac{\partial w(t,x)}{\partial t}=\mb{E}_{p_{t}(y|x)}\left[\mcal{H}(t,x,u^{*},w)\right], 
	\label{eq: OSC-HJB eq}
\end{align}
where the terminal condition is given by $w(T,x)=g(x)$. 
In the optimal control function of OSC [Eq. (\ref{eq: OSC-OCF})], conditional expectation $\mb{E}_{p_{t}(x|y)}\left[\cdot\right]$ appears, which is consistent with the fact that the agent needs to estimate state $x$ from observation $y$. 
Conditional probability density function $p_{t}(x|y)$ is calculated as follows: 
\begin{align}
	p_{t}(x|y)=\frac{p_{t}(y|x)p_{t}(x)}{\int p_{t}(y|x)p_{t}(x)dx}.
\end{align}
Probability density function $p_{t}(x):=p(t,x)$ is obtained by solving the following time-forward partial differential equation, Fokker-Planck (FP) equation: 
\begin{align}
	\frac{\partial p(t,x)}{\partial t}=\check{\mcal{L}}_{u^{*}}^{\dag}p(t,x), 
	\label{eq: OSC-FP eq}
\end{align}
where the initial condition is given by $p(0,x)=p_{0}(x)$. 
$\check{\mcal{L}}_{u}^{\dag}$ is the expected time-forward diffusion operator for state $x$, which is defined by
\begin{align}
	\check{\mcal{L}}_{u}^{\dag}p(t,x)&:=-\sum_{i=1}^{d_{x}}\frac{\partial\left(\mb{E}_{p_{t}(y|x)}\left[b_{i}(t,x,u(t,y))\right]p(t,x)\right)}{\partial x_{i}}
	+\frac{1}{2}\sum_{i,j=1}^{d_{x}}\frac{\partial^{2}\left(\mb{E}_{p_{t}(y|x)}\left[D_{ij}(t,x,u(t,y))\right]p(t,x)\right)}{\partial x_{i}\partial x_{j}}. 
	\label{eq: OSC-expected time-forward diffusion operator}
\end{align}

The optimal control function of OSC [Eq. (\ref{eq: OSC-OCF})] is obtained by solving the time-backward HJB equation [Eq. (\ref{eq: OSC-HJB eq})] and the time-forward FP equation [Eq. (\ref{eq: OSC-FP eq})]. 
Since HJB and FP equations are mutually dependent through the optimal control function, they must be solved simultaneously, which corresponds to a two-point boundary value problem of partial differential equations. 
This problem can be solved using the forward-backward sweep method, which computes the time-backward HJB equation and the time-forward FP equation alternately \cite{tottori_forward-backward_2023}. 
While the convergence of the forward-backward sweep method is not guaranteed in deterministic optimal control problems \cite{hackbusch_numerical_1978,mcasey_convergence_2012,lenhart_optimal_2007,sharp_implementation_2021} and mean-field stochastic optimal control problems \cite{carlini_semi-lagrangian_2013,carlini_fully_2014,carlini_semi-lagrangian_2015,lauriere_numerical_2021}, it is guaranteed in OSC because the coupling of HJB and FP equations is limited to the optimal control function.  
This proof is almost the same as our previous work \cite{tottori_forward-backward_2023}. 

We note that Pontryagin's minimum principle is a necessary condition for the optimal control function, but not a sufficient condition \cite{tottori_forward-backward_2023}. 
This means that while the optimal control function always satisfies Eq. (\ref{eq: OSC-OCF}), a control function that satisfies Eq. (\ref{eq: OSC-OCF}) is not necessarily the optimal control function. 
This is evident from the fact that Pontryagin's minimum principle corresponds to the method of Lagrange multipliers, as shown in Appendix \ref{subsec: OSC-MLM}. 
In fact, our minimal models yield not only optimal solutions but also non-optimal solutions based on Pontryagin's minimum principle (See Fig. \ref{fig: ABIP-TMS}(d--f)(blue, brown, and orange dots) as an example). 

If observation $y_{t}$ is identical to state $x_{t}$, i.e., $y_{t}=x_{t}$, OSC reduces to the conventional stochastic control, and Eq. (\ref{eq: OSC-OCF}) becomes 
\begin{align}
	u^{*}(t,x)=\arg\min_{u}\mcal{H}(t,x,u,w). 
	\label{eq: SC-OCF}
\end{align}
In this case, the optimal control function is obtained by solving only the time-backward HJB equation [Eq. (\ref{eq: OSC-HJB eq})], which is consistent with Bellman's dynamic programming principle on the state space \cite{bellman_dynamic_1958,kushner_optimal_1962,fleming_deterministic_1975,yong_stochastic_1999,nisio_stochastic_2015,bensoussan_estimation_2018}. 
Therefore, Pontryagin's minimum principle on the probability density function space can be interpreted as a generalization of Bellman's dynamic programming principle on the state space to cases of incomplete observation \cite{tottori_memory-limited_2022,tottori_forward-backward_2023,tottori_decentralized_2023}. 

\subsection{Derivation}\label{subsec: OSC-PMP}
In this subsection, we derive the optimal control function of OSC [Eq. (\ref{eq: OSC-OCF})] using Pontryagin's minimum principle on the probability density function space. 
In this paper, we derive this result in three steps, following our previous works \cite{tottori_memory-limited_2022,tottori_decentralized_2023}. 
First, we convert OSC on the state space into the deterministic control on the probability density function space (Appendix \ref{subsec: OSC PMP deterministic control}). 
Second, we apply Bellman's dynamic programming principle on the probability density function space, not on the state space (Appendix \ref{subsec: OSC PMP DPP}). 
Finally, we convert Bellman's dynamic programming principle on the probability density function space into Pontryagin's minimum principle on the same space (Appendix \ref{subsec: OSC PMP PMP}). 
It is worth noting that Pontryagin's minimum principle is directly applicable without going through Bellman's dynamic programming principle. 
We omit this proof because it is nearly identical to our previous work \cite{tottori_forward-backward_2023}. 

\subsubsection{Conversion to deterministic control on probability density function space}\label{subsec: OSC PMP deterministic control}
We first convert OSC on the state space into the deterministic control on the probability density function space. 
The state SDE [Eq. (\ref{eq: OSC-state SDE})] is converted into the following FP equation: 
\begin{align}
	\frac{\partial p(t,x)}{\partial t}=\check{\mcal{L}}_{u}^{\dag}p(t,x), 
	\label{eq: OSC-PMP-FP eq}
\end{align}
where $p(0,x)=p_{0}(x)$ and $\check{\mcal{L}}_{u}^{\dag}$ is the expected time-forward diffusion operator [Eq. (\ref{eq: OSC-expected time-forward diffusion operator})]. 
The objective function of OSC [Eq. (\ref{eq: OSC-OF})] is also calculated as follows: 
\begin{align}
	J[u]=\int_{0}^{T}\bar{f}(t,p_{t},u_{t})dt+\bar{g}(p_{T}), 
	\label{eq: OSC-PMP-OF}
\end{align}
where $p_{t}(x):=p(t,x)$, $\bar{f}(t,p,u):=\mb{E}_{p(x)}[\mb{E}_{p_{t}(y|x)}[f(t,x,u(t,y))]]$, and $\bar{g}(p):=\mb{E}_{p(x)}[g(x)]$. 
From Eqs. (\ref{eq: OSC-PMP-FP eq}) and (\ref{eq: OSC-PMP-OF}), OSC on the state space is converted into the deterministic control on the probability density function space. 
While OSC on the state space controls state SDE [Eq. (\ref{eq: OSC-state SDE})], the deterministic control on the probability density function space controls FP equation [Eq. (\ref{eq: OSC-PMP-FP eq})].  

\subsubsection{Bellman's dynamic programming principle on probability density function space}\label{subsec: OSC PMP DPP}
While OSC cannot completely observe state $x_{t}$, it can obtain complete information about probability density function $p_{t}$ because probability density function $p_{t}$ is deterministic. 
As a result, OSC can be solved by Bellman's dynamic programming principle on the probability density function space, not on the state space. 
In the following, we show Bellman's dynamic programming principle on the probability density function space. 

The minimization of the objective function is calculated as follows: 
\begin{align}	
	\min_{u}J[u]
	&=\min_{u_{0:T-dt}}\left[\int_{0}^{T}\bar{f}(\tau,p_{\tau},u_{\tau})d\tau+\bar{g}(p_{T})\right]\nonumber\\
	&=\min_{u_{0:t-dt}}\left[\int_{0}^{t-dt}\bar{f}(\tau,p_{\tau},u_{\tau})d\tau+\min_{u_{t:T-dt}}\left[\int_{t}^{T}\bar{f}(t,p_{\tau},u_{\tau})d\tau+\bar{g}(p_{T})\right]\right]\nonumber\\
	&=\min_{u_{0:t-dt}}\left[\int_{0}^{t-dt}\bar{f}(\tau,p_{\tau},u_{\tau})d\tau+V(t,p_{t})\right], 
\end{align}
where $V(t,p)$ is the value function, which is defined as follows: 
\begin{align}
	V(t,p):=\min_{u_{t:T-dt}}\left[\int_{t}^{T}\bar{f}(t,p_{\tau},u_{\tau})d\tau+\bar{g}(p_{T})\right],
\end{align}
where $\{p_{\tau}|\tau\in[t,T]\}$ is the solution of FP equation [Eq. (\ref{eq: OSC-PMP-FP eq})] with $p_{t}=p$. 
The optimal control function is obtained by calculating the value function. 

The value function is calculated as follows: 
\begin{align}
	V(t,p)
	&=\min_{u_{t:T}}\left[\int_{t}^{T}\bar{f}(\tau,p_{\tau},u_{\tau})d\tau+\bar{g}(p_{T})\right]\nonumber\\
	&=\min_{u}\left[\bar{f}(t,p,u)dt+\min_{u_{t+dt:T}}\left[\int_{t+dt}^{T}\bar{f}(\tau,p_{\tau},u_{\tau})d\tau+\bar{g}(p_{T})\right]\right]\nonumber\\
	&=\min_{u}\left[\bar{f}(t,p,u)dt+V(t+dt,p+\check{\mcal{L}}_{u}^{\dag}pdt)\right]\nonumber\\
	&=\min_{u}\left[\bar{f}(t,p,u)dt+V(t,p)+\frac{\partial V(t,p)}{\partial t}dt+\left(\int \frac{\delta V(t,p)}{\delta p}(x)\check{\mcal{L}}_{u}^{\dag}p(x)dx\right)dt\right].
\end{align}
By rearranging the equation, the following equation is obtained: 
\begin{align}
	-\frac{\partial V(t,p)}{\partial t}
	&=\min_{u}\left[\bar{f}(t,p,u)+\int \frac{\delta V(t,p)}{\delta p}(x)\check{\mcal{L}}_{u}^{\dag}p(x)dx\right], 
\end{align}
where $V(T,p)=\bar{g}(p)$. 
We define expected time-backward diffusion operator $\check{\mcal{L}}_{u}$ as follows: 
\begin{align}
	\check{\mcal{L}}_{u}w(t,x)&:=\sum_{i=1}^{d_{x}}\mb{E}_{p_{t}(y|x)}\left[b_{i}(t,x,u(t,y))\right]\frac{\partial w(t,x)}{\partial x_{i}}+\frac{1}{2}\sum_{i,j=1}^{d_{x}}\mb{E}_{p_{t}(y|x)}\left[D_{ij}(t,x,u(t,y))\right]\frac{\partial^{2} w(t,x)}{\partial x_{i}\partial x_{j}}. 
	\label{eq: expected backward diffusion operator}
\end{align}
Since 
\begin{align}
	\int \frac{\delta V(t,p)}{\delta p}(x)\check{\mcal{L}}_{u}^{\dag}p(x)dx=\int p(x)\check{\mcal{L}}_{u}\frac{\delta V(t,p)}{\delta p}(x)dx, 
\end{align}
the following equation is obtained: 
\begin{align}
	-\frac{\partial V(t,p)}{\partial t}
	&=\min_{u}\left[\bar{f}(t,p,u)+\int p(x)\check{\mcal{L}}_{u}\frac{\delta V(t,p)}{\delta p}(x)dx\right]\nonumber\\
	&=\min_{u}\mb{E}_{p_{t}(y|x)p(x)}\left[f(t,x,u)+\mcal{L}_{u}\frac{\delta V(t,p)}{\delta p}(x)\right], 
\end{align}
where $\mcal{L}_{u}$ is the time-backward diffusion operator [Eq. (\ref{eq: OSC-time-backward diffusion operator})]. 
From the definition of the Hamiltonian [Eq. (\ref{eq: OSC-Hamiltonian})], the following equation is obtained:  
\begin{align}
	-\frac{\partial V(t,p)}{\partial t}
	&=\min_{u}\mb{E}_{p_{t}(y|x)p(x)}\left[\mcal{H}\left(t,x,u,\frac{\delta V(t,p)}{\delta p}(x)\right)\right], 
	\label{eq: Bellman eq pre}
\end{align}
where $V(T,p)=\mathbb{E}_{p(x)}[g(x)]$. 
This time-backward functional differential equation is called Bellman equation. 

Optimal control function $u^{*}$ is given by the right-hand side of Bellman equation [Eq. (\ref{eq: Bellman eq pre})]. 
Since control $u$ is determined based on observation $y$ in OSC, the minimization by control $u$ can be exchanged with the expectation by observation $y$ as follows: 
\begin{align}
	-\frac{\partial V(t,p)}{\partial t}
	&=\mb{E}_{p_{t}(y)}\left[\min_{u}\mb{E}_{p_{t}(x|y)}\left[\mcal{H}\left(t,x,u,\frac{\delta V(t,p)}{\delta p}(x)\right)\right]\right], 
\end{align}
where 
\begin{align}
	p_{t}(y)&:=\int p_{t}(y|x)p(x) dx,\\
	p_{t}(x|y)&:=\frac{p_{t}(y|x)p(x)}{\int p_{t}(y|x)p(x) dx}. 
\end{align}
As a result, optimal control function $u^{*}$ is given by minimizing the conditional expected Hamiltonian as follows: 
\begin{align}	
	u^{*}(t,y,p)&=\arg\min_{u}\mb{E}_{p_{t}(x|y)}\left[\mcal{H}\left(t,x,u,\frac{\delta V(t,p)}{\delta p}(x)\right)\right], 
	\label{eq: optimal control of OSC DPP}
\end{align}
where $V(t,p)$ is the solution of the following Bellman equation: 
\begin{align}
	-\frac{\partial V(t,p)}{\partial t}
	&=\mb{E}_{p_{t}(y|x)p(x)}\left[\mcal{H}\left(t,x,u^{*},\frac{\delta V(t,p)}{\delta p}(x)\right)\right], 
	\label{eq: Bellman eq}
\end{align}
where $V(T,p)=\mathbb{E}_{p(x)}[g(x)]$. 
Therefore, from Bellman's dynamic programming principle on the probability density function space, the optimal control function of OSC [Eq. (\ref{eq: optimal control of OSC DPP})] is obtained by solving Bellman equation [Eq. (\ref{eq: Bellman eq})] backward in time from terminal condition $V(T,p)=\mathbb{E}_{p(x)}[g(x)]$. 

In Bellman's dynamic programming principle on the probability density function space, the optimal control function of OSC [Eq. (\ref{eq: optimal control of OSC DPP})] depends on probability density function $p$ as $u^{*}(t,y,p)$, while it is defined by $u^{*}(t,y)$. 
In OSC, probability density function $p$ is given by the solution of FP equation [Eq. (\ref{eq: OSC-PMP-FP eq})], which is denoted by $p_{t}$. 
Therefore, in Bellman's dynamic programming principle on the probability density function space, we first compute Bellman equation [Eq. (\ref{eq: Bellman eq})] backward in time from terminal condition $V(T,p)=\mathbb{E}_{p(x)}[g(x)]$ and obtain optimal control function $u^{*}(t,y,p)$. 
We then compute FP equation [Eq. (\ref{eq: OSC-PMP-FP eq})] forward in time from initial condition $p(0,x)=p_{0}(x)$ and obtain optimal control function $u^{*}(t,y):=u^{*}(t,y,p_{t})$. 
The agent determines optimal control $u_{t}^{*}$ based on real-time observation $y_{t}$ as $u_{t}^{*}:=u^{*}(t,y_{t})$. 

\subsubsection{Conversion to Pontryagin's minimum principle on probability density function space}\label{subsec: OSC PMP PMP}
Bellman's dynamic programming principle on the probability density function space is less practical because Bellman equation [Eq. (\ref{eq: Bellman eq})] is a functional differential equation, not a partial differential equation. 
We circumvent this problem by converting Bellman's dynamic programming principle on the probability density function space to Pontryagin's minimum principle on the same space. 
This conversion has also been used in mean-field stochastic optimal control problems \cite{bensoussan_master_2015,bensoussan_interpretation_2017,carmona_probabilistic_2018,carmona_probabilistic_2018-1} and in our previous works \cite{tottori_memory-limited_2022,tottori_decentralized_2023}. 

We first define 
\begin{align}
	W(t,p,x):=\frac{\delta V(t,p)}{\delta p}(x), 
	\label{eq: def of large w}
\end{align}
which satisfies $W(T,p,x)=g(x)$. 
Differentiating Bellman equation [Eq. (\ref{eq: Bellman eq})] with respect to probability density function $p$, the following equation is obtained: 
\begin{align}
	-\frac{\partial W(t,p,x)}{\partial t}
	&=\mb{E}_{p_{t}(y|x)}\left[\mcal{H}\left(t,x,u^{*},W\right)\right]\nonumber\\
	&\ \ \ +\mb{E}_{p_{t}(y|x')p(x')}\left[\frac{\partial\mcal{H}\left(t,x',u^{*},W\right)}{\partial u}\frac{\delta u^{*}(t,y,p)}{\delta p}(x)\right]
	+\mb{E}_{p_{t}(y|x')p(x')}\left[\mcal{L}_{u^{*}}\frac{\delta W(t,p,x')}{\delta p}(x)\right]\nonumber\\
	&=\mb{E}_{p_{t}(y|x)}\left[\mcal{H}\left(t,x,u^{*},W\right)\right]\nonumber\\
	&\ \ \ +\mb{E}_{p_{t}(y)p_{t}(x'|y)}\left[\frac{\partial\mcal{H}\left(t,x',u^{*},W\right)}{\partial u}\frac{\delta u^{*}(t,y,p)}{\delta p}(x)\right]
	+\mb{E}_{p(x')}\left[\check{\mcal{L}}_{u^{*}}\frac{\delta W(t,p,x')}{\delta p}(x)\right]\nonumber\\
	&=\mb{E}_{p_{t}(y|x)}\left[\mcal{H}\left(t,x,u^{*},W\right)\right]\nonumber\\
	&\ \ \ +\mb{E}_{p_{t}(y)}\left[\frac{\partial\mb{E}_{p_{t}(x'|y)}\left[\mcal{H}\left(t,x',u^{*},W\right)\right]}{\partial u}\frac{\delta u^{*}(t,y,p)}{\delta p}(x)\right]
	+\int\frac{\delta W(t,p,x)}{\delta p}(x')\check{\mcal{L}}_{u^{*}}^{\dag}p(x')dx'.
\end{align}
From the optimal control function of OSC [Eq. (\ref{eq: optimal control of OSC DPP})], the following stationary condition is satisfied: 
\begin{align}
	\frac{\partial\mb{E}_{p_{t}(x'|y)}\left[\mcal{H}\left(t,x',u^{*},W\right)\right]}{\partial u}=0. 
\end{align}
Therefore, the following equation is obtained: 
\begin{align}
	-\frac{\partial W(t,p,x)}{\partial t}
	&=\mb{E}_{p_{t}(y|x)}\left[\mcal{H}\left(t,x,u^{*},W\right)\right]
	+\int\frac{\delta W(t,p,x)}{\delta p}(x')\check{\mcal{L}}_{u^{*}}^{\dag}p(x')dx'.
	\label{eq: Master eq}
\end{align}

We then define 
\begin{align}
	w(t,x):=W(t,p_{t},x), 
	\label{eq: def of w}
\end{align}
where $p_{t}$ is the solution of FP equation [Eq. (\ref{eq: OSC-PMP-FP eq})]. 
The time derivative of $w(t,x)$ is calculated as follows: 
\begin{align}
	\frac{\partial w(t,x)}{\partial t}
	&=\frac{\partial W(t,p_{t},x)}{\partial t}+\int\frac{\delta W(t,p_{t},x)}{\delta p}(x')\frac{\partial p_{t}(x')}{\partial t}dx'. 
	\label{eq: the partial derivative of w(t,s)}
\end{align}
By substituting Eq. (\ref{eq: Master eq}) into Eq. (\ref{eq: the partial derivative of w(t,s)}), the following equation is obtained: 
\begin{align}
	-\frac{\partial w(t,x)}{\partial t}
	&=\mb{E}_{p_{t}(y|x)}\left[\mcal{H}\left(t,x,u^{*},w\right)\right]-\int\frac{\delta W(t,p_{t},x)}{\delta p}(x')\underbrace{\left(\frac{\partial p_{t}(x')}{\partial t}-\check{\mcal{L}}_{u^{*}}^{\dag}p_{t}(x')\right)}_{(*)}dx'. 
\end{align}
From FP equation [Eq. (\ref{eq: OSC-PMP-FP eq})], $(*)=0$ holds and Bellman equation [Eq. (\ref{eq: Bellman eq})] is reduced to the following time-backward partial differential equation, HJB equation:  
\begin{align}
	-\frac{\partial w(t,x)}{\partial t}
	&=\mb{E}_{p_{t}(y|x)}\left[\mcal{H}\left(t,x,u^{*},w\right)\right], 
	\label{eq: OSC-PMP-HJB eq}
\end{align}
where $w(T,x)=g(x)$. 
The optimal control function of OSC [Eq. (\ref{eq: optimal control of OSC DPP})] is given as follows: 
\begin{align}	
	u^{*}(t,y)&=\arg\min_{u}\mb{E}_{p_{t}(x|y)}\left[\mcal{H}\left(t,x,u,w\right)\right],
	\label{eq: OSC-PMP-optimal control function}
\end{align}
where $p_{t}(x|y):=p_{t}(y|x)p_{t}(x)/\int p_{t}(y|x)p_{t}(x) dx$ and $p_{t}(x):=p(t,x)$ is the solution of FP equation [Eq. (\ref{eq: OSC-PMP-FP eq})]. 
Therefore, the optimal control function of OSC [Eq. (\ref{eq: OSC-PMP-optimal control function})] is obtained by solving the time-backward HJB equation  [Eq. (\ref{eq: OSC-PMP-HJB eq})] and the time-forward FP equation [Eq. (\ref{eq: OSC-PMP-FP eq})]. 
This optimality condition corresponds to Pontryagin's minimum principle on the probability density function space [Eq. (\ref{eq: OSC-OCF})]. 

In Pontryagin's minimum principle on the state space, the time-forward state equation and the time-backward adjoint equation can be expressed by the derivatives of the Hamiltonian \cite{pontryagin_mathematical_1987,vinter_optimal_2010,lewis_optimal_2012}. 
Similarly, in Pontryagin's minimum principle on the probability density function space, the time-forward FP equation [Eq. (\ref{eq: OSC-PMP-FP eq})] and the time-backward HJB equation [Eq. (\ref{eq: OSC-PMP-HJB eq})] can be expressed by the variations of expected Hamiltonian 
\begin{align}	
	\bar{\mcal{H}}(t,p,u,w)&:=\mb{E}_{p(x)}\left[\mb{E}_{p_{t}(y|x)}\left[\mcal{H}(t,x,u,w)\right]\right]
	\label{eq: expected Hamiltonian}
\end{align}
as follows:
\begin{align}
	\frac{\partial p(t,x)}{\partial t}&=\frac{\delta\bar{\mcal{H}}(t,p,u^{*},w)}{\delta w}(x),\label{eq: FP eq PMP}\\
	-\frac{\partial w(t,x)}{\partial t}&=\frac{\delta\bar{\mcal{H}}(t,p,u^{*},w)}{\delta p}(x), \label{eq: HJB eq PMP}
\end{align}
where $p(0,x)=p_{0}(x)$ and $w(T,x)=g(x)$ \cite{tottori_forward-backward_2023}. 

While Bellman's dynamic programming principle on the probability density function space is a terminal value problem of a functional differential equation, Bellman equation  [Eq. (\ref{eq: Bellman eq})], Pontryagin's minimum principle on that space is a two-point boundary value problem of partial differential equations, HJB and FP equations [Eqs. (\ref{eq: FP eq PMP}) and (\ref{eq: HJB eq PMP})], which simplifies computations. 
We note that while Bellman's dynamic programming principle is a necessary and sufficient condition for the optimal control function, Pontryagin's minimum principle is a necessary condition, not a sufficient condition \cite{tottori_forward-backward_2023}. 
Pontryagin's minimum principle becomes a necessary and sufficient condition when expected Hamiltonian $\bar{\mcal{H}}(t,p,u,w)$ is convex with respect to $p$ and $u$, which is shown in our previous work \cite{tottori_forward-backward_2023}.

\subsection{Relationship with the Method of Lagrange Multipliers}\label{subsec: OSC-MLM}
In this subsection, we show that Pontryagin's minimum principle corresponds to the method of Lagrange multipliers for readers who may be unfamiliar with Pontryagin's minimum principle. 
The method of Lagrange multipliers is a widely used technique for solving constrained optimization problems. 
The deterministic control on the probability density function space can be interpreted as the following constrained optimization problem: 
\begin{align}
	{\rm minimize}\ \ \ &J[u,p]=\int_{0}^{T}\bar{f}(t,p_{t},u_{t})dt+\bar{g}(p_{T}),\\
	{\rm subject\ to}\ \ \ &\frac{\partial p(t,x)}{\partial t}=\check{\mcal{L}}_{u}^{\dag}p(t,x),\quad p(0,x)=p_{0}(x). 
\end{align}
As a result, the method of Lagrange multipliers is an effective approach to solving this problem. 

In the method of Lagrange multipliers, we consider the following Lagrange function: 
\begin{align}
	\tilde{J}:=\int_{0}^{T}\bar{f}(t,p_{t},u_{t})dt+\bar{g}(p_{T})+\int_{0}^{T}\int w(t,x)\left(\check{\mcal{L}}_{u}^{\dag}p(t,x)-\frac{\partial p(t,x)}{\partial t}\right)dxdt,
\end{align}
where $w(t,x)$ is a Lagrange multiplier. 
Lagrange function $\tilde{J}$ is calculated as follows: 
\begin{align}
	\tilde{J}
	&=\int_{0}^{T}\bar{f}(t,p_{t},u_{t})dt+\bar{g}(p_{T})
	+\int_{0}^{T}\int p(t,x)\check{\mcal{L}}_{u}w(t,x)dxdt\nonumber\\
	&\ \ \ -\int \left(w(T,x)p(T,x)-w(0,x)p(0,x)\right)dx
	+\int_{0}^{T}\int p(t,x)\frac{\partial w(t,x)}{\partial t}dxdt\nonumber\\
	&=\int_{0}^{T}\int\int p(t,x)p_{t}(y|x)\mcal{H}(t,x,u,w)dxdydt+\int p(T,x)g(x)dx\nonumber\\
	&\ \ \ -\int \left(p(T,x)w(T,x)-p(0,x)w(0,x)\right)dx
	+\int_{0}^{T}\int p(t,x)\frac{\partial w(t,x)}{\partial t}dxdt.
\end{align}
We variate Lagrange function $\tilde{J}$ with respect to $u$ and $p$ as follows: 
\begin{align}
	\delta\tilde{J}
	&=\int_{0}^{T}\int\int p_{t}(y|x)\mcal{H}(t,x,u,w)\delta p(t,x)dxdydt
	+\int_{0}^{T}\int\int p(t,x)p_{t}(y|x)\frac{\partial\mcal{H}(t,x,u,w)}{\partial u}\delta u(t,y)dxdydt\nonumber\\
	&\ \ \ +\int g(x)\delta p(T,x)dx
	-\int w(T,x)\delta p(T,x)dx
	+\int_{0}^{T}\int \frac{\partial w(t,x)}{\partial t}\delta p(t,x)dxdt
	\nonumber\\
	&=\int_{0}^{T}\int\left(\mb{E}_{p_{t}(y|x)}\left[\mcal{H}(t,x,u,w)\right]+\frac{\partial w(t,x)}{\partial t}\right)\delta p(t,x)dxdt
	+\int\left(g(x)-w(T,x)\right)\delta p(T,x)dx\nonumber\\
	&\ \ \ +\int_{0}^{T}\int p(t,y)\mb{E}_{p_{t}(x|y)}\left[\frac{\partial\mcal{H}(t,x,u,w)}{\partial u}\right]\delta u(t,y)dydt.
\end{align}
From stationary condition $\delta\tilde{J}=0$, we obtain the following optimality condition: 
\begin{align}
	&\mb{E}_{p_{t}(x|y)}\left[\frac{\partial\mcal{H}(t,x,u,w)}{\partial u}\right]=0,\label{eq: OSC-MLM-OCF}\\
	&-\frac{\partial w(t,x)}{\partial t}=\mb{E}_{p_{t}(y|x)}\left[\mcal{H}(t,x,u,w)\right],\quad w(T,x)=g(x),\label{eq: OSC-MLM-HJB eq}\\
	&\frac{\partial p(t,x)}{\partial t}=\check{\mcal{L}}_{u}^{\dag}p(t,x),\quad p(0,x)=p_{0}(x).\label{eq: OSC-MLM-FP eq}
\end{align}
Equations (\ref{eq: OSC-OCF}), (\ref{eq: OSC-HJB eq}), and  (\ref{eq: OSC-FP eq}) correspond to Eqs. (\ref{eq: OSC-MLM-OCF}), (\ref{eq: OSC-MLM-HJB eq}), and  (\ref{eq: OSC-MLM-FP eq}), respectively. 
Therefore, Pontryagin's minimum principle corresponds to the method of Lagrange multipliers. 

\section{Observation-and-Memory-based Stochastic Control}\label{sec: OMSC}
In this section, we derive the optimal solution of OMSC (Fig. \ref{fig: OSC-OMSC}(b)). 
OMSC corresponds to the optimal estimation and control problem with resource limitations presented in Sec. \ref{sec: NTOEC}. 
Additionally, the optimal estimation problem with resource limitations presented in Sec. \ref{sec: NTOE} is a special case of OMSC.  
Therefore, deriving the optimal solution of OMSC corresponds to deriving the optimal solutions presented in Sec. \ref{sec: NTOE} [Eq. (\ref{eq: NTOE-optimal control function})] and \ref{sec: NTOEC} [Eq. (\ref{eq: NTOEC-optimal control function})]. 

We note that OMSC is similar to our previous works \cite{tottori_memory-limited_2022,tottori_forward-backward_2023,tottori_decentralized_2023}, but the problem formulation is slightly modified, allowing for more biologically plausible discussions. 
A comparison between OMSC and our previous works \cite{tottori_memory-limited_2022,tottori_forward-backward_2023,tottori_decentralized_2023} is provided in Appendix \ref{sec: ML-POSC}. 

This section is organized as follows: 
In Appendix \ref{subsec: OMSC-PF}, we formulate OMSC. 
In Appendix \ref{subsec: OMSC-OSC}, we show that OMSC can converted into OSC on an extended space (Fig. \ref{fig: OMSC-OSC}). 
In Appendix \ref{subsec: OMSC-OS}, we derive the optimal solution of OMSC based on Appendix \ref{subsec: OMSC-OSC}. 

\subsection{Problem Formulation}\label{subsec: OMSC-PF}
In this subsection, we formulate the optimal estimation and control problem with memory, observation-and-memory-based stochastic control (OMSC) (Fig. \ref{fig: OSC-OMSC}(b)). 
OMSC considers the same state $x_{t}\in\mb{R}^{d_{x}}$ and observation $y_{t}\in\mb{R}^{d_{y}}$ as OSC, which are given by
\begin{align}
	dx_{t}&=b(t,x_{t},u_{t})dt+\sigma(t,x_{t},u_{t})d\omega_{t}, \label{eq: OMSC-state SDE}\\
	y_{t}&\sim p_{t}(y_{t}|x_{t}),\label{eq: OMSC-observation PDF}
\end{align}
where initial state $x_{0}$ follows $p_{0}(x_{0})$, $u_{t}\in\mb{R}^{d_{u}}$ is the state control of the agent, and $\omega_{t}\in\mb{R}^{d_{\omega}}$ is a standard Wiener process. 
In contrast, OMSC formulates the memory of the agent, $z_{t}\in\mb{R}^{d_{z}}$, which evolves by the following SDE: 
\begin{align}
	dz_{t}=c(t,z_{t},v_{t})dt+\eta(t,z_{t},v_{t})d\xi_{t}, 
	\label{eq: OMSC-memory SDE}
\end{align}
where initial memory $z_{0}$ follows $p_{0}(z_{0})$, $v_{t}\in\mb{R}^{d_{v}}$ is the memory control of the agent, and $\xi_{t}\in\mb{R}^{d_{\xi}}$ is a standard Wiener process. 
OMSC determines state control $u_{t}$ and memory control $v_{t}$ based on observation $y_{t}$ and memory $z_{t}$ as follows: 
\begin{align}
	u_{t}=u(t,y_{t},z_{t}),\label{eq: OMSC-state control}\\
	v_{t}=v(t,y_{t},z_{t}).\label{eq: OMSC-memory control}
\end{align}
The objective function is given by the following expected cumulative cost function: 
\begin{align}
	J[u]:=\mb{E}\left[\int_{0}^{T}f(t,x_{t},z_{t},u_{t},v_{t})dt+g(x_{T},z_{T})\right],
	\label{eq: OMSC-OF}
\end{align}
where $f$ and $g$ are the running and terminal cost functions, respectively. 
OMSC considers a problem to find optimal state control function $u^{*}$ and optimal memory control function $v^{*}$ that minimize objective function $J[u]$: 
\begin{align}
	u^{*}:=\arg\min_{u}J[u]. 
	\label{eq: OMSC}
\end{align}

OMSC is identical to the optimal estimation and control problem under resource limitations presented in Sec. \ref{sec: NTOEC}. 
Additionally, the optimal estimation problem under resource limitations presented in Sec. \ref{sec: NTOE} is a special case of OMSC, where $b(t,x_{t},u_{t})=b(t,x_{t})$ and $\sigma(t,x_{t},u_{t})=\sigma(t,x_{t})$, and state control $u_{t}=u(t,y_{t},z_{t})$ is replaced by state estimator $\hat{x}_{t}=\hat{x}(t,y_{t},z_{t})$. 

\subsection{Conversion to Observation-based Stochastic Control on Extended Space}\label{subsec: OMSC-OSC}
 \begin{figure*}
 	\includegraphics[width=170mm]{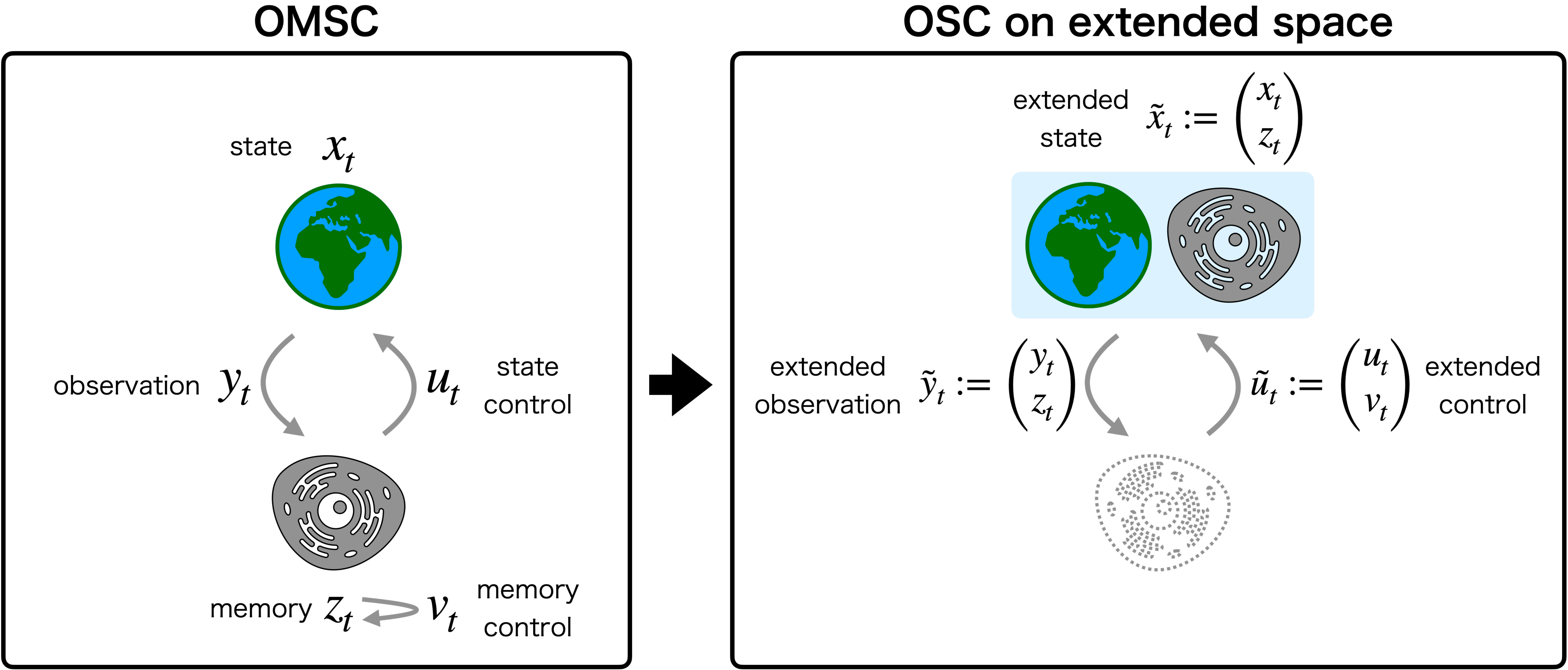}
 	\caption{\label{fig: OMSC-OSC}
 	Schematic diagram of the conversion from observation-and-memory-based stochastic control (OMSC) to observation-based stochastic control (OSC) on the extended space. 
 	}
 \end{figure*}
In this subsection, we show that OMSC can be converted into OSC on an extended space (Fig. \ref{fig: OMSC-OSC}) as a preparatory step to obtaining the optimal solution of OMSC. 
We define extended state $\tilde{x}_{t}$, extended observation $\tilde{y}_{t}$, and extended control $\tilde{u}_{t}$ as follows: 
\begin{align}
	\tilde{x}_{t}:=\left(\begin{array}{c}
		x_{t}\\
		z_{t}\\
	\end{array}
	\right),\quad 
	\tilde{y}_{t}:=\left(\begin{array}{c}
		y_{t}\\
		z_{t}'\\
	\end{array}
	\right),\quad 
	\tilde{u}_{t}:=\left(\begin{array}{c}
		u_{t}\\
		v_{t}\\
	\end{array}
	\right). 
\end{align}
We denote the memory in extended state $\tilde{x}_{t}$ by $z_{t}$ and that in extended observation $\tilde{y}_{t}$ by $z_{t}'$ to distinguish between them. 
In practice, $z_{t}=z_{t}'$ holds. 
Extended state $\tilde{x}_{t}$ evolves by the following SDE: 
\begin{align}
	d\tilde{x}_{t}=\tilde{b}(t,\tilde{x}_{t},\tilde{u}_{t})dt+\tilde{\sigma}(t,\tilde{x}_{t},\tilde{u}_{t})d\tilde{\omega}_{t}, 
\end{align}
where 
\begin{align}
	\tilde{b}(t,\tilde{x}_{t},\tilde{u}_{t}):=\left(\begin{array}{c}
		b(t,x_{t},u_{t})\\
		c(t,z_{t},v_{t})\\
	\end{array}
	\right),\quad
	\tilde{\sigma}(t,\tilde{x}_{t},\tilde{u}_{t}):=\left(\begin{array}{cc}
		\sigma(t,x_{t},u_{t})&O\\
		O&\eta(t,z_{t},v_{t})\\
	\end{array}
	\right),\quad
	\tilde{\omega}_{t}:=\left(\begin{array}{c}
		\omega_{t}\\
		\xi_{t}\\
	\end{array}
	\right). 
\end{align}
Extended observation $\tilde{y}_{t}$ is generated from the following probability density function: 
\begin{align}
	\tilde{p}_{t}(\tilde{y}_{t}|\tilde{x}_{t}):=p_{t}(y_{t}|x_{t})\delta(z_{t}'-z_{t}). 
\end{align}
Extended control $\tilde{u}_{t}$ is determined based on extended observation $\tilde{y}_{t}$ as follows: 
\begin{align}
	\tilde{u}_{t}=\tilde{u}(t,\tilde{y}_{t})
	:=\left(\begin{array}{c}
		u(t,y_{t},z_{t})\\
		v(t,y_{t},z_{t})\\
	\end{array}
	\right). 
\end{align}
The objective function is given by the following expected cumulative cost function: 
\begin{align}
	J[\tilde{u}]:=\mb{E}\left[\int_{0}^{T}\tilde{f}(t,\tilde{x}_{t},\tilde{u}_{t})dt+\tilde{g}(\tilde{x}_{T})\right], 
\end{align}
where 
\begin{align}
	\tilde{f}(t,\tilde{x}_{t},\tilde{u}_{t}):=f(t,x_{t},z_{t},u_{t},v_{t}),\quad 
	\tilde{g}(\tilde{x}_{T}):=g(x_{T},z_{T}).
\end{align}
Therefore, OMSC can be interpreted as OSC on the extended space. 

\subsection{Optimal Solution}\label{subsec: OMSC-OS}
The optimal solution of OMSC is can be obtained in a similar way to that of OSC because OMSC can be interpreted as OSC on the extended space.  
In the following, we present the optimal solution of OMSC, which corresponds to the optimal solutions presented in Sec. \ref{sec: NTOE} [Eq. (\ref{eq: NTOE-optimal control function})] and \ref{sec: NTOEC} [Eq. (\ref{eq: NTOEC-optimal control function})]. 
We note that the following result is identical to the optimal solution of OSC except for $\tilde{\cdot}$. 

The optimal extended control function of OMSC is given by minimizing the conditional expectation of the extended Hamiltonian: 
\begin{align}
	\tilde{u}^{*}(t,\tilde{y})=\arg\min_{\tilde{u}}\mb{E}_{\tilde{p}_{t}(\tilde{x}|\tilde{y})}\left[\tilde{\mcal{H}}(t,\tilde{x},\tilde{u},\tilde{w})\right]. 
	\label{eq: OMSC-OCF}
\end{align}
Extended Hamiltonian $\tilde{\mcal{H}}$ is defined by
\begin{align}
	\tilde{\mcal{H}}(t,\tilde{x},\tilde{u},\tilde{w}):=\tilde{f}(t,\tilde{x},\tilde{u})+\tilde{\mcal{L}}_{\tilde{u}}\tilde{w}(t,\tilde{x}), 
	\label{eq: OMSC-Hamiltonian}
\end{align}
where $\tilde{f}(t,\tilde{x},\tilde{u})$ is the current cost and $\tilde{\mcal{L}}_{\tilde{u}}\tilde{w}(t,\tilde{x})$ represents the future cost. 
$\tilde{\mcal{L}}_{\tilde{u}}$ is the time-backward diffusion operator for extended state $\tilde{x}$, which is defined by
\begin{align}
	\tilde{\mcal{L}}_{\tilde{u}}\tilde{w}(t,\tilde{x}):=\sum_{i=1}^{d_{\tilde{x}}}\tilde{b}_{i}(t,\tilde{x},\tilde{u})\frac{\partial \tilde{w}(t,\tilde{x})}{\partial \tilde{x}_{i}}
	+\frac{1}{2}\sum_{i,j=1}^{d_{\tilde{x}}}\tilde{D}_{ij}(t,\tilde{x},\tilde{u})\frac{\partial^{2} \tilde{w}(t,\tilde{x})}{\partial \tilde{x}_{i}\partial \tilde{x}_{j}},
	\label{eq: OMSC-time-backward diffusion operator}
\end{align}
where $\tilde{D}(t,\tilde{x},\tilde{u}):=\tilde{\sigma}(t,\tilde{x},\tilde{u})\tilde{\sigma}^{\top}(t,\tilde{x},\tilde{u})$. 
$\tilde{w}(t,\tilde{x})$ is the value function, which represents the expected cumulative cost from the starting point at time $t$ and extended state $\tilde{x}$ to the terminal time $T$. 
Value function $\tilde{w}(t,\tilde{x})$ is obtained by solving the following time-backward partial differential equation, HJB equation: 
\begin{align}
	-\frac{\partial \tilde{w}(t,\tilde{x})}{\partial t}=\mb{E}_{\tilde{p}_{t}(\tilde{y}|\tilde{x})}\left[\tilde{\mcal{H}}(t,\tilde{x},\tilde{u}^{*},\tilde{w})\right], 
	\label{eq: OMSC-HJB eq}
\end{align}
where the terminal condition is given by $\tilde{w}(T,\tilde{x})=\tilde{g}(\tilde{x})$. 
In the optimal extended control function of OMSC [Eq. (\ref{eq: OMSC-OCF})], conditional expectation $\mb{E}_{\tilde{p}_{t}(\tilde{x}|\tilde{y})}\left[\cdot\right]$ appears. 
Conditional probability density function $\tilde{p}_{t}(\tilde{x}|\tilde{y})$ is calculated as follows: 
\begin{align}
	\tilde{p}_{t}(\tilde{x}|\tilde{y})=\frac{\tilde{p}_{t}(\tilde{y}|\tilde{x})\tilde{p}_{t}(\tilde{x})}{\int \tilde{p}_{t}(\tilde{y}|\tilde{x})\tilde{p}_{t}(\tilde{x})d\tilde{x}}.
\end{align}
Probability density function $\tilde{p}_{t}(\tilde{x}):=\tilde{p}(t,\tilde{x})$ is obtained by solving the following time-forward partial differential equation, FP equation: 
\begin{align}
	\frac{\partial \tilde{p}(t,\tilde{x})}{\partial t}=\check{\tilde{\mcal{L}}}_{\tilde{u}^{*}}^{\dag}\tilde{p}(t,\tilde{x}), 
	\label{eq: OMSC-FP eq}
\end{align}
where the initial condition is given by $\tilde{p}(0,\tilde{x})=\tilde{p}_{0}(\tilde{x})$. 
$\check{\tilde{\mcal{L}}}_{\tilde{u}}^{\dag}$ is the expected time-forward diffusion operator for extended state $\tilde{x}$, which is defined by
\begin{align}
	\check{\tilde{\mcal{L}}}_{\tilde{u}}^{\dag}\tilde{p}(t,\tilde{x})&:=-\sum_{i=1}^{d_{\tilde{x}}}\frac{\partial\left(\mb{E}_{\tilde{p}_{t}(\tilde{y}|\tilde{x})}\left[\tilde{b}_{i}(t,\tilde{x},\tilde{u}(t,\tilde{y}))\right]\tilde{p}(t,\tilde{x})\right)}{\partial \tilde{x}_{i}}
	+\frac{1}{2}\sum_{i,j=1}^{d_{\tilde{x}}}\frac{\partial^{2}\left(\mb{E}_{\tilde{p}_{t}(\tilde{y}|\tilde{x})}\left[\tilde{D}_{ij}(t,\tilde{x},\tilde{u}(t,\tilde{y}))\right]\tilde{p}(t,\tilde{x})\right)}{\partial \tilde{x}_{i}\partial \tilde{x}_{j}}. 
	\label{eq: OMSC-expected time-forward diffusion operator}
\end{align}
Therefore, the optimal extended control function of OMSC [Eq. (\ref{eq: OMSC-OCF})] is obtained by solving HJB equation [Eq. (\ref{eq: OMSC-HJB eq})] and FP equation [Eq. (\ref{eq: OMSC-FP eq})]. 

\section{Linear-Quadratic-Gaussian Problem}\label{sec: LQG}
In this section, we apply OSC and OMSC to the Linear-Quadratic-Gaussian (LQG) problem, where the dynamics are linear Gaussian and the costs are quadratic. 
The LQG problem is a well-known problem in stochastic optimal control theory, which allows for semi-analytical calculations \cite{bensoussan_stochastic_1992,yong_stochastic_1999,nisio_stochastic_2015,bensoussan_estimation_2018}. 
We demonstrate that this advantageous property also holds in OSC and OMSC. 

The minimal models discussed in Sec. \ref{sec: ABIP} and  \ref{sec: ABDM} correspond to the infinite-horizon LQG problem. 
The relationship between these models and the infinite-horizon LQG problem is discussed in Appendix \ref{sec: ABIPA} and \ref{sec: ABDMA}. 

This section is organized as follows: 
In Appendix \ref{subsec: LQG-PF}, we formulate the LQG problem. 
In Appendix \ref{subsec: LQG-OS}, we present the optimal solution. 
In Appendix \ref{subsec: LQG-D}, we provide the derivation of the optimal solution. 
In Appendix \ref{subsec: LQG-infinite}, we extend the LQG problem from a finite-horizon setting to an infinite-horizon setting. 
We mainly discuss the case of OSC for the notational simplicity, but the same discussion is possible for the case of OMSC by adding $\tilde{\cdot}$, which represents extended variables and parameters. 

\subsection{Problem Formulation}\label{subsec: LQG-PF}
In this subsection, we formulate the LQG problem in OSC. 
In the LQG problem, the state of the environment, $x_{t}\in\mb{R}^{d_{x}}$, evolves by the following linear Gaussian dynamics: 
\begin{align}
	dx_{t}=\left(Ax_{t}+Bu_{t}\right)dt+\sigma d\omega_{t},
\end{align}
where initial state $x_{0}$ follows Gaussian distribution $\mcal{N}(x_{0}| \mu_{0},\Sigma_{0})$, $u_{t}\in\mb{R}^{d_{u}}$ is the state control of the agent, and $\omega_{t}\in\mb{R}^{d_{\omega}}$ is a standard Wiener process. 
The agent receives observation $y_{t}$ generated from the following Gaussian distribution: 
\begin{align}
	y_{t}\sim\mcal{N}(y_{t}|Hx_{t},E).
\end{align}
The agent determines control $u_{t}$ based on observation $y_{t}$ as $u_{t}=u(t,y_{t})$. 
The objective function is defined as the following quadratic function: 
\begin{align}
	J[u]:=\mb{E}\left[\int_{0}^{T}\left(x_{t}^{\top}Qx_{t}-2x_{t}^{\top}Su_{t}+u_{t}^{\top}Ru_{t}\right)dt+x_{T}^{\top}Px_{T}\right],\nonumber
\end{align}
where $Q\succeq O$, $R\succ O$, and $P\succeq O$. 
The LQG problem is to find optimal control function $u^{*}$ that minimizes objective function $J[u]$ as follows: 
\begin{align}
	u^{*}:=\arg\min_{u}J[u].
\end{align}
We note that the LQG problem in OMSC is formulated by adding $\tilde{\cdot}$. 

\subsection{Optimal Solution}\label{subsec: LQG-OS}
In this subsection, we present the optimal solution in the LQG problem of OSC. 
The derivation of the optimal solution is provided in the following subsection, Appendix \ref{subsec: LQG-D}. 

The optimal control function in the LQG problem of OSC is given by 
\begin{align}
	u^{*}(t,y)
	&=-R^{-1}\left(\left(\Psi B-S\right)^{\top}\mu+\left(\Pi B-S\right)^{\top}K\left(y-H\mu\right)\right), 
	\label{eq: LQG-OS-optimal control in LQG}
\end{align}
where $K:=\Sigma H^{\top}(E+H\Sigma H^{\top})^{-1}$. 
$\mu$ and $\Sigma$ are the mean vector and the covariance matrix of state $x$, respectively, which are the solutions of the following time-forward ordinary differential equations:  
\begin{align}
	\frac{d\mu}{dt}&=\left(A-BR^{-1}\left(\Psi B-S\right)^{\top}\right)\mu,\label{eq: LQG-OS-ODE of mu}\\
	\frac{d\Sigma}{dt}&=\sigma\sigma^{\top}+\left(A-BR^{-1}\left(\Pi B-S\right)^{\top}KH\right)\Sigma+\Sigma\left(A-BR^{-1}\left(\Pi B-S\right)^{\top}KH\right)^{\top},\label{eq: LQG-OS-ODE of Sigma}
\end{align}
where the initial conditions are $\mu(0)=\mu_{0}$ and $\Sigma(0)=\Sigma_{0}$. 
$\Psi$ and $\Pi$ are the solutions of the following time-backward ordinary differential equations:  
\begin{align}
	-\frac{d\Psi}{dt}&=Q+A^{\top}\Psi+\Psi A -\left(\Psi B-S\right)R^{-1}\left(\Psi B-S\right)^{\top},\label{eq: LQG-OS-ODE of Psi}\\
	-\frac{d\Pi}{dt}&=Q+A^{\top}\Pi+\Pi A-\left(\Pi B-S\right)R^{-1}\left(\Pi B-S\right)^{\top}+\left(I-KH\right)^{\top}\left(\Pi B-S\right)R^{-1}\left(\Pi B-S\right)^{\top}\left(I-KH\right),\label{eq: LQG-OS-ODE of Pi}
\end{align}
where the terminal conditions are $\Psi(T)=\Pi(T)=P$. 

Equation (\ref{eq: LQG-OS-ODE of Psi}) can be solved backward in time independently. 
Equation (\ref{eq: LQG-OS-ODE of mu}) can be solved forward in time based on the solution of Eq. (\ref{eq: LQG-OS-ODE of Psi}). 
Equations (\ref{eq: LQG-OS-ODE of Sigma}) and (\ref{eq: LQG-OS-ODE of Pi}) are mutually dependent and need to be solved simultaneously, which correspond to a two-point boundary value problem of ordinary differential equations. 
The LQG problem reduces the two-point boundary value problem of partial differential equations, HJB and FP equations, to that of ordinary differential equations, Eqs. (\ref{eq: LQG-OS-ODE of Sigma}) and (\ref{eq: LQG-OS-ODE of Pi}), allowing for efficient computation. 
This two-point boundary value problem is solved by using the forward-backward sweep method, which computes Eqs. (\ref{eq: LQG-OS-ODE of Sigma}) and (\ref{eq: LQG-OS-ODE of Pi}) alternately \cite{tottori_forward-backward_2023}. 

The optimal control function in the LQG problem of OSC [Eq. (\ref{eq: LQG-OS-optimal control in LQG})] is decomposed into a deterministic first term and a stochastic second term. 
The deterministic first term of Eq. (\ref{eq: LQG-OS-optimal control in LQG}) also appears in conventional optimal control problems, and Eq. (\ref{eq: LQG-OS-ODE of Psi}) is known as Riccati equation \cite{yong_stochastic_1999,nisio_stochastic_2015,bensoussan_estimation_2018}. 
In contrast, the stochastic second term of Eq. (\ref{eq: LQG-OS-optimal control in LQG}) is unique to OSC, and we refer to Eq. (\ref{eq: LQG-OS-ODE of Pi}) as observation-based Riccati equation. 
Although Riccati equation and observation-based Riccati equation are similar, the latter includes an additional term that accounts for the observation uncertainty in OSC. 
In fact, if observation $y_{t}$ is identical to state $x_{t}$, i.e., $y_{t}=x_{t}$, $H=I$ and $E=O$ hold, and the last term of observation-based Riccati equation vanishes, reducing it to Riccati equation. 

\subsection{Derivation}\label{subsec: LQG-D}
In this subsection, we derive the optimal control function in the LQG problem of OSC. 
From Pontryagin's minimum principle on the probability density function space, the optimal control function of OSC is given by
\begin{align}	
	u^{*}(t,y)&=\arg\min_{u}\mb{E}_{p_{t}(x|y)}\left[\mcal{H}\left(t,x,u,w\right)\right]. 
 \label{eq: optimal control in LQG ver0}
\end{align}
In the LQG problem, Hamiltonian $\mcal{H}\left(t,x,u,w\right)$ is given by
\begin{align}
	\mcal{H}(t,x,u,w)
	&=x^{\top}Qx-2x^{\top}Su+u^{\top}Ru+\left(\frac{\partial w(t,x)}{\partial x}\right)^{\top}\left(Ax+Bu\right)
	+\frac{1}{2}{\rm tr}\left(\frac{\partial}{\partial x}\left(\frac{\partial w(t,x)}{\partial x}\right)^{\top}\sigma\sigma^{\top}\right).
\end{align}
From 
\begin{align}
	\frac{\partial \mb{E}_{p_{t}(x|y)}\left[\mcal{H}(t,x,u,w)\right]}{\partial u}
	=-2S^{\top}\mb{E}_{p_{t}(x|y)}\left[x\right]+2Ru+B^{\top}\mb{E}_{p_{t}(x|y)}\left[\frac{\partial w(t,x)}{\partial x}\right],
\end{align}
the optimal control function of OSC is given by
\begin{align}
	u^{*}(t,y)
	&=-\frac{1}{2}R^{-1}\left(B^{\top}\mb{E}_{p_{t}(x|y)}\left[\frac{\partial w(t,x)}{\partial x}\right]-2S^{\top}\mb{E}_{p_{t}(x|y)}\left[x\right]\right).
	\label{eq: optimal control in LQG ver1}
\end{align}

We first assume that $w(t,x)$ is given by the following quadratic function: 
\begin{align}
	w(t,x)=x^{\top}\Pi(t)x+\alpha^{\top}(t)x+\beta(t), 
	\label{eq: assumption of HJB LQG}
\end{align}
where $\Pi(t)$ is a symmetric matrix. 
Under this assumption, HJB equation 
\begin{align}
	-\frac{\partial w(t,x)}{\partial t}
	&=\mb{E}_{p_{t}(y|x)}\left[\mcal{H}\left(t,x,u^{*},w\right)\right]
	\label{eq: HJB eq LQG ver0}
\end{align}
is calculated as follows: 
\begin{align}
	&-\left(x^{\top}\frac{d\Pi}{dt}x+\left(\frac{d\alpha}{dt}\right)^{\top}x+\frac{d\beta}{dt}\right)\nonumber\\
	&=x^{\top}\left(Q+A^{\top}\Pi+\Pi A\right)x
	+\left(\left(A-BR^{-1}\left(\Pi B-S\right)^{\top}\right)^{\top}\alpha\right)^{\top}x
	+{\rm tr}\left(\Pi\sigma\sigma^{\top}\right)
	-\frac{1}{4}\alpha^{\top}BR^{-1}B\alpha\nonumber\\
	&+\mb{E}_{p_{t}(y|x)}\left[\mb{E}_{p_{t}(x|y)}\left[x\right]^{\top}(\Pi B-S)R(\Pi B-S)^{\top}\mb{E}_{p_{t}(x|y)}\left[x\right]\right]
	-2x^{\top}(\Pi B-S)R(\Pi B-S)^{\top}\mb{E}_{p_{t}(y|x)}\left[\mb{E}_{p_{t}(x|y)}\left[x\right]\right],
	\label{eq: HJB eq LQG ver1}
\end{align}
where $\Pi(T)=P$, $\alpha(T)=0$, and $\beta(T)=0$. 
We next assume that $p(t,x)$ is given by the following Gaussian distribution: 
\begin{align}
	p(t,x)=\mcal{N}(x|\mu(t),\Sigma(t)), 
	\label{eq: assumption of FP LQG}
\end{align}
where $\mu(t)$ and $\Sigma(t)$ are the mean vector and the covariance matrix of state $x$, respectively. 
Under this assumption, $\mb{E}_{p_{t}(x|y)}\left[x\right]$ is calculated as follows: 
\begin{align}
	\mb{E}_{p_{t}(x|y)}\left[x\right]=\mu+K\left(y-H\mu\right), 
	\label{eq: LQG-conditional expectation}
\end{align}
where $K$ is defined by 
\begin{align}
	K:=\Sigma H^{\top}\left(E+H\Sigma H^{\top}\right)^{-1}.
\end{align}
Similarly, the following equations hold: 
\begin{align}
	&\mb{E}_{p_{t}(y|x)}\left[\mb{E}_{p_{t}(x|y)}\left[x\right]\right]
	=\mu+KH\left(x-\mu\right),\\
	&\mb{E}_{p_{t}(y|x)}\left[\mb{E}_{p_{t}(x|y)}\left[x\right]^{\top}(\Pi B-S)R(\Pi B-S)^{\top}\mb{E}_{p_{t}(x|y)}\left[x\right]\right]\nonumber\\
	&={\rm tr}\left(K^{\top}(\Pi B-S)R(\Pi B-S)^{\top}KE\right)
	+\left(\mu+KH\left(x-\mu\right)\right)^{\top}(\Pi B-S)R(\Pi B-S)^{\top}\left(\mu+KH\left(x-\mu\right)\right).
\end{align}
As a result, HJB equation [Eq. (\ref{eq: HJB eq LQG ver1})] is reduced to the following ordinary differential equations: 
\begin{align}
	-\frac{d\Pi}{dt}&=Q+A^{\top}\Pi+\Pi A-\left(\Pi B-S\right)R^{-1}\left(\Pi B-S\right)^{\top}+\left(I-KH\right)^{\top}\left(\Pi B-S\right)R^{-1}\left(\Pi B-S\right)^{\top}\left(I-KH\right),\label{eq: ODE of Pi tmp}\\
	-\frac{d\alpha}{dt}&=\left(A-BR^{-1}\left(\Pi B-S\right)^{\top}\right)^{\top}\alpha+2\left(I-KH\right)^{\top}\left(\Pi B-S\right)R^{-1}\left(\Pi B-S\right)^{\top}\left(I-KH\right)\mu,\label{eq: ODE of alpha}\\
	-\frac{d\beta}{dt}&={\rm tr}\left(\Pi\sigma\sigma^{\top}+K^{\top}(\Pi B-S)R(\Pi B-S)^{\top}KE\right)\nonumber\\
	&\ \ \ -\frac{1}{4}\alpha^{\top}BR^{-1}B^{\top}\alpha+\mu^{\top}\left(I-KH\right)^{\top}\left(\Pi B-S\right)R^{-1}\left(\Pi B-S\right)^{\top}\left(I-KH\right)\mu,\label{eq: ODE of beta}
\end{align}
where $\Pi(T)=P$, $\alpha(T)=0$, and $\beta(T)=0$. 
If $\Pi(t)$, $\alpha(t)$, and $\beta(t)$ satisfy Eqs. (\ref{eq: ODE of Pi tmp}), (\ref{eq: ODE of alpha}), and (\ref{eq: ODE of beta}), respectively, HJB equation [Eq. (\ref{eq: HJB eq LQG ver0})] is satisfied, which indicates that our first assumption [Eq. (\ref{eq: assumption of HJB LQG})] is consistent. 
The optimal control function of OSC [Eq. (\ref{eq: optimal control in LQG ver1})] is calculated as follows: 
\begin{align}
	u^{*}(t,y)
	&=-\frac{1}{2}R^{-1}\left(2\left(\Pi B-S\right)^{\top}\left(\mu+K\left(y-H\mu\right)\right)+B^{\top}\alpha\right).
	\label{eq: optimal control in LQG ver2}
\end{align}
Since the optimal control function of OSC [Eq. (\ref{eq: optimal control in LQG ver2})] is linear with respect to $y$, probability density function $p(t,x)$ becomes a Gaussian distribution, which is consistent with our second assumption [Eq. (\ref{eq: assumption of FP LQG})]. 

We assume that $\alpha(t)=2\Upsilon(t)\mu(t)$ holds, where $\Upsilon(t)$ is a symmetric matrix. 
Under this assumption, the optimal control function of OSC [Eq. (\ref{eq: optimal control in LQG ver2})] is calculated as follows: 
\begin{align}
	u^{*}(t,y)
	&=-R^{-1}\left(\left(\left(\Pi+\Upsilon\right) B-S\right)^{\top}\mu+\left(\Pi B-S\right)^{\top}K\left(y-H\mu\right)\right).
	\label{eq: optimal control in LQG ver3}
\end{align}
From 
\begin{align}
	\frac{d\alpha}{dt}=2\frac{d\Upsilon}{dt}\mu+2\Upsilon\frac{d\mu}{dt},\quad 
	\frac{d\mu}{dt}=\left(A-BR^{-1}\left(\left(\Pi+\Upsilon\right) B-S\right)^{\top}\right)\mu, 
\end{align}
and Eq. (\ref{eq: ODE of alpha}), $\Upsilon(t)$ evolves by the following ordinary differential equation: 
\begin{align}
	-\frac{d\Upsilon}{dt}&=\left(A-BR^{-1}\left(\Pi B-S\right)\right)^{\top}\Upsilon+\Upsilon\left(A-BR^{-1}\left(\Pi B-S\right)\right)\nonumber\\
	&\ \ \ -\Upsilon BR^{-1}B^{\top}\Upsilon
	-\left(I-KH\right)^{\top}\Upsilon BR^{-1}B^{\top}\Upsilon\left(I-KH\right),\label{eq: ODE of Upsilon}
\end{align}
where $\Upsilon(T)=O$. 
From Eq. (\ref{eq: ODE of Upsilon}), $\Upsilon(t)$ is a symmetric matrix, which is consistent with our assumption. 
By defining $\Psi(t):=\Pi(t)+\Upsilon(t)$, the optimal control function of OSC [Eq. (\ref{eq: optimal control in LQG ver3})] is calculated as follows: 
\begin{align}
	u^{*}(t,y)
	&=-R^{-1}\left(\left(\Psi B-S\right)^{\top}\mu+\left(\Pi B-S\right)^{\top}K\left(y-H\mu\right)\right).
	\label{eq: optimal control in LQG ver4}
\end{align}
From Eqs. (\ref{eq: ODE of Pi tmp}), and (\ref{eq: ODE of Upsilon}), $\Psi(t)$ evolves by the following ordinary differential equation: 
\begin{align}
	-\frac{d\Psi}{dt}&=Q+A^{\top}\Psi+\Psi A -\left(\Psi B-S\right)R^{-1}\left(\Psi B-S\right)^{\top},\label{eq: ODE of Psi tmp}
\end{align}
where $\Psi(T)=P$. 
Eqs. (\ref{eq: optimal control in LQG ver4}), (\ref{eq: ODE of Psi tmp}), and (\ref{eq: ODE of Pi tmp}) are Eqs. (\ref{eq: LQG-OS-optimal control in LQG}), (\ref{eq: LQG-OS-ODE of Psi}), and (\ref{eq: LQG-OS-ODE of Pi}), respectively. 
Given the optimal control function [Eq. (\ref{eq: optimal control in LQG ver4})], the mean vector and the covariance matrix of state $x$, $\mu(t)$ and $\Sigma(t)$, are given by the solutions of Eqs. (\ref{eq: LQG-OS-ODE of mu}) and (\ref{eq: LQG-OS-ODE of Sigma}), respectively. 

\subsection{Infinite-Horizon Case}\label{subsec: LQG-infinite}
In Appendix \ref{subsec: LQG-PF}, \ref{subsec: LQG-OS}, and \ref{subsec: LQG-D}, we discuss the finite-horizon case, in which the objective function is defined by 
\begin{align}
	J[u]:=\mb{E}\left[\int_{0}^{T}\left(x_{t}^{\top}Qx_{t}-2x_{t}^{\top}Su_{t}+u_{t}^{\top}Ru_{t}\right)dt+x_{T}^{\top}Px_{T}\right]. 
	\label{eq: OF of LQG finite-OSC}
\end{align}
In this subsection, we discuss the infinite-horizon case, where the objective function is defined by 
\begin{align}
	J[u]:=\lim_{T\to\infty}\frac{1}{T}\mb{E}\left[\int_{0}^{T}\left(x_{t}^{\top}Qx_{t}-2x_{t}^{\top}Su_{t}+u_{t}^{\top}Ru_{t}\right)dt\right]. 
	\label{eq: OF of LQG infinite-OSC}
\end{align}
Additionally, since the infinite-horizon case is related with our minimal models in Sec. \ref{sec: ABIP} and \ref{sec: ABDM}, we mainly consider OMSC rather than OSC by adding $\tilde{\cdot}$ as follows: 
\begin{align}
	J[\tilde{u}]:=\lim_{T\to\infty}\frac{1}{T}\mb{E}\left[\int_{0}^{T}\left(\tilde{x}_{t}^{\top}\tilde{Q}\tilde{x}_{t}-2\tilde{x}_{t}^{\top}\tilde{S}\tilde{u}_{t}+\tilde{u}_{t}^{\top}\tilde{R}\tilde{u}_{t}\right)dt\right], 
	\label{eq: OF of LQG infinite-OMSC}
\end{align}
where $\tilde{\cdot}$ indicates the extended variables and parameters. 
We note that we can make the same discussion for OSC by removing $\tilde{\cdot}$. 

The optimal control function in the infinite-horizon case is the same as that in the finite-horizon case, which is given by 
\begin{align}
	\tilde{u}^{*}(t,\tilde{y})
	&=-\tilde{R}^{-1}\left(\left(\tilde{\Psi}\tilde{B}-\tilde{S}\right)^{\top}\tilde{\mu}+\left(\tilde{\Pi}\tilde{B}-\tilde{S}\right)^{\top}\tilde{K}\left(\tilde{y}-\tilde{H}\tilde{\mu}\right)\right), 
	\label{eq: optimal control in LQG infinite}
\end{align}
where $\tilde{K}:=\tilde{\Sigma}\tilde{H}^{\top}(\tilde{E}+\tilde{H}\tilde{\Sigma}\tilde{H}^{\top})^{-1}$. 
In contrast, in the infinite-horizon case, the ordinary differential equations of $\tilde{\mu}$, $\tilde{\Sigma}$, $\tilde{\Psi}$, and $\tilde{\Pi}$ become the steady-state equations as follows: 
\begin{align}
	0&=\left(\tilde{A}-\tilde{B}\tilde{R}^{-1}\left(\tilde{\Psi}\tilde{B}-\tilde{S}\right)^{\top}\right)\tilde{\mu},\label{eq: ODE of mu infinite}\\
	O&=\tilde{\sigma}\tilde{\sigma}^{\top}+\left(\tilde{A}-\tilde{B}\tilde{R}^{-1}\left(\tilde{\Pi}\tilde{B}-\tilde{S}\right)^{\top}\tilde{K}\tilde{H}\right)\tilde{\Sigma}+\tilde{\Sigma}\left(\tilde{A}-\tilde{B}\tilde{R}^{-1}\left(\tilde{\Pi}\tilde{B}-\tilde{S}\right)^{\top}\tilde{K}\tilde{H}\right)^{\top},\label{eq: ODE of Sigma infinite}\\
	O&=\tilde{Q}+\tilde{A}^{\top}\tilde{\Psi}+\tilde{\Psi}\tilde{A} -\left(\tilde{\Psi}\tilde{B}-\tilde{S}\right)\tilde{R}^{-1}\left(\tilde{\Psi}\tilde{B}-\tilde{S}\right)^{\top},\label{eq: ODE of Psi infinite}\\
	O&=\tilde{Q}+\tilde{A}^{\top}\tilde{\Pi}+\tilde{\Pi}\tilde{A}-\left(\tilde{\Pi}\tilde{B}-\tilde{S}\right)\tilde{R}^{-1}\left(\tilde{\Pi}\tilde{B}-\tilde{S}\right)^{\top}+\left(I-\tilde{K}\tilde{H}\right)^{\top}\left(\tilde{\Pi}\tilde{B}-\tilde{S}\right)\tilde{R}^{-1}\left(\tilde{\Pi}\tilde{B}-\tilde{S}\right)^{\top}\left(I-\tilde{K}\tilde{H}\right).\label{eq: ODE of Pi infinite}
\end{align}
$\tilde{\Psi}$ is obtained by solving Eq. (\ref{eq: ODE of Psi infinite}). 
Given $\tilde{\Psi}$, $\tilde{\mu}$ is obtained by solving Eq. (\ref{eq: ODE of mu infinite}). 
$\tilde{\Sigma}$ and $\tilde{\Pi}$ are obtained by solving Eqs. (\ref{eq: ODE of Sigma infinite}) and (\ref{eq: ODE of Pi infinite}) simultaneously. 

However, Eq. (\ref{eq: ODE of Sigma infinite}) is impractical for our discussions in Sec. \ref{sec: ABIP} and \ref{sec: ABDM}. 
When $\Pi_{zx}=\Pi_{zz}=0$, the variance of the memory diverges to infinity as $\Sigma_{zz}\to\infty$, and Eq. (\ref{eq: ODE of Sigma infinite}) does not hold. 
In order to avoid this problem, we convert the steady-state equation for covariance matrix $\tilde{\Sigma}$ [Eq. (\ref{eq: ODE of Sigma infinite})] to that for precision matrix $\tilde{\Lambda}:=\tilde{\Sigma}^{-1}$ as follows: 
\begin{align}
	O&=-\left(\tilde{A}-\tilde{B}\tilde{R}^{-1}\left(\tilde{\Pi}\tilde{B}-\tilde{S}\right)^{\top}\tilde{K}\tilde{H}\right)^{\top}\tilde{\Lambda}
	-\tilde{\Lambda}\left(\tilde{A}-\tilde{B}\tilde{R}^{-1}\left(\tilde{\Pi}\tilde{B}-\tilde{S}\right)^{\top}\tilde{K}\tilde{H}\right)
	-\tilde{\Lambda}\tilde{\sigma}\tilde{\sigma}^{\top}\tilde{\Lambda}. 
	\label{eq: ODE of Lambda infinite}
\end{align}
When $\Sigma_{zz}\to\infty$, $\Lambda_{zz}\to0$. 
Therefore, Eq. (\ref{eq: ODE of Lambda infinite}) holds even when $\Pi_{zx}=\Pi_{zz}=0$. 
In Sec. \ref{sec: ABIP} and \ref{sec: ABDM}, we obtain control matrix $\tilde{\Pi}$ and precision matrix $\tilde{\Lambda}$ by numerically solving Eqs. (\ref{eq: ODE of Pi infinite}) and (\ref{eq: ODE of Lambda infinite}). 

Given $\tilde{\mu}$, $\tilde{\Sigma}$, $\tilde{\Psi}$, and $\tilde{\Pi}$, the objective function in the infinite-horizon case [Eq. (\ref{eq: OF of LQG infinite-OMSC})] is calculated as follows: 
\begin{align}
	J[\tilde{u}^{*}]
	&=\lim_{T\to\infty}\frac{1}{T}\mb{E}\left[\int_{0}^{T}\left(\tilde{x}_{t}^{\top}\tilde{Q}\tilde{x}_{t}-2\tilde{x}_{t}^{\top}\tilde{S}\tilde{u}_{t}^{*}+(\tilde{u}_{t}^{*})^{\top}\tilde{R}\tilde{u}_{t}^{*}\right)dt\right]\nonumber\\
	&=\mb{E}\left[\tilde{x}^{\top}\tilde{Q}\tilde{x}-2\tilde{x}^{\top}\tilde{S}\tilde{u}^{*}+(\tilde{u}^{*})^{\top}\tilde{R}\tilde{u}^{*}\right]\nonumber\\
	&=\tilde{\mu}^{\top}\tilde{Q}\tilde{\mu}+2\tilde{\mu}^{\top}\tilde{S}\tilde{R}^{-1}\left(\tilde{\Psi}\tilde{B}-\tilde{S}\right)^{\top}\tilde{\mu}
	+\tilde{\mu}^{\top}\left(\tilde{\Psi}\tilde{B}-\tilde{S}\right)\tilde{R}\left(\tilde{\Psi}\tilde{B}-\tilde{S}\right)^{\top}\tilde{\mu}\nonumber\\
	&\ \ \ +{\rm tr}\left\{\left(\tilde{Q}+2\tilde{S}\tilde{R}^{-1}\left(\tilde{\Pi}\tilde{B}-\tilde{S}\right)^{\top}\tilde{K}\tilde{H}+\tilde{K}^{\top}\left(\tilde{\Pi}\tilde{B}-\tilde{S}\right)\tilde{R}\left(\tilde{\Pi}\tilde{B}-\tilde{S}\right)^{\top}\tilde{K}\right)\tilde{\Sigma}
	\right\}\nonumber\\
	&\ \ \ +{\rm tr}\left\{\tilde{K}^{\top}\left(\tilde{\Pi}\tilde{B}-\tilde{S}\right)\tilde{R}\left(\tilde{\Pi}\tilde{B}-\tilde{S}\right)^{\top}\tilde{K}\tilde{E}
	\right\}.
\end{align}
In Sec. \ref{sec: ABIP} and \ref{sec: ABDM}, we use this result to obtain $J$ given $\tilde{\mu}$, $\tilde{\Sigma}$, $\tilde{\Psi}$, and $\tilde{\Pi}$. 

\section{Details of Application to Biological Information Processing}\label{sec: ABIPA}
In this section, we provide a detailed explanation of the minimal model of biological information processing, the target estimation problem, presented in Sec. \ref{sec: ABIP}. 
This section is organized as follows: 
In Appendix \ref{subsec: tep-lqg}, we show that the target estimation problem corresponds to the infinite-horizon LQG problem of OMSC. 
In Appendix \ref{subsec: tep-ocf}, we derive the optimal solution of the target estimation problem [Eqs. (\ref{eq: ABIP-optimal state estimator}) and (\ref{eq: ABIP-optimal memory control})]. 
In Appendix \ref{subsec: tep-ptm}, we show the discontinuous phase transition with respect to the other parameters. 

\subsection{Correspondence to Infinite-Horizon Linear-Quadratic-Gaussian Problem}\label{subsec: tep-lqg}
The target estimation problem and the LQG problem are discussed in Sec. \ref{sec: ABIP} and  Appendix. \ref{sec: LQG}, respectively. 
In this subsection, we show that the target estimation problem corresponds to the infinite-horizon LQG problem of OMSC. 
We define extended state $\tilde{x}_{t}$, extended observation $\tilde{y}_{t}$, extended control $\tilde{u}_{t}$, and extended standard Wiener process $\tilde{\omega}_{t}$ as follows: 
\begin{align}
	\tilde{x}_{t}:=\left(\begin{array}{c}
		x_{t}\\
		z_{t}\\
	\end{array}
	\right),\quad 
	\tilde{y}_{t}:=\left(\begin{array}{c}
		y_{t}\\
		z_{t}\\
	\end{array}
	\right),\quad 
	\tilde{u}_{t}:=\left(\begin{array}{c}
		\hat{x}_{t}\\
		v_{t}\\
	\end{array}
	\right),\quad
	\tilde{\omega}_{t}:=\left(\begin{array}{c}
		\omega_{t}\\
		\xi_{t}\\
	\end{array}
	\right).
\end{align}
Extended state $\tilde{x}_{t}$ evolves by the following SDE: 
\begin{align}
	d\tilde{x}_{t}=\left(\tilde{A}\tilde{x}_{t}+\tilde{B}\tilde{u}_{t}\right)dt+\tilde{\sigma}d\tilde{\omega}_{t},\nonumber
\end{align}
where 
\begin{align}
	\tilde{A}:=\left(\begin{array}{cc}
		-1&0\\
		0&0\\
	\end{array}\right),\quad 
	\tilde{B}:=\left(\begin{array}{cc}
		0&0\\
		0&1\\
	\end{array}\right),\quad 
	\tilde{\sigma}:=\left(\begin{array}{cc}
		\sqrt{D}&0\\
		0&\sqrt{F}\\
	\end{array}\right). 
\end{align}
Extended observation $\tilde{y}_{t}$ is generated from the following Gaussian distribution: 
\begin{align}
	\tilde{p}(\tilde{y}_{t}|\tilde{x}_{t})=\mcal{N}(\tilde{y}_{t}|\tilde{H}\tilde{x}_{t},\tilde{E}),
\end{align}
where 
\begin{align}
	\tilde{H}:=\left(\begin{array}{cc}
		1&0\\
		0&1\\
	\end{array}
	\right),\quad 
	\tilde{E}:=\left(\begin{array}{cc}
		E&0\\
		0&0\\
	\end{array}
	\right). 
\end{align}
The objective function is given by the following cost function: 
\begin{align}
	J[\tilde{u}]:=\lim_{T\to\infty}\frac{1}{T}\mb{E}\left[\int_{0}^{T}\left(\tilde{x}_{t}^{\top}\tilde{Q}\tilde{x}_{t}-2\tilde{x}_{t}^{\top}\tilde{S}\tilde{u}_{t}+\tilde{u}_{t}^{\top}\tilde{R}\tilde{u}_{t}\right)dt\right],
\end{align}
where
\begin{align}
	\tilde{Q}:=\left(\begin{array}{cc}
		Q&0\\
		0&0\\
	\end{array}\right),\quad 
	\tilde{S}:=\left(\begin{array}{cc}
		Q&0\\
		0&0\\
	\end{array}\right),\quad 
	\tilde{R}:=\left(\begin{array}{cc}
		Q&0\\
		0&M\\
	\end{array}\right). 
	\label{eq: oc-tep}
\end{align}
Therefore, the target estimation problem corresponds to the infinite-horizon LQG problem of OMSC. 

\subsection{Derivation of Optimal Solution}\label{subsec: tep-ocf}
In this subsection, we derive the optimal solution of the target estimation problem, which is given by Eqs. (\ref{eq: ABIP-optimal state estimator}) and (\ref{eq: ABIP-optimal memory control}). 
Since the target estimation problem corresponds to the infinite-horizon LQG problem of OMSC, the optimal solution of the target estimation problem is given by 
\begin{align}
	\tilde{u}^{*}(t,\tilde{y})&=-\tilde{R}^{-1}\left(\left(\tilde{\Psi}\tilde{B}-\tilde{S}\right)^{\top}\tilde{\mu}+\left(\tilde{\Pi}\tilde{B}-\tilde{S}\right)^{\top}\tilde{K}\left(\tilde{y}-\tilde{H}\tilde{\mu}\right)\right), 
\end{align}
where $\tilde{K}:=\tilde{\Lambda}^{-1}\tilde{H}^{\top}(\tilde{E}+\tilde{H}\tilde{\Lambda}^{-1}\tilde{H}^{\top})^{-1}$. 
$\tilde{\mu}$, $\tilde{\Psi}$, $\tilde{\Pi}$, and $\tilde{\Lambda}$ are the solutions of Eqs.  (\ref{eq: ODE of mu infinite}), (\ref{eq: ODE of Psi infinite}),  (\ref{eq: ODE of Pi infinite}), and (\ref{eq: ODE of Lambda infinite}) (See Appendix \ref{subsec: LQG-infinite}). 

Since $\tilde{\mu}=0$, the optimal solution of the target estimation problem is calculated as follows: 
\begin{align}
	\tilde{u}^{*}(t,\tilde{y})&=-\tilde{R}^{-1}\left(\tilde{\Pi}\tilde{B}-\tilde{S}\right)^{\top}\tilde{K}\tilde{y}, 
\end{align}
where $\tilde{K}:=\tilde{\Lambda}^{-1}\tilde{H}^{\top}(\tilde{E}+\tilde{H}\tilde{\Lambda}^{-1}\tilde{H}^{\top})^{-1}$. 
$\tilde{\Pi}$ and $\tilde{\Lambda}$ are the solutions of Eqs. (\ref{eq: ODE of Pi infinite}) and (\ref{eq: ODE of Lambda infinite}). 

Since $\mb{E}_{\tilde{p}_{t}(\tilde{x}|\tilde{y})}\left[\tilde{x}\right]=\tilde{\mu}+\tilde{K}(\tilde{y}-\tilde{H}\tilde{\mu})$ [Eq. (\ref{eq: LQG-conditional expectation})] and $\tilde{\mu}=0$, 
\begin{align}
	\tilde{K}\tilde{y}
	=\mb{E}_{\tilde{p}_{t}(\tilde{x}|\tilde{y})}\left[\tilde{x}\right]
	=\left(\begin{array}{c}
		\mb{E}_{p_{t}(x|y,z)}\left[x\right]\\
		z\\
	\end{array}\right)
\end{align}
holds. 
Thus, the optimal solution of the target estimation problem is calculated as follows: 
\begin{align}
	\tilde{u}^{*}(t,\tilde{y})
	&=-\left(\begin{array}{cc}
		Q&0\\
		0&M\\
	\end{array}\right)^{-1}\left(\left(\begin{array}{cc}
		\Pi_{xx}&\Pi_{xz}\\
		\Pi_{zx}&\Pi_{zz}\\
	\end{array}\right)\left(\begin{array}{cc}
		0&0\\
		0&1\\
	\end{array}\right)-\left(\begin{array}{cc}
		Q&0\\
		0&0\\
	\end{array}\right)\right)^{\top}
	\left(\begin{array}{c}
		\mb{E}_{p_{t}(x|y,z)}\left[x\right]\\
		z\\
	\end{array}\right)\nonumber\\
	&=\left(\begin{array}{c}
		\mb{E}_{p_{t}(x|y,z)}\left[x\right]\\
		-M^{-1}\Pi_{zx}\mb{E}_{p_{t}(x|y,z)}\left[x\right]-M^{-1}\Pi_{zz}z\\
	\end{array}\right). 
\end{align}
Therefore, Eqs. (\ref{eq: ABIP-optimal state estimator}) and (\ref{eq: ABIP-optimal memory control}) are obtained. 

\subsection{Phase Transition with respect to Other Parameters}\label{subsec: tep-ptm}
\begin{figure*}
	\includegraphics[width=170mm]{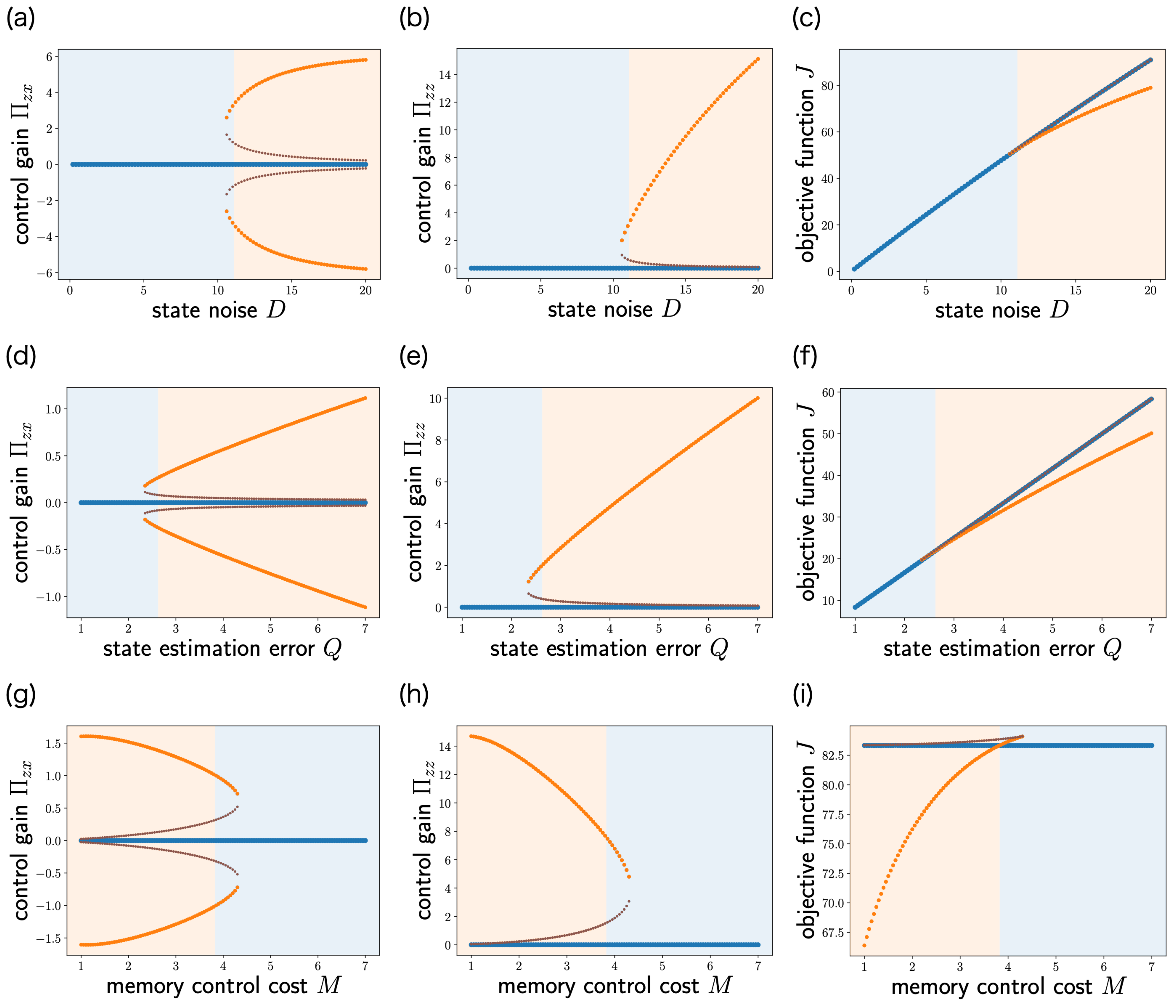}
	\caption{\label{fig: ABIPA-TMS}
	Memory control gains $\Pi_{zx}$ and $\Pi_{zz}$ and associated objective function $J$ with respect to state noise $D$ (a--c), state estimation error $Q$ (d--f), and memory control cost $M$ (g--i) in the target estimation problem. 
	The blue, brown, and orange dots are the solutions of Pontryagin's minimum principle, corresponding to zero, intermediate, and high memory controls, respectively. 
	The blue dots are optimal in the blue region, while the orange dots are optimal in the orange region. 
	The parameters are $E=100$, $F=1$, $Q=10$, and $M=1$ (a--c), $D=100$, $E=1$, $F=1$, and $M=1$ (d--f), and $D=100$, $E=10$, $F=1$, and $Q=10$ (g--i). 
	}
\end{figure*}

The discontinuous phase transition also occurs for the other parameters, i.e., the intensity of the state noise $D$, the weight of the state estimation error $Q$, and the weight of the memory control cost $M$ (Fig. \ref{fig: ABIPA-TMS}). 
The phase transition for state noise $D$ and state estimation error $Q$ is similar to that for observation noise $E$ (Fig. \ref{fig: ABIPA-TMS}(a--c) and (d--f)). 
This result suggests that the estimation strategy with memory is optimal when the state estimation is either challenging or significant. 
In contrast, the phase transition for memory control cost $M$ is similar to that for memory noise $F$, indicating that resource limitations, such as the noise and cost associated with intrinsic information processing, render the estimation strategy without memory optimal (Fig. \ref{fig: ABIPA-TMS}(g--i)). 

\section{Details of Application to Biological Information Processing and Decision-Making}\label{sec: ABDMA}
In this section, we provide a detailed explanation of the minimal model of biological information processing and decision-making, the target tracking problem, presented in Sec. \ref{sec: ABDM}. 
This section is organized as follows: 
In Appendix \ref{subsec: ttp-lqg}, we show that the target tracking problem corresponds to the infinite-horizon LQG problem of OMSC. 
In Appendix \ref{subsec: ttp-ocf}, we derive the optimal solution of the target tracking problem [Eqs. (\ref{eq: ABDM-optimal state control}) and (\ref{eq: ABDM-optimal memory control})]. 
In Appendix \ref{subsec: ttp-ptm}, we show the discontinuous phase transition with respect to the other parameters. 

\subsection{Correspondence to Infinite-Horizon Linear-Quadratic-Gaussian Problem}\label{subsec: ttp-lqg}
The target tracking problem and the LQG problem are discussed in Sec. \ref{sec: ABDM} and  Appendix. \ref{sec: LQG}, respectively. 
In this subsection, we show that the target tracking problem corresponds to the infinite-horizon LQG problem of OMSC. 
We define extended state $\tilde{x}_{t}$, extended observation $\tilde{y}_{t}$, extended control $\tilde{u}_{t}$, and extended standard Wiener process $\tilde{\omega}_{t}$ as follows: 
\begin{align}
	\tilde{x}_{t}:=\left(\begin{array}{c}
		x_{t}\\
		z_{t}\\
	\end{array}
	\right),\quad 
	\tilde{y}_{t}:=\left(\begin{array}{c}
		y_{t}\\
		z_{t}\\
	\end{array}
	\right),\quad 
	\tilde{u}_{t}:=\left(\begin{array}{c}
		u_{t}\\
		v_{t}\\
	\end{array}
	\right),\quad
	\tilde{\omega}_{t}:=\left(\begin{array}{c}
		\omega_{t}\\
		\xi_{t}\\
	\end{array}
	\right).
\end{align}
Extended state $\tilde{x}_{t}$ evolves by the following SDE: 
\begin{align}
	d\tilde{x}_{t}=\left(\tilde{A}\tilde{x}_{t}+\tilde{B}\tilde{u}_{t}\right)dt+\tilde{\sigma}d\tilde{\omega}_{t},\nonumber
\end{align}
where 
\begin{align}
	\tilde{A}:=\left(\begin{array}{cc}
		0&0\\
		0&0\\
	\end{array}\right),\quad
	\tilde{B}:=\left(\begin{array}{cc}
		1&0\\
		0&1\\
	\end{array}\right),\quad
	\tilde{\sigma}:=\left(\begin{array}{cc}
		\sqrt{D}&0\\
		0&\sqrt{F}\\
	\end{array}\right). 
\end{align}
Extended observation $\tilde{y}_{t}$ is generated from the following Gaussian distribution: 
\begin{align}
	\tilde{p}(\tilde{y}_{t}|\tilde{x}_{t})=\mcal{N}(\tilde{y}_{t}|\tilde{H}\tilde{x}_{t},\tilde{E}),
\end{align}
where 
\begin{align}
	\tilde{H}:=\left(\begin{array}{cc}
		1&0\\
		0&1\\
	\end{array}
	\right),\quad 
	\tilde{E}:=\left(\begin{array}{cc}
		E&0\\
		0&0\\
	\end{array}
	\right). 
\end{align}
The objective function is given by the following cost function: 
\begin{align}
	J[\tilde{u}]:=\lim_{T\to\infty}\frac{1}{T}\mb{E}\left[\int_{0}^{T}\left(\tilde{x}_{t}^{\top}\tilde{Q}\tilde{x}_{t}-2\tilde{x}_{t}^{\top}\tilde{S}\tilde{u}_{t}+\tilde{u}_{t}^{\top}\tilde{R}\tilde{u}_{t}\right)dt\right],
\end{align}
where
\begin{align}
	\tilde{Q}:=\left(\begin{array}{cc}
		Q&0\\
		0&0\\
	\end{array}\right),\quad 
	\tilde{S}:=\left(\begin{array}{cc}
		0&0\\
		0&0\\
	\end{array}\right),\quad 
	\tilde{R}:=\left(\begin{array}{cc}
		R&0\\
		0&M\\
	\end{array}\right). 
\end{align}
Therefore, the target tracking problem corresponds to the infinite-horizon LQG problem of OMSC. 

\subsection{Derivation of Optimal Solution}\label{subsec: ttp-ocf}
In this subsection, we derive the optimal solution of the target tracking problem, which is given by Eqs. (\ref{eq: ABDM-optimal state control}) and (\ref{eq: ABDM-optimal memory control}). 
Since the target tracking problem corresponds to the infinite-horizon LQG problem of OMSC, the optimal solution of the target tracking problem is given by 
\begin{align}
	\tilde{u}^{*}(t,\tilde{y})&=-\tilde{R}^{-1}\left(\left(\tilde{\Psi}\tilde{B}-\tilde{S}\right)^{\top}\tilde{\mu}+\left(\tilde{\Pi}\tilde{B}-\tilde{S}\right)^{\top}\tilde{K}\left(\tilde{y}-\tilde{H}\tilde{\mu}\right)\right), 
\end{align}
where $\tilde{K}:=\tilde{\Lambda}^{-1}\tilde{H}^{\top}(\tilde{E}+\tilde{H}\tilde{\Lambda}^{-1}\tilde{H}^{\top})^{-1}$. 
$\tilde{\mu}$, $\tilde{\Psi}$, $\tilde{\Pi}$, and $\tilde{\Lambda}$ are the solutions of Eqs.  (\ref{eq: ODE of mu infinite}), (\ref{eq: ODE of Psi infinite}),  (\ref{eq: ODE of Pi infinite}), and (\ref{eq: ODE of Lambda infinite}) (See Appendix \ref{subsec: LQG-infinite}). 

Since $\tilde{\mu}=0$, the optimal solution of the target estimation problem is calculated as follows: 
\begin{align}
	\tilde{u}^{*}(t,\tilde{y})&=-\tilde{R}^{-1}\left(\tilde{\Pi}\tilde{B}-\tilde{S}\right)^{\top}\tilde{K}\tilde{y}, 
\end{align}
where $\tilde{K}:=\tilde{\Lambda}^{-1}\tilde{H}^{\top}(\tilde{E}+\tilde{H}\tilde{\Lambda}^{-1}\tilde{H}^{\top})^{-1}$. 
$\tilde{\Pi}$ and $\tilde{\Lambda}$ are the solutions of Eqs. (\ref{eq: ODE of Pi infinite}) and (\ref{eq: ODE of Lambda infinite}). 

Since $\mb{E}_{\tilde{p}_{t}(\tilde{x}|\tilde{y})}\left[\tilde{x}\right]=\tilde{\mu}+\tilde{K}(\tilde{y}-\tilde{H}\tilde{\mu})$ [Eq. (\ref{eq: LQG-conditional expectation})] and $\tilde{\mu}=0$, 
\begin{align}
	\tilde{K}\tilde{y}
	=\mb{E}_{\tilde{p}_{t}(\tilde{x}|\tilde{y})}\left[\tilde{x}\right]
	=\left(\begin{array}{c}
		\mb{E}_{p_{t}(x|y,z)}\left[x\right]\\
		z\\
	\end{array}\right)
\end{align}
holds. 
Thus, the optimal solution of the target tracking problem is calculated as follows: 
\begin{align}
	\tilde{u}^{*}(t,\tilde{y})	
	&=-\left(\begin{array}{cc}
		R&0\\
		0&M\\
	\end{array}\right)^{-1}\left(\left(\begin{array}{cc}
		\Pi_{xx}&\Pi_{xz}\\
		\Pi_{zx}&\Pi_{zz}\\
	\end{array}\right)\left(\begin{array}{cc}
		1&0\\
		0&1\\
	\end{array}\right)-\left(\begin{array}{cc}
		0&0\\
		0&0\\
	\end{array}\right)\right)^{\top}
	\left(\begin{array}{c}
		\mb{E}_{p_{t}(x|y,z)}\left[x\right]\\
		z\\
	\end{array}\right)\nonumber\\
	&=\left(\begin{array}{c}
		-R^{-1}\Pi_{xx}\mb{E}_{p_{t}(x|y,z)}\left[x\right]-R^{-1}\Pi_{xz}z\\
		-M^{-1}\Pi_{zx}\mb{E}_{p_{t}(x|y,z)}\left[x\right]-M^{-1}\Pi_{zz}z\\
	\end{array}\right). 
	\label{eq: oc-ttp}
\end{align}
Therefore, Eqs. (\ref{eq: ABDM-optimal state control}) and (\ref{eq: ABDM-optimal memory control}) are obtained. 

\subsection{Phase Transition with respect to Other Parameters}\label{subsec: ttp-ptm}
\begin{figure*}
	\includegraphics[width=170mm]{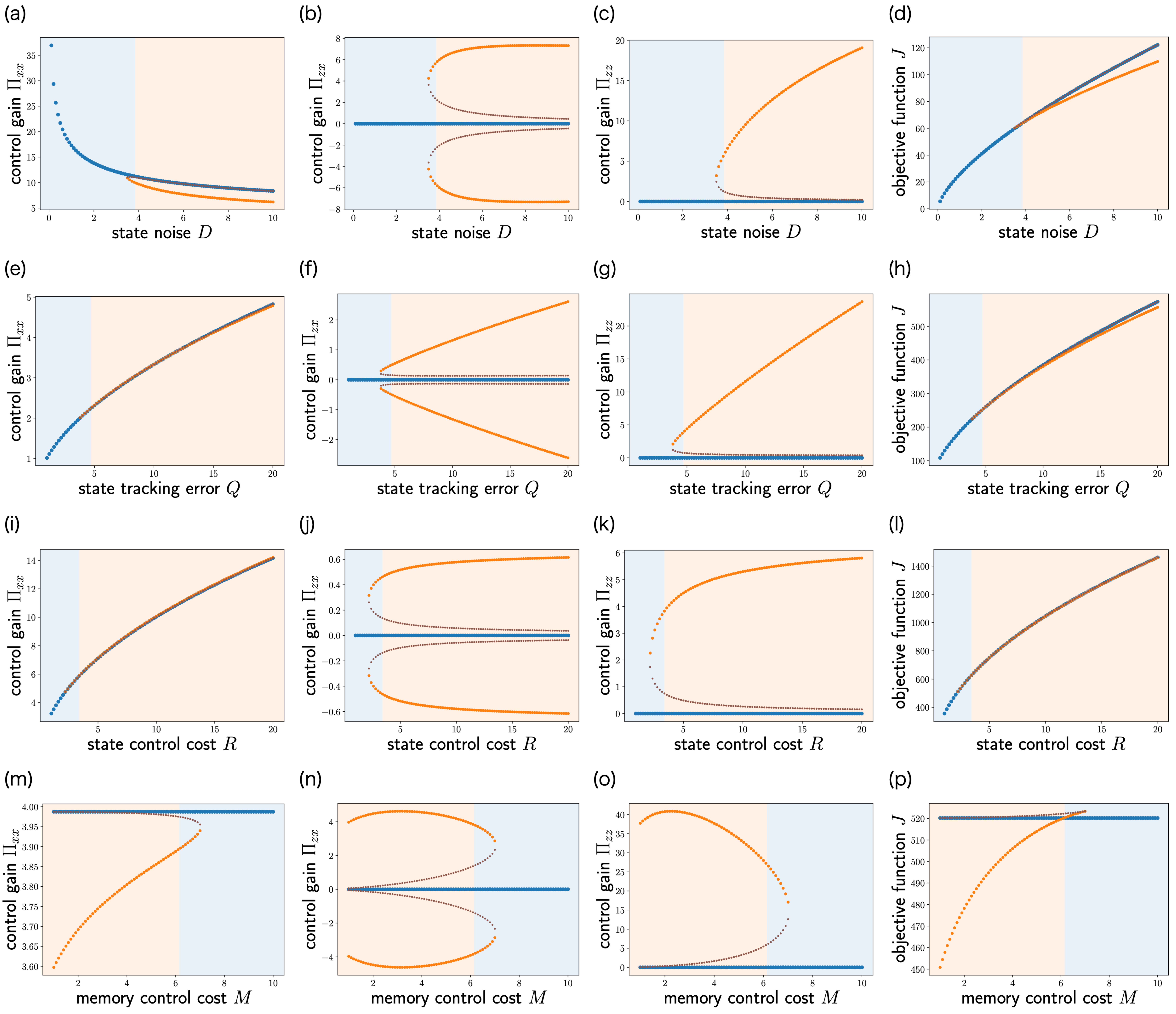}
	\caption{\label{fig: ABDMA-TMS}
	Control gains $\Pi_{xx}$, $\Pi_{zx}$, and $\Pi_{zz}$ and associated objective function $J$ with respect to state noise $D$ (a--d), state tracking error $Q$ (e--h), state control cost $R$ (i--l), and memory control cost $M$ (m--p) in the target tracking problem. 
	Control gain $\Pi_{xz}$ is not visualized because $\Pi_{xz}=\Pi_{zx}$ holds. 
	The blue, brown, and orange dots are the solutions of Pontryagin's minimum principle, corresponding to zero, intermediate, and high memory controls, respectively. 
	The blue dots are optimal in the blue region, while the orange dots are optimal in the orange region. 
	The parameters are $E=100$, $F=1$, $Q=10$, $R=1$, and $M=1$ (a--d), $D=100$, $E=1$, $F=1$, $R=1$, and $M=1$ (e--h), $D=100$, $E=5$, $F=1$, $Q=10$, and $M=1$ (i--l), and $D=100$, $E=50$, $F=1$, $Q=10$, and $R=1$ (m--p). 
	}
\end{figure*}

The discontinuous phase transition also occurs for the other parameters, i.e., the intensity of the state noise $D$, the weight of the state estimation error $Q$, the weight of the state control cost $R$, and the weight of the memory control cost $M$ (Fig. \ref{fig: ABDMA-TMS}). 
While the phase transition for state noise $D$, state estimation error $Q$, and state control cost $R$ is similar to that for observation noise $E$ (Fig. \ref{fig: ABDMA-TMS}(a--l)), the phase transition for memory control cost $M$ is similar to that for memory noise $F$ (Fig. \ref{fig: ABDMA-TMS}(m--p)).

\section{Memory-Limited Partially Observable Stochastic Control}\label{sec: ML-POSC}
In this section, we briefly review our theoretical framework proposed previously, memory-limited partially observable stochastic control (ML-POSC) \cite{tottori_memory-limited_2022,tottori_forward-backward_2023,tottori_decentralized_2023}, and compare it with OSC and OMSC introduced in this work. 
OSC and OMSC are introduced in Appendix \ref{sec: OSC} and \ref{sec: OMSC}, respectively. 
This section is organized as follows: 
In Appendix \ref{subsec: ML-POSC-PF}, we formulate ML-POSC, comparing it with OMSC. 
In Appendix \ref{subsec: ML-POSC-OCF}, we show that ML-POSC can be converted into OSC on the extended space, similar to OMSC. 

\subsection{Problem Formulation}\label{subsec: ML-POSC-PF}
In this subsection, we formulate ML-POSC \cite{tottori_memory-limited_2022,tottori_forward-backward_2023,tottori_decentralized_2023}, comparing it with OMSC. 
The dynamics of state $x_{t}\in\mb{R}^{d_{x}}$ in ML-POSC is the same with that in OMSC [Eq. (\ref{eq: OMSC-state SDE})], which is given by the following SDE: 
\begin{align}
	dx_{t}&=b(t,x_{t},u_{t})dt+\sigma(t,x_{t},u_{t})d\omega_{t},\label{eq: ML-POSC-state SDE}
\end{align}
where initial state $x_{0}$ follows $p_{0}(x_{0})$, $u_{t}\in\mb{R}^{d_{u}}$ is the control for the state, $\omega_{t}\in\mb{R}^{d_{\omega}}$ is a standard Wiener process. 
In contrast, the generation rule of observation $y_{t}\in\mb{R}^{d_{y}}$ in ML-POSC is different from that in OMSC. 
While observation $y_{t}$ is generated from probability density function $p_{t}(y_{t}|x_{t})$ as $y_{t}\sim p_{t}(y_{t}|x_{t})$ in OMSC [Eq. (\ref{eq: OMSC-observation PDF})], it is generated from the following SDE in ML-POSC: 
\begin{align}
	dy_{t}&=h(t,x_{t})dt+\gamma(t)d\nu_{t},\label{eq: ML-POSC-observation SDE}
\end{align}
where $\nu_{t}\in\mb{R}^{d_{\nu}}$ is a standard Wiener process. 
From this difference, the dynamics of memory $z_{t}\in\mb{R}^{d_{z}}$ in ML-POSC is formulated by 
\begin{align}
	dz_{t}=c(t,z_{t},v_{t})dt+\kappa(t,z_{t},v_{t})dy_{t}+\eta(t,z_{t},v_{t})d\xi_{t},\label{eq: ML-POSC-memory SDE}
\end{align}
where initial memory $z_{0}$ follows $p_{0}(z_{0})$, $v_{t}\in\mb{R}^{d_{v}}$ is the control for the memory, and $\xi_{t}\in\mb{R}^{d_{\xi}}$ is a standard Wiener process. 
Compared with the memory dynamics in OMSC [Eq. (\ref{eq: OMSC-memory SDE})], $\kappa(t,z_{t},v_{t})dy_{t}$ is an additional term in ML-POSC. 
While OMSC determines state control $u_{t}$ and memory control $v_{t}$ based on observation $y_{t}$ and memory $z_{t}$ [Eqs. (\ref{eq: OMSC-state control}) and (\ref{eq: OMSC-memory control})], ML-POSC determines them based solely on memory $z_{t}$ as follows: 
\begin{align}
	u_{t}&=u(t,z_{t}),\label{eq: ML-POSC-state control}\\
	v_{t}&=v(t,z_{t}).\label{eq: ML-POSC-memory control}
\end{align}
The rest of the formulation in ML-POSC is the same with that in OMSC. 
The objective function is given by the following expected cumulative cost function: 
\begin{align}	
	J[u,v]:=\mb{E}\left[\int_{0}^{T}f(t,x_{t},z_{t},u_{t},v_{t})dt+g(x_{T},z_{T})\right]. 
	\label{eq: OF of ML-POSC}
\end{align}
where $f$ and $g$ are the running and terminal cost functions, respectively. 
ML-POSC is a problem to find optimal state control function $u^{*}$ and optimal memory control function $v^{*}$ that minimize objective function $J[u,v]$ as follows: 
\begin{align}	
	u^{*},v^{*}:=\arg\min_{u,v}J[u,v].
	\label{eq: ML-POSC}
\end{align}

While OMSC encodes observation information into memory through memory control $v_{t}=v(t,y_{t},z_{t})$, ML-POSC encodes it through $\kappa(t,z_{t},v_{t})dy_{t}$. 
Since the optimization problem of $\kappa(t,z_{t},v_{t})dy_{t}$ lies outside the scope of the LQG problem \cite{tottori_memory-limited_2022}, ML-POSC cannot explicitly and efficiently investigate how the agent optimally encodes observation information into memory. 
OMSC addresses this issue by encoding observation information into memory through memory control $v_{t}=v(t,y_{t},z_{t})$. 

\subsection{Conversion to Observation-based Stochastic Control on Extended Space}\label{subsec: ML-POSC-OCF}
In this subsection, we show that ML-POSC can be converted into OSC on the extended space, which is similar to OMSC. 
We define extended state $\tilde{x}_{t}$, extended observation $\tilde{y}_{t}$, and extended control $\tilde{u}_{t}$ as follows: 
\begin{align}
	\tilde{x}_{t}:=\left(\begin{array}{c}
		x_{t}\\
		z_{t}\\
	\end{array}
	\right),\quad 
	\tilde{y}_{t}:=z_{t}',\quad 
	\tilde{u}_{t}:=\left(\begin{array}{c}
		u_{t}\\
		v_{t}\\
	\end{array}
	\right). 
\end{align}
We denote the memory in extended state $\tilde{x}_{t}$ by $z_{t}$ and that in extended observation $\tilde{y}_{t}$ by $z_{t}'$ to distinguish between them. 
In practice, $z_{t}=z_{t}'$ holds. 
Extended state $\tilde{x}_{t}$ evolves by the following SDE: 
\begin{align}
	d\tilde{x}_{t}=\tilde{b}(t,\tilde{x}_{t},\tilde{u}_{t})dt+\tilde{\sigma}(t,\tilde{x}_{t},\tilde{u}_{t})d\tilde{\omega}_{t}, 
\end{align}
where 
\begin{align}
	\tilde{b}(t,\tilde{x}_{t},\tilde{u}_{t})&:=\left(\begin{array}{c}
		b(t,x_{t},u_{t})\\
		c(t,z_{t},v_{t})+\kappa(t,z_{t},v_{t})h(t,x_{t})\\
	\end{array}
	\right),\\ 
	\tilde{\sigma}(t,\tilde{x}_{t},\tilde{u}_{t})&:=\left(\begin{array}{ccc}
		\sigma(t,x_{t},u_{t})&O&O\\
		O&\kappa(t,z_{t},v_{t})\gamma(t)&\eta(t,z_{t},v_{t})\\
	\end{array}
	\right),\\ 
	\tilde{\omega}_{t}&:=\left(\begin{array}{c}
		\omega_{t}\\
		\nu_{t}\\
		\xi_{t}\\
	\end{array}
	\right). 
\end{align}
Extended observation $\tilde{y}_{t}$ is generated from the following probability: 
\begin{align}
	\tilde{p}_{t}(\tilde{y}_{t}|\tilde{x}_{t}):=\delta(z_{t}'-z_{t}). 
\end{align}
Extended control $\tilde{u}_{t}$ is determined based on extended observation $\tilde{y}_{t}$ as follows: 
\begin{align}
	\tilde{u}_{t}=\tilde{u}(t,\tilde{y}_{t})
	:=\left(\begin{array}{c}
		u(t,z_{t})\\
		v(t,z_{t})\\
	\end{array}
	\right). 
\end{align}
The objective function is given by the following expected cumulative cost function: 
\begin{align}
	J[\tilde{u}]:=\mb{E}\left[\int_{0}^{T}\tilde{f}(t,\tilde{x}_{t},\tilde{u}_{t})dt+\tilde{g}(\tilde{x}_{T})\right], 
\end{align}
where 
\begin{align}
	\tilde{f}(t,\tilde{x}_{t},\tilde{u}_{t}):=f(t,x_{t},z_{t},u_{t},v_{t}),\quad 
	\tilde{g}(\tilde{x}_{T}):=g(x_{T},z_{T}).
\end{align}
Therefore, ML-POSC can be converted into OSC on the extended space, and its optimal solution can be obtained through OSC, similar to OMSC. 

\twocolumngrid
\bibliography{230601_MPT_ref}
\end{document}